\documentclass[11pt,a4paper]{article}
\usepackage{cite}

\usepackage{amsmath, amsthm,mathtools, setspace, url,amssymb,upgreek,slashed}
\usepackage[mathscr]{euscript}
\usepackage{ifpdf}
\ifpdf
  \usepackage[pdftex]{graphicx}
  \usepackage{epstopdf}
\else
  \usepackage[dvips]{graphicx}
\fi
\parskip=\baselineskip

\usepackage{hyperref}

\begin{document}
\begin{titlepage}
\begin{flushright}

\end{flushright}

\vskip 1.5in
\begin{center}
{\bf\Large{Probing Quantization Via Branes}}

\vskip
0.5cm { Davide Gaiotto} \vskip 0.05in {\small{ \textit{Perimeter Institute}
\vskip -.4cm
{\textit{Waterloo, Ontario, Canada N2L 2Y5}}
}}\vskip
0.5cm { Edward Witten} \vskip 0.05in {\small{ \textit{Institute for Advanced Study}\vskip -.4cm
{\textit{Einstein Drive, Princeton, NJ 08540 USA}}}
}
\end{center}
\vskip 0.5in
\baselineskip 16pt
\begin{abstract}  We re-examine quantization via branes with the goal of understanding its relation to geometric quantization.
If a symplectic manifold $M$ can be quantized in geometric quantization using a polarization ${\mathcal P}$, and in brane quantization using a complexification
$Y$, then the two quantizations agree if ${\mathcal P}$ can be analytically continued to a holomorphic polarization of $Y$.   We also show, roughly,
that the automorphism group of $M$ that is realized as a group of symmetries in brane quantization of $M$ is the group of symplectomorphisms of $M$
that can be analytically continued to holomorphic symplectomorphisms of $Y$. We describe from the point of view of brane quantization several
examples in which geometric quantization with different polarizations gives equivalent results.    \end{abstract}
\date{May, 2020}
\end{titlepage}
\def\SO{{\mathrm{SO}}}
\def\G{{\text{\sf G}}}
\def\la{\langle}
\def\ra{\rangle}
\def\sW{{\mathscr W}}
\def\dR{{\mathrm{dR}}}
\def\g{{\mathfrak g}}
\def\Re{{\mathrm{Re}}}
\def\Im{{\mathrm{Im}}}
\def\SU{{\mathrm{SU}}}
\def\SL{{\mathrm{SL}}}
\def\U{{\mathrm U}}
\def\M{{\mathcal M}}
\def\O{{\mathcal O}}
\def\Spinc{{\mathrm{Spin}}_c}
\def\sA{{\sf A}}
\def\sI{{\mathscr I}}
\def\sT{{\sf T}}
\def\sa{{a}}
\def\c{c}
\def\CPT{{\sf{CPT}}}
\def\GL{{\mathrm{GL}}}
\def\v{v}
\def\frak{\mathfrak}
\def\sff{{\sf f}}
\def\sG{{\sf G}}
\def\d{{\mathrm d}}
\def\CS{{\mathrm{CS}}}
\def\Z{{\Bbb Z}}
\def\veps{\varepsilon}
\def\zZ{{\mathcal Z}}
\def\DD{{\mathcal D}}
\def\R{{\Bbb R}}
\def\I{{\mathcal I}}
\def\F{{\mathcal F}}
\def\op{{\mathrm{op}}}
\def\cF{{\mathcal F}}
\def\J{{\mathcal J}}
\def\Bbb{\mathbb}
\def\Tr{{\rm Tr}}
\def\sF{{\sf F}}
\def\sB{{\sf B}}
\def\j{{\sf j}}
\def\16{{\bf 16}}
\def\1{{(1)}}
\def\bCP{{\Bbb{CP}}}
\def\2{{(2)}}
\def\3{{\bf 3}}
\def\4{{\bf 4}}
\def\frL{{\mathfrak L}}
\def\free{{\mathrm{free}}}
\def\sg{{\mathrm g}}
\def\J{{\mathcal J}}
\def\i{{\mathrm i}}
\def\h{\widehat}
\def\b{\overline}
\def\u{u}
\def\D{{\mathrm D}}
\def\Rf{{R}}
\def\K{{\sf K}}
\def\sp{{\sigma}}
\def\E{{\mathcal E}}
\def\O{{\mathcal O}}
\def\PF{{\mathit{P}\negthinspace\mathit{F}}}
\def\tr{{\mathrm{tr}}}
\def\be{\begin{equation}}
\def\ee{\end{equation}}
 \def\Sp{{\mathrm{Sp}}}
 \def\Spin{{\mathrm{Spin}}}
 \def\SL{{\mathrm{SL}}}
 \def\SU{{\mathrm{SU}}}
 \def\SO{{\mathrm{SO}}}
 \def\PGL{{\mathrm{PGL}}}
 \def\ll{\langle\langle}
 \def\RR{{\mathcal R}}
\def\rr{\rangle\rangle}
\def\la{\langle}
\def\CP{{\mathrm{CP}}}
\def\sCP{{\sf{CP}}}
\def\I{{\mathcal I}}
\def\ra{\rangle}
\def\T{{\mathcal T}}
\def\V{{\mathcal V}}
\def\bar{\overline}
\def\spinc{{\mathrm{spin}_c}}
\def\dim{{\mathrm{dim}}}
\def\v{v}
\def\Pic{{\mathrm{Pic}}}
\def\gq{{\mathrm{gq}}}
\def\RP{{\Bbb{RP}}}

\def\tilde{\widetilde}
\def\t{\widetilde}
\def\R{{\Bbb{R}}}\def\Z{{\Bbb{Z}}}
\def\N{{\mathcal N}}
\def\B{{\mathcal B}}
\def\H{{\mathcal H}}
\def\hat{\widehat}
\def\Pf{{\mathrm{Pf}}}
\def\bM{{\overline\M}}
\def\PSL{{\mathrm{PSL}}}
\def\Im{{\mathrm{Im}}}
\font\teneurm=eurm10 \font\seveneurm=eurm7 \font\fiveeurm=eurm5
\newfam\eurmfam
\textfont\eurmfam=\teneurm \scriptfont\eurmfam=\seveneurm
\scriptscriptfont\eurmfam=\fiveeurm
\def\eurm#1{{\fam\eurmfam\relax#1}}
\font\tencmmib=cmmib10 \skewchar\tencmmib='177
\font\sevencmmib=cmmib7 \skewchar\sevencmmib='177
\font\fivecmmib=cmmib5 \skewchar\fivecmmib='177
\newfam\cmmibfam
\textfont\cmmibfam=\tencmmib \scriptfont\cmmibfam=\sevencmmib
\scriptscriptfont\cmmibfam=\fivecmmib
\def\cmmib#1{{\fam\cmmibfam\relax#1}}
\numberwithin{equation}{section}

\def\neg{\negthinspace}
\def\d{\mathrm d}
\def\C{{\Bbb C}}
\def\cc{{\mathrm{cc}}}
\def\HH{{\mathbb H}}
\def\P{{\mathcal P}}
\def\NS{{\sf{NS}}}
\def\Ra{{\sf{R}}}
\def\sV{{\sf V}}
\def\Hom{{\mathrm{Hom}}}
\def\Z{{\Bbb Z}}

\def\A{{\mathscr A}}
\def\cA{{\mathcal A}}
\def\ltwo{{\mathrm L}^2}
\def\SS{{\mathcal S}}
\def\bar{\overline}
\def\sc{{\mathrm{sc}}}
\def\Max{{\mathrm{Max}}}
\def\CS{{\mathrm{CS}}}
\def\ga{\gamma}
\def\bg{\bar\ga}
\def\pP{{\mathscr P}}
\def\S{{\mathfrak L}}
\def\W{{\mathcal W}}
\def\M{{\mathcal M}}
\def\bM{{\overline \M}}
\def\L{{\mathcal L}}
\def\sM{{\sf M}}
\def\gst{\mathrm{g}_{\mathrm{st}}}
\def\gstt{\widetilde{\mathrm{g}}_{\mathrm{st}}}
\def\hbbar{\pmb{\hbar}}
\def\G{{\mathcal G}}
\def\U{{\mathcal U}}
\def\UU{{\mathrm U}}
\def\ad{{\mathrm{ad}}}
\def\TAut{{\mathrm{TAut}}}
\def\Aut{{\mathrm{Aut}}}

\def\be{\begin{equation}}
\def\ee{\end{equation}}

\tableofcontents

\section{Introduction}\label{intro} 

There is no entirely satisfactory general theory of how to quantize a classical symplectic manifold, because there is an anomaly in the passage from 
classical mechanics to quantum mechanics. The commutators of quantum operators do not agree with the Poisson brackets of classical functions on phase
space, except to lowest order in $\hbar$ \cite{G46,V51,Todorov}.  
In practice, quantization is usually carried out by, roughly, splitting the phase space coordinates into positions
and momenta and defining a Hilbert space of functions of the positions.   In geometric quantization \cite{Ko,So,Wo,Snia,Hall}, this process is formalized in the choice of a
``polarization'' of the phase space.  The resulting quantum theory does generically depend on the choice of polarization.    In special cases when different
polarizations do lead to equivalent quantum theories, this often leads to important results.   

An alternative to geometric quantization is quantization via branes.  This was proposed in \cite{GW} following prior developments
that had indicated a relationship between the $A$-model of two-dimensional topological field theory and deformation quantization \cite{KO,BrS,Ka,Pe,Gu} and an analysis 
of some informative examples \cite{AZ}.   
Deformation quantization \cite{Many,WL, Fedosov, Kon,CF,CF2,SL}, which is an important part of the $A$-model story,
 is a cousin of quantization
that  associates to a classical phase space a 
quantum-deformed algebra of functions, but not a Hilbert space on which the algebra acts.  As formulated in \cite{Fedosov}, deformation quantization
involves the choice of a  symplectic connection rather than a polarization. 

In quantization via branes, the starting point, rather than the choice of a polarization of a classical phase space $M$, is the choice of a suitable complexification $Y$
of $M$.   $Y$ should be a complex symplectic manifold, with an antiholomorphic involution that has $M$ as a component of its fixed point set, and furthermore
$Y$ should have a well-defined $A$-model.   Then a quantum Hilbert space of $M$ can be defined as the space of $A$-model states of
$(\B,\B_\cc)$ strings, where $\B$ is a Lagrangian $A$-brane supported on $M$ and $\B_\cc$ is a certain coisotropic brane of the $A$-model.
Coisotropic branes were  originally introduced in \cite{KO} and studied in a number of later papers \cite{Ka,Pe,Gu,AZ,KW,Herbst}.    

 In geometric quantization of $M$ based on a polarization $\P$, the functions on $M$ that can be naturally quantized to quantum operators are the ones that via Poisson
brackets generate a canonical transformation of $M$ that preserves the polarization.   In quantization via branes using a complexification $Y$ of $M$, the functions on
$M$ that can be naturally quantized are the ones that can be analytically continued to holomorphic functions on $Y$.  Usually $Y$ is not compact and one
places a condition on the growth of a function at infinity on $Y$.  

We were led to reconsider quantization by branes because of  recent developments involving an analytic version of the geometric Langlands correspondence
\cite{EFK,EFK2,EFK3,T}.    As we describe elsewhere \cite{GaW}, quantization by branes is the additional ingredient that is needed to 
extend the gauge theory approach
to geometric Langlands \cite{KW} to the analytic version of this subject.

In Section \ref{two} of this article, we briefly review the difficulty in making a general theory of quantization, and then briefly recall  geometric quantization and 
deformation quantization.  Then we describe in more detail
the idea of quantization by branes.   We explain some matters more precisely than in \cite{GW}, but on some topics the reader will find more detail in that previous article.  

In Section \ref{comparing}, we explore the relation between geometric quantization and  quantization by branes. 
If the same phase space $M$ can be quantized in geometric quantization using a polarization $\P$ and in brane quantization using a complexification $Y$ of $M$,
can one state conditions on $\P$ and $Y$ such that the two methods of quantization will agree?  We formulate conditions under which this is the case;
roughly $\P$ should analytically continue to a holomorphic polarization $\varPi$ of $Y$.

In Section \ref{symq}, we explore the use of Lagrangian correspondences between $M$ and itself, rather than functions on $M$, to define operators on the Hilbert
space obtained by brane quantization of $M$.   A Lagrangian correspondence is  simply a Lagrangian submanifold of $M\times M$ (where the symplectic structures
on the two factors are equal and opposite).  There are at least two reasons to define operators by quantizing correspondences rather than functions.
First, this gives an effective framework to study the symmetries of brane quantization.  Second, in  some important examples, including the case
relevant to geometric Langlands,  the complexification $Y$ of $M$ has relatively few 
holomorphic functions  and instead there
are very important operators associated to correspondences.

In Section \ref{qcomp}, we consider the special case of quantization of a complex symplectic manifold $Y$ viewed as a real symplectic manifold.   This case
is important in an application to geometric Langlands \cite{GaW}.

\section{Review Of Quantization Via Branes}\label{two}

We begin with a review of quantization via branes, referring the reader to \cite{GW} for a somewhat different exposition and more detail on some points.

\subsection{The Problem}\label{problem}

Our starting point is a real symplectic manifold $M$ of  dimension $2n$, with symplectic form $\omega$.   We assume that  $M$, which is known
as the phase space,  is endowed with\footnote{We here postpone the discussion of a ``metaplectic correction'' that necessitates
a slight refinement of this statement.}  a ``prequantum line
bundle,'' which is a complex line bundle $\frL\to M$   that has a unitary connection with curvature $\omega$.   
Such an $\frL$ exists if and only if the periods of $\omega$ are valued in $2\pi\Z$. $\frL$ is unique up to isomorphism
if in addition $H^1(M,\UU(1))=0$.   Quantization will depend on the choice of $\frL$.

 Roughly speaking, the idea of quantization is to associate to this data a Hilbert space $\H$, such that functions on $M$ 
 become operators on $\H$, with Poisson brackets of functions 
corresponding to commutators of operators, in the sense that if $\h f$ is the operator corresponding to a function $f$,
then
\be\label{goal} [\h f,\h g]=-\i\hbar \h {\{f,g\} }. \ee  
 (here $\{~,~\}$ is the Poisson bracket of functions, $[~,~]$ is the commutator of operators, and $\hbar$ is
Planck's constant).   Real functions on $M$ are supposed to correspond to hermitian operators on $\H$.  
The dimension of $\H$ is supposed to be roughly the symplectic volume 
$\frac{1}{(2\pi)^n}\int_M e^\omega$, and in particular this dimension is infinite if and only if $M$ has infinite volume.

Without additional structure, the problem of quantization does not have a natural solution.  It is not possible to satisfy
eqn. (\ref{goal}) (in a Hilbert space of appropriate ``size'') except for special classes of functions.   In quantum mechanics textbooks, this is called
the ``operator ordering problem,'' but it is not always emphasized that this problem does not have a general solution.
There is an anomaly in the passage from classical mechanics to quantum mechanics. 

The natural symmetries of classical mechanics are  generated by real functions on $M$, viewed as Hamiltonian functions.
A  function $f$ generates a symplectic transformation of $M$ via the vector field $v^i=\omega^{ij}\frac{\partial f}{\partial x^j}$
(here the $x^i$ are local coordinates on $M$, $\omega=\frac{1}{2}\omega_{ij}\d x^i \d x^j$, and $\omega^{ij}\omega_{jk}=\delta^i{}_k$).
Such vector fields are called Hamiltonian vector fields and 
generate a group $\G_0$.   Elements of this group are called exact symplectomorphisms.\footnote{If
$M$ has a positive first Betti number, it also has sympectomorphisms that are not exact; their generators are multi-valued Hamiltonians.   Note
that elements of $\G_0$ are automatically homotopic to the identity, since such an element is obtained, roughly, by exponentiating the action of a Hamiltonian function.
Symplectomorphisms of $M$ that are not homotopic to the identity will be included in a moment when we replace $\G_0$ with $\G$.}
In passing from $f$ to $v^i=\omega^{ij}\partial_j f$, we have lost some information, because $v^i$ is invariant under adding a constant to $f$.
To match classical mechanics with quantum mechanics as accurately as possible, one should keep track of this constant, which one
can do by considering the symmetries not of $M$ but of $M$ together with its prequantum line bundle $\frL$.   The symmetries of the 
pair $(M,\frL)$ make a group $\G$ that we will view as the symmetry group of classical mechanics.  
Concretely, an element of $\G$ is a symplectomorphism $\varphi:M\to M$ together with an isomorphism $\beta:\frL\to \varphi^*(\frL)$.
Including the isomorphism as part of the definition may be unfamiliar, but actually 
 leads to the closest possible -- though still highly imperfect -- match between the symmetries of classical and quantum mechanics.
The component $\G_1$ of  $\G$ that contains the identity 
is a central extension by $\UU(1)$ of the group $\G_0$ of exact symplectomorphisms:
\be\label{centext}1\to \UU(1)\to \G_1\to \G_0\to 1. \ee
Here $\UU(1)$ acts trivially on $M$ and acts on the fiber of $\frL\to M$ as multiplication by a complex constant $e^{\i\alpha}$ of modulus 1.
$\G$ in general also has components that are not continuously connected to the identity.

By contrast, the natural symmetry group of quantum mechanics is the group
$\U$ of unitary  transformations of the Hilbert space $\H$.   The groups $\G$ and $\U$  do not coincide and the discrepancy between them
is not just a matter of a central extension.  $\G$ contains information on 
the topology of the phase  space $M$ and $\U$ does not.  This is an anomaly in the usual sense that the term is used in quantum field theory: 
the symmetry group of a classical system is modified when one quantizes the system.     

For example, consider $M=\R^{2n}$ with its standard symplectic structure.   Suppose one is given a notion of what is meant
by a function on $\R^{2n}$ that is a polynomial of degree $\leq 1$.   It is important that there is no natural notion of this; having
such a notion is equivalent to being given a distinguished set of coordinates 
 $\vec x =(x^1,x^2,\cdots, x^{2n})$, defined up to a transformation $\vec x
\to A\vec x +\vec b$ with a symplectic matrix $A$ and constant $b$.   Such a transformation is called an affine linear transformation,
and if $M$ is given a set of coordinates that is uniquely defined up to an affine linear transformation, we say that $M$ has an affine
linear structure.  If we are given an affine linear structure on $\R^{2n}$, then any nonlinear  symplectomorphism of $\R^{2n}$ will
map this affine linear structure to a different one. So the space of affine linear structures on $\R^{2n}$ is infinite-dimensional.

In a given affine linear coordinate system, the
 Hamiltonian functions that generate affine linear transformations are the polynomials of degree $\leq 2$.   For polynomials
of degree $\leq 2$, there is no anomaly.   This is a well-known fact, possibly not with precisely this phrasing.   It is equivalent to saying
that affine linear changes of the phase space coordinates are realized as symmetries in quantum mechanics.\footnote{\label{cext} There is a double cover
involved.   The linear transformations of phase space $\vec x\to A\vec x$ form a group $\Sp(2n,\R)$; the corresponding quantum symmetry
group is a double cover of this.  In $\R^2$, an example of a quadratic Hamiltonian is the harmonic oscillator Hamiltonian $H=\frac{1}{2}(p^2+q^2)$,
which generates a rotation of $\R^2$.
The quantum symmetry group is a double cover of the classical symmetry group because
 the energy levels of $H$ are half-integers, so $\exp(2\pi\i H)$,
which corresponds to the identity element of $\G_0$, acts as $-1$ on the quantum Hilbert space.}  (For example, the Fourier transform exchanges coordinates and momenta.)  However, an anomaly arises
for polynomials of degree $>2$; no operator ordering or addition of lower order terms can make their quantum commutators agree
with their classical Poisson brackets.   This rather old result  \cite{G46,V51,Todorov} is demonstrated explicitly in Appendix \ref{anomaly}
for the case $n=2$.

Once an affine linear structure is given, one can quantize $\R^{2n}$  in a familiar way, for example by splitting the $x$'s into $p$'s and $q$'s
(chosen so that  $\omega = \sum_i \d p_i \d q^i$) and
defining a Hilbert space consisting of $\ltwo$  functions of the $q$'s.   Standard arguments shows that it does not really matter how
one makes this splitting; there are natural identifications between the resulting Hilbert spaces (up to an overall factor $\pm 1$, mentioned
in footnote \ref{cext}).  This statement is actually equivalent to the statement that there is no anomaly in the commutators of
polynomials of degree $\leq 2$.   However, there are no natural equivalences, even up to $c$-number factors,
 between the Hilbert spaces that are constructed  starting
with different affine linear structures.   If there were, one would end up proving that the symmetry groups $\G$ and $\U$ of classical
and quantum mechanics are the same, which is not the case.

So any method of quantization requires additional structure on $M$, beyond its symplectic structure and prequantum line bundle.
That is why Ludwig Faddeev used to say  that quantization is ``an art, not a science.''   However, there is a good reason that these
matters may not be familiar.    Though there are some
notable exceptions,  in physics the phase space usually has enough additional 
structure (such as a natural affine linear structure) 
that the issues described here are not prominent.  

\subsection{Polarizations}\label{gq}

In practice,  quantization is usually carried out by separating the phase space variables into coordinates and momenta.   Then one defines
a Hilbert space consisting of functions of the coordinates.   We discussed this already in the case of $\R^{2n}$.
In geometric quantization \cite{Ko,So,Wo,Snia,Hall}, this process is formalized in terms of a choice of ``polarization.''   Though there are more 
general possibilities, two basic cases, and the primary cases that we will consider in this article, are a real polarization and a 
complex polarization.

In the simplest form of a real polarization, one identifies $M$ as the total space of a  cotangent bundle $T^*N$, for some $N$.
One assumes that the symplectic structure of $M$ is the natural symplectic structure of $T^*N$ as a cotangent bundle, or the
sum of this and the pullback of a closed two-form on $N$.   In terms of local coordinates $q^i$, $i=1,\cdots,n$ on $N$,
this means that the symplectic form is $\omega =\sum_i \d p_i \d q^i +\frac{1}{2}\sum_{i,j}\alpha_{ij}(q) \d q^i\d q^j$, where $p_i$ are suitable linear 
functions on the fibers of $\pi:T^*N\to N$, and $\alpha=\frac{1}{2}\sum_{i,j}\alpha_{ij}(q) \d q^i\d q^j$ is a closed two-form on $N$.   More briefly,
$\omega=\sum_i \d p_i \d q^i+\pi^*(\alpha)$ where $\pi:M=T^*N\to N$ is the natural projection.
In this situation, the $q^i$ are called coordinates and the $p_i$ are called momenta. 
The fibers of the cotangent bundle are called the leaves of the polarization.
 One defines a Hilbert space $\H$
consisting of half-densities\footnote{In a  system of local coordinates $\vec q=(q^1,q^2,\cdots, q^n)$ on a real manifold $N$, a density is just a real-valued
function $\sigma(\vec q)$.  Under  a change of local coordinates, $\sigma$ transforms in such a way that the measure $|\d q^1\d q^2 \cdots \d q^n |\sigma(\vec q)$ is invariant.
A half-density is an object $\psi(\vec q)$ such that $\sigma=\psi^2$ is a density.     Thus, in a given coordinate system, $\psi(\vec q)$ is just a function, and in 
a change of  local coordinates, $\psi(\vec q)$ transforms in such as way that the measure $|\d q^1 \d q^2 \cdots \d q^n| \,\psi(\vec q)^2$ is invariant.   There is
a trivial real line bundle $\K\to N$ whose sections are (real) densities; $\K$ has a square root $\K^{1/2}$, also trivial, whose sections are half-densities.   
We generally write $\K$ for the real bundle of densities on a real manifold $N$, $K$ for the canonical bundle of a complex manifold $X$, and $\bar K$ for the complex
conjugate of $K$ (equivalently, the canonical bundle of $X$ viewed as a complex manifold with opposite complex structure).   If a complex manifold
$X$ is viewed as a real manifold,
then its bundle of densities $\K$ is related to its canonical bundle $K$ by $\K=K\otimes \bar K=|K|^2$.}
on $N$ with values in $\frL$, the prequantum line bundle.   In other words, if $\K^{1/2}$ is the real line bundle
over $N$ whose sections are half-densities, then a vector in $\H$ is represented by a section $\psi$ of $\K^{1/2}\otimes \frL\to N$, with the inner
product $\la\psi,\chi\ra=\int_N \bar\psi \chi$.     This definition is sometimes restated as follows.
 First pull back $\K^{1/2}\to N$  to a real line bundle over $T^*N$,
which we also call $\K^{1/2}$.   Then a vector in $\H$ can be described as a section of $\K^{1/2}\otimes \frL$ over $T^*N$ that
is covariantly constant when restricted to any fiber of $T^*N\to N$.   This makes sense because (since its curvature is $\omega$) $\frL$
is flat when restricted to any leaf of the polarization, and $\K^{1/2}$, as it is a pullback from $N$, also has a natural flat connection along each leaf.  The Hilbert
space inner product is then described as an integral $\la\psi,\chi\ra=\int_\sigma \bar\psi\chi$, where $\sigma$ is  any section of $T^*N\to N$.

Now we discuss complex polarizations.
A complex polarization of $M$ is defined as a choice of complex structure  $J$ on $M$ such that the symplectic form $\omega$ is of type $(1,1)$, so
that the prequantum line bundle $\frL$ becomes a holomorphic line bundle.    If $\omega$ is positive with respect to $J$ as well as of type $(1,1)$, and so represents
a Kahler form on $M$, then the polarization is called a Kahler polarization.  This is the most frequently considered case in practice, though we do not
wish to restrict to this case.   
Let $K$ be the canonical line bundle of $M$ and let $K^{1/2}$ be a square root of $K$, assuming momentarily that such a square root exists.
  Usually in Kahler quantization, one considers a situation in which  $\frL$ is ``sufficiently ample''
so that $H^r(M,K^{1/2}\otimes \frL)=0$ for $r>0$.   (This is always true sufficiently near the classical limit, in other words if the periods of $\omega$
are sufficiently large.)    
Then one defines a Hilbert space of holomorphic sections of\footnote{In the case of a real polarization, the bundle $\K^{1/2}$ of real half-densities is part of the
definition of a quantum state because otherwise the definition $\la\psi,\chi\ra=\int_N\bar\psi\chi$ of the Hilbert space inner product would not make sense.
In the case of a Kahler polarization, the canonical bundle $K$ has a hermitian metric that comes from the Kahler metric of $M$, so a Hilbert space inner product
could be defined whether or not one includes the factor of $K^{1/2}$ in the definition (\ref{cph}).   This factor is usually included for two reasons: (1) this makes
possible a more uniform description of geometric quantization for polarizations of different types; (2) it leads to more natural results in some cases, such as
the computation of the ground state energy of a harmonic oscillator quantized with a Kahler polarization of the phase space.   See for example Section 23.7 of \cite{Hall}.} $K^{1/2}\otimes \frL$,
\be\label{cph}\H=H^0(M,K^{1/2}\otimes \frL),\ee  the Hilbert space inner product being
\be\label{inpro}\la s,s'\ra=\int_M \d\mu \, \bar s s',\ee
where $\d\mu$ is the symplectic measure on $M$ and the Kahler metric on $K$ and unitary structure of $\frL$ are used to turn $\bar s s'$ into a real-valued
function.  The generalization of this if  $\omega$ is not Kahler or $\frL$ is not sufficiently ample is discussed in Section
\ref{branecp}.

Built into the last paragraph is that there is an anomaly in geometric quantization, related to what is known as the metaplectic correction.   In general
the object $K^{1/2}$ that was used in the definition of the Hilbert space does not exist as a complex line bundle, and it may not be  unique if it exists.   
If $K^{1/2}$ exists, it defines a spin structure on $M$; conversely, a spin structure determines a choice of $K^{1/2}$.  The obstruction to $M$ admitting
a spin structure is $w_2(M)$, the second Stieffel-Whitney class of $M$.
However, on
any complex manifold,
$K^{1/2}$ always exists canonically as a $\spinc$ structure.   A $\spinc$ structure in general is the data required to define a Dirac operator
acting on a spin 1/2 field of ``charge 1.''    On a complex manifold of dimension $n$, the Dirac operator that corresponds to the canonical $\spinc$
structure $K^{1/2}$ is the $\bar\partial$ operator acting on $\oplus_{r=0}^n\Omega^{0,r}(M)$, where $\Omega^{0,r}(M)$ is the bundle of $(0,r)$-forms.
   Since the obstruction to spin is of order 2,
the tensor product of two $\spinc$ structures is an ordinary complex line bundle.   Hence if the ``prequantum line bundle $\frL$'' of geometric quantization is
really a $\spinc$ structure, then $K^{1/2}\otimes \frL$ is an ordinary complex line bundle and the definition (\ref{cph}) of the quantum Hilbert space makes sense.

More generally, if $\SS$ is any $\spinc$ structure  on a manifold $M$ (we do not assume here that $M$ is a complex manifold
or that $\SS$ is related to a prequantum line bundle), then $\SS^2$ is an ordinary line bundle.
This line bundle has a first Chern class $x=c_1(\SS^2)$ and this is an integer lift of $w_2(M)$, in the sense that the reduction mod 2
of $x\in H^2(M,\Z)$ is $w_2(M)\in H^2(M,\Z_2)$.    If $\SS$ is a $\spinc$ structure with connection, then the ordinary line bundle $\SS^2$ has a connection
and curvature; we define the curvature of $\SS$ to be one-half the curvature of $\SS^2$.   We call a  $\spinc$ structure {\it flat} if this curvature vanishes.
With this definition, the condition for $\SS$ to admit a flat connection is simply that $x$ should be torsion; in other words, $M$ has a flat $\spinc$ structure
if $w_2(M)$ has an integer lift that is torsion.  

With these definitions, a ``prequantum line bundle'' $\frL$ over a Kahler manifold $M$  with Kahler form $\omega$ should  really be defined as
a $\spinc$ structure on $M$ of curvature $\omega$.

Once we reinterpret the ``prequantum line bundle'' of geometric quantization as a $\spinc$ structure, we should ask if this creates any difficulty in the
discussion of the quantization of a cotangent bundle $M=T^*N$.  In fact, the anomaly does not affect the classification of prequantum line bundles
over $T^*N$, because $T^*N$ always has a canonical flat $\spinc$ structure $\SS_0$, and therefore a $\spinc$ structure on $T^*N$ with curvature $\omega$
is simply $\SS_0\otimes \L$, where $\L$ is an ordinary complex line bundle with a unitary connection of curvature $\omega$.

 To explain this, let us first understand  topologically why $T^*N$ admits a flat $\spinc$ structure.
The tangent bundle to $T^*N$ is topologically the pullback to $T^*N$ of the direct sum of two copies of $TN$,  the tangent bundle to $N$.   The Whitney sum formula
gives $w_2(TN\oplus TN)= w_1(TN)^2$, where $w_1(TN)$, usually just denoted as $w_1(N)$, is the obstruction to an orientation of $N$.    So $w_2(T^*N)$
is the pullback to $T^*N$ of $w_1(N)^2$.  
Let $\varepsilon=\det \,TN$ be the orientation bundle of $N$, a real line bundle.  Then $w_1(N)=w_1(\varepsilon)$.
The object  $\h\varepsilon=\varepsilon\oplus \varepsilon\cong \varepsilon\otimes_\R\C$
is a complex line bundle over $N$.   Its first Chern class $x=c_1(\h\varepsilon)\in H^2(N,\Z)$ reduces mod 2 to  $w_2(\h\varepsilon)=w_1(\varepsilon)^2=w_2(T^*N)\in
H^2(N,\Z_2)$.
$x$ is certainly a torsion class, since $\h\varepsilon$, as the direct sum of real line bundles, admits a flat connection with finite structure group.   So $x$ is
an integer lift of $w_2(T^*N)$ which moreover is torsion, and therefore $T^*N$ admits a flat $\spinc$ structure.

To explicitly construct a flat $\spinc$ structure on $T^*N$, we construct a flat $\spinc$ structure on the bundle $TN\oplus TN\to N$; its pullback to $T^*N$
is the desired flat $\spinc$ structure on $T^*N$.    Picking a local orthonormal frame of $TN$, we introduce two sets of gamma matrices $\psi_k, \chi_k$,
$k=1,\cdots,n$, satisfying $\{\psi_k,\psi_l\}=\{\chi_k,\chi_l\}=2\delta_{kl}$, $\{\psi_k,\chi_l\}=0$.   A $\spinc$ structure on $TN\oplus TN$
is a vector bundle on which this Clifford algebra acts irreducibly.    Introduce annihilation and creation operators
$a_k=(\psi_k+\i \chi_k)/2,$ $a^\dagger_k=(\psi_k-\i\chi_k)/2$, so $\{a_k,a^\dagger_l\}=\delta_{kl}$,
$\{a_k,a_l\}=\{a^\dagger_k,a^\dagger_l\}=0$.  Locally we can construct a module for the Clifford algebra by starting with a state $|\negthinspace\downarrow\ra$ annihilated by the
$a_k$ and adding states $a^\dagger_{k_1}a^\dagger_{k_2}\cdots a^\dagger_{k_r}|\negthinspace\downarrow\ra$, $r=1,\cdots, n$.  In particular, the state
$|\negthinspace\uparrow\ra=a_1^\dagger a_2^\dagger\cdots a_n^\dagger|\negthinspace\downarrow\ra$ is annihilated by the $a^\dagger_k$.  If $N$ is orientable, this
construction gives a canonical spin structure on $T^*N$.    If $N$ is unorientable, there is an asymmetry globally
between the two states $|\negthinspace\downarrow\ra$ and $|\negthinspace\uparrow\ra$; if one of them is  a section of a trivial bundle over $N$, then the other is a section of $\h\varepsilon$.
The asymmetry implies\footnote{The structure group of $TN\oplus TN$ is $\SO(2n)$, and an associated spin bundle will have structure group $\Spin(2n)$.
Such a spin bundle will have a nondegenerate $\Spin(2n$)-invariant quadratic form (symmetric or antisymmetric depending on $n$).  This invariant form
will restrict to a nondegenerate pairing $|\negthinspace\uparrow\ra \otimes |\negthinspace\downarrow\ra\to \C$.   If $|\negthinspace\downarrow\ra$ is
a section of a line bundle $\ell$, then $|\negthinspace\uparrow\ra$ is a section of $\ell\otimes \varepsilon$, so the existence of the pairing implies
that $\ell^2\otimes \varepsilon$ is trivial, so that $2 c_1(\ell)+x=0$.   This implies that the mod 2 reduction of $x$ vanishes; since the mod 2 reduction of $x$
is $w_2(T^*N)$, we learn, as expected, that the construction in the text gives a spin bundle only if $w_2(T^*N)=0$. Otherwise it gives a flat $\spinc$ bundle.  } that what we have constructed is a $\spinc$ bundle rather than a spin bundle; however, because $\h\varepsilon$ has a natural flat
structure, this  $\spinc$ structure is flat.   So the general statement is that $T^*N$ always has a canonical flat $\spinc$ structure.

The main limitation of geometric quantization is that, if a polarization of $M$ exists, then there are many different polarizations, since any generic
element of the symmetry group $\G$
will map a polarization $\P$ to a different polarization $\P'$.  Under reasonable conditions, 
one can define relatively natural maps between the Hilbert spaces defined using different polarizations.\footnote{\label{BKS}This is done 
using the Blattner-Kostant-Sternberg kernel of geometric quantization. The definition 
 is most straightforward for two polarizations that are suitably transverse.  See Section 23.8 of \cite{Hall} for
a brief introduction and \cite{Snia} for much more detail. See also footnote \ref{BKStwo} below.}   
But these maps do not obey the conditions
that one would want in order to establish equivalences between the different quantizations.   What one would want is the following.
Given polarizations $\P_i$ that lead to Hilbert spaces $\H_i$, one would like a family of unitary maps $\Phi_{ij}:\H_i\to \H_j$,  with
$\Phi_{ii}=1$, satisfying $\Phi_{ij}\Phi_{ji}=1$ (so that the $\Phi_{ij}$ are isomorphisms) and $\Phi_{ij}\Phi_{jk}=\Phi_{ik}$ (so
that these isomorphisms between different Hilbert spaces are compatible).    But these conditions are generically not satisfied; that
statement is another expression of the anomaly in the passage from classical to quantum physics.

The condition for compatibility between the different Hilbert spaces could be slightly relaxed to allow 
$\Phi_{ij}\Phi_{jk}=c_{ijk} \Phi_{ik}$, where $c_{ijk}$ is a central factor, valued in $\UU(1)$.  In such a case, one would say that the $\H_i$
are ``projectively equivalent,'' meaning that there are equivalences between them that are naturally defined up to a 
scalar factor of modulus 1. This slightly more relaxed condition  for equivalence between different polarizations is 
also generically not satisfied, since  
the anomaly in the passage from classical to quantum mechanics does not simply involve a central extension. 

There is no general theory of when different polarizations lead to equivalent quantizations, but there are a number of important cases in which this
does occur.     The most familiar and important example arises in quantizing $M=\R^{2n}$.  In this case, as we have already remarked,
once an affine linear structure is given,
one does get a projective
equivalence between the Hilbert spaces defined with different linear polarizations.
   A linear polarization of $\R^{2n}$ is, in the real case,
a separation of the linear functions on $\R^{2n}$ into coordinates and momenta, or in the complex case, a choice of which complex-valued  linear
functions on $\R^{2n}$ should be considered ``holomorphic.''     In this article, we will see a few additional examples in which different
polarizations  lead to quantizations that are at least projectively equivalent.   In each case this leads to important results.

A more exotic limitation of geometric quantization is that there are symplectic manifolds that do not admit any polarization \cite{Gotay}.    Can such a symplectic manifold
be quantized in any sensible way, at least in some cases?   That question is one motivation for seeking an alternative to geometric quantization.  

\subsection{A Formal Procedure And Its Uses}\label{uses}

In this section, we discuss a formal recipe for quantization that does not avoid the  
anomaly, which indeed is inescapable, but nonetheless provides useful orientation for arguments we give later.   
If we are given a prequantum line bundle $\frL$ with connection $\sT/\hbar$ and curvature $\omega=\d \sT /\hbar$ over a classical phase space $M$, 
then it is possible to write a classical action whose associated path integral would formally solve the problem of quantization of $M$.
For a path $\gamma\subset M$ between points $r,r'\in M$, the action is simply
\be\label{wiff}\sI=\frac{1}{\hbar}\int_\gamma \sT    .\ee
Alternatively, let $I=[0,1]$ be the unit interval and view $\gamma$ as the image of a map $x:I\to M$.   Then the action can be written
\be\label{iff} \sI =\frac{1}{\hbar}\int_I  x^*(\sT ). \ee
Note that $\mathscr I$ is invariant under reparametrization of the interval $I$.
If the endpoints $0,1$ of $I$ are mapped to points $r,r'\in M$, then 
to define the action $\sI$ as a number, one needs to trivialize the fibers of $\frL$ at those points.  We denote these fibers as $\frL_r$, $\frL_{r'}$.
Let  $\pP_{r,r'}$ be the space of maps $x:I\to M$ with the endpoints mapped to $r,r'$. 
Consider the Feynman integral
\be\label{hiff} Z(r,r')=\int_{\pP_{r,r'}}\D x \,\exp(\i \sI). \ee
For a fixed $x$ (before trying to integrate over $x$) the integrand  $e^{\i \sI}$ of the path integral is a map from 
$\frL_r$ to $\frL_{r'}$, and one could
hope that the integral over $x$ would give a  map $Z_{r,r'}:\frL_r\to \frL_{r'}$ 
from which the Hilbert space that quantizes the phase space
$M$ could be constructed.   

This cannot really be expected to work as stated because the points $r$ and $r'$ would correspond here to initial and final states, but initial and 
final quantum states should only depend on one-half of the phase space variables.   It should
not be possible without more input to construct a map $Z_{r,r'}$ associated to two arbitrary points in $M$.
In fact the symmetry group $\G$ of classical mechanics (exact symplectomorphisms with a lift to an action on $\frL$, as described in Section \ref{problem})
is so large that there can be no $\G$-invariant definition of $Z_{r,r'}$.   The group $\G$ contains elements that map any pair $r,r'$ of distinct points to any other
such pair, with arbitrary lifts to the fibers of $\frL$. So in order to be $\G$-invariant, $ Z_{r,r'}$ would have to be $\delta(r,r')$ times a constant map on the fibers
of $\frL$.  This is not a sensible answer.  Formally, one would interpret $Z_{r,r'}$ as
\be\label{nutto} Z_{r,r'}\stackrel{?}{=} \sum_{\alpha\in S} \bar\psi_\alpha(r) \psi_\alpha(r'), \ee
where the states $\psi_\alpha$, $\alpha\in S$, form a basis of the Hilbert space.   To reconcile such a formula with $Z_{r,r'}\sim\delta(r,r')$,
one needs a Hilbert space of infinite dimension, even if the phase space $M$ is compact.  This is not the expected behavior of quantum theory.   Moreover for $M=\R^2$
where we know what quantization should mean, there is no formula like (\ref{nutto}) (bearing in mind that $r,r'$ are phase space points, labeled by both position
and momentum).    Concretely, if we try to compute $Z_{r,r'}$ in perturbation theory, the first thing we find is that the classical Euler-Lagrange equations
for the action $\sI$ just tell us that $\dot x=0$, that is, the map $x:I\to M$ is constant.   To compute the coefficient of the delta function in $Z_{r,r'}$
at one-loop level, we need the determinant of the differential operator that we obtain by linearizing the action around a constant map.   But this
differential operator, with a boundary condition that specifies the phase space points at the ends, is not self-adjoint, and it is problematic to make sense of its determinant.
In the example of $\R^2$ with coordinates $p,q$ and symplectic form $\omega=\d p\d q$, the action is $\sI=\int \d t\, p \dot q$.   The linearization around the
solution $p=q=0$ gives the
operator $D=\begin{pmatrix} 0 & \frac{\d}{\d t}\cr -\frac{\d}{\d t}& 0\end{pmatrix}$.   The eigenvalue problem $D\psi=\lambda\psi$ does not have any nonzero
solutions with the boundary conditions that $p,q$ vanish at the endpoints.   The problem is  that this boundary condition does not make $D$ self-adjoint.
To get a self-adjoint problem, one should specify only one-half of the boundary values at each end, and more specifically the variables that are fixed
should be Poisson-commuting.   In particular, if $\P$ is a real polarization of $M$ in the sense described in Section \ref{gq}, one can require the endpoint
values $x(0)$ and $x(1)$ to lie in specified leaves $F,F'$ of\footnote{\label{BKStwo}More generally, one can take $F$ to be a leaf of one polarization $\P$ and $F'$ to be
a leaf of another polarization $\P'$.   Computing in that situation will lead to the Blattner-Kostant-Sternberg kernel, which was mentioned in 
footnote \ref{BKS}.} $\P$. 
  Computing
the path integral with this boundary condition will lead back to the recipe that geometric quantization gives for quantizing with the polarization
$\P$.  The result one gets will depend on $\P$ in an essential way. 

Even though contemplating the path integral (\ref{hiff}) does not give a way to avoid the anomaly in the passage from classical mechanics to quantum mechanics,
this formulation of the problem of quantization is nonetheless
important.  Suppose that we have a well-defined microscopic 
theory of some kind that somehow produces a Hilbert space $\H$
 associated to $M$ (and possibly to some other data that enters the definition of the theory in question).  
 Suppose moreover that the path integral of this microscopic theory on the interval $I$ can be reduced to (\ref{wiff}) plus some
 ``massive'' degrees of freedom that one can formally argue to be irrelevant.   
 This actually means, assuming the microscopic theory in question really is well-defined, 
that going to this microscopic theory is giving us a way to embed the  ill-defined problem of quantizing the theory (\ref{wiff}) in a larger,
well-defined problem.     In other words, the microscopic theory is giving us a way to make sense of the original
ill-defined problem.     Thus, under these conditions, it is sensible to define $\H$ to be a quantization of $M$. 
This is how we will actually
proceed when we come to quantization by branes.

Now let us discuss what we can do with this theory {\it without} introducing a polarization.
The reason that, in the preceding discussion, we considered a path integral on a 1-manifold with boundary, namely the unit interval $I$,
is that in general, in a Feynman path integral, an initial or final quantum state appears on the boundary
of the space on which the path integral is performed,  in this case $I$.    One can avoid having to choose
a polarization by replacing $I$ with  a circle $S^1$.   Then one considers the same path integral, except that one integrates
over the space $\pP$ of  loops $x:S^1\to M$.    No boundary condition is needed, because the circle has no boundary.
As long as we do not introduce boundaries, whatever we compute cannot be related to states in a quantum Hilbert space.
What can we hope to compute without being able to define a Hilbert space?   In deformation quantization, as proposed in  \cite{Many} and
implemented rigorously on an arbitrary symplectic manifold \cite{WL,Fedosov}, one aims to deform the commutative associative algebra $\A_0$
of functions on $M$ to a noncommutative but still associative algebra $\A$, to all orders in a power series in $\hbar$.    To first order in  $\hbar$, 
the deformation
should be given by the Poisson bracket.   There is no construction of a Hilbert space that the algebra $\A$ acts on.\footnote{To be more precise,
there is no construction of a Hilbert space of appropriate size.  $\A$ itself can be given the structure of a Hilbert space, using the trace that is introduced
presently.  This Hilbert space is infinite-dimensional even when $M$ is compact.}

Though deformation quantization is not necessarily tied to path integrals, the path integral on a circle gives a natural framework for deformation quantization.
  The formal framework is as follows.  (This framework has been justified rigorously to all orders in $\hbar$, using a machinery of BV quantization \cite{SL}.)
Let  $f$ be a function on $M$ and $r$ a point in $S^1$.   Given a loop $x:S^1\to M$,
one can pull back $f$ to a function on $S^1$ and evaluate this function at the point $r$ to get what we will call $f(r)$, a function on the loop space $\pP$ of $M$.
Now let $f_1,f_2,\cdots, f_k$ be functions on $M$ and $r_1,r_2,\cdots, r_k$ a cyclically ordered set of points on $S^1$, and consider the path integral
\be\label{cycl}\bigl\langle f_1 f_2 \cdots f_k\bigr\rangle =\int_\pP\D x \,e^{\i \sI} \,f_1(r_1) f_2(r_2)\cdots f_k(r_k). \ee
Because the classical action is invariant under orientation-preserving reparametrization of $S^1$, the quantity
$\bigl\langle f_1 f_2 \cdots f_k\bigr\rangle$ should formally only depend on the cyclic ordering of the points $r_1,r_2,\dots,r_k$, not the choice of specific
points.    The ``correlation function''  $\bigl\langle f_1 f_2 \cdots f_k\bigr\rangle$ has the information needed for deformation
quantization of the symplectic manifold $M$.    The basic relation between the correlation function and deformation quantization is that if
 \be\label{woppo} \bigl\langle f_1 f_2 f_3\cdots f_k\bigr\rangle =\bigl\langle g f_3 \cdots f_k\bigr\rangle \ee
(where $g$ depends on $f_1$ and $f_2$ but not on $f_3,\cdots,f_k$), then one declares that in the algebra $\A$,
one has $\h f_1\h f_2=\h g$ (where $\h f$ is an element of $\A$ that reduces to  $f\in \A_0$ in the limit $\hbar\to 0$).    This relationship
is a version of the operator product expansion, adapted to topological field theory.
Cyclic symmetry means that the correlation function $ \bigl\langle f_1 f_2 f_3\cdots f_k\bigr\rangle$ has the algebraic
properties of a trace (since for instance $\bigl\langle fg\bigr\rangle =\bigl\langle gf \bigr\rangle$), though it is not actually a trace in an
$\A$-module.  

At least as presently understood, deformation quantization breaks or at least deforms the symmetries of classical mechanics, somewhat similarly to
what happens with quantization, though the details are different.   In Fedosov's approach \cite{Fedosov}, the construction of the quantum-deformed algebra $\A$
depends on the choice of a symplectic connection on the manifold $M$ (a connection on the tangent bundle of $M$ such that the symplectic form is covariantly
constant).   Symplectic connections exist on every $M$ (unlike polarizations \cite{Gotay}), but a symplectic connection is far from unique.    A generic symplectomorphism
maps the algebra $\A_\Gamma$ constructed with one symplectic connection $\Gamma$ to the algebra $\A_{\Gamma'}$ constructed using another symplectic
connection ${\Gamma'}$.   $\A_\Gamma$ and $\A_{\Gamma'}$ are  isomorphic, but not canonically so (see Corollary 4.5 in \cite{Fedosov}).    The absence
of canonical isomorphisms between the different possible $\A$'s means that the symplectomorphism group of $M$ does not act on any one of them.   If $G$ is
any {\it compact} group of symplectomorphisms, then there is a $G$-invariant symplectic connection, and therefore any such $G$ can be realized as a symmetry
group of the quantum algebra $\A$.   By contrast, geometric quantization in general cannot be defined to be invariant
 under a compact group of symplectomorphisms.\footnote{For example, if $M$ is a two-torus with translation-invariant symplectic form $\omega$,
then it is possible to pick on $M$ a (flat) symplectic connection that is translation invariant.   Therefore, deformation quantization can be defined to preserve the
translation invariance of the torus.   But geometric quantization  of the torus requires the choice of a prequantum line bundle $\frL$ with connection,
and any choice breaks the translation symmetries.   That is because the curvature of $\frL$ makes it impossible for the holonomy of $\frL$ around a noncontractible
loop in $M$ to be translation invariant.}

Unlike quantization,
deformation quantization depends only on the symplectic structure $\omega$ and not on the existence or choice of a prequantum line bundle $\frL$.   In fact, 
the action $\sI=\oint_{S^1}x^*(\sT )/\hbar$ can, for purposes of deformation quantization, be rewritten in a way that refers only to $\omega$ and not $\sT $.   We simply
observe that perturbatively in $\hbar$, we can assume that  $x:S^1\to M$ is almost constant. Let $D$ be a disc with boundary $S^1$.
  An almost constant map $x:S^1\to M$ can be extended to a map $x:D\to M$, by shrinking the image of $x$ to a point in $M$.  
  This extension is not unique, but it is unique up to homotopy.
  Then we can replace $\frac{1}{\hbar}\oint_{S_1}x^*(\sT )$
  by $\int_D  x^*(\omega)$, and rewrite (\ref{cycl}) in the form 
\be\label{pycl}\bigl\langle f_1 f_2 \cdots f_k\bigr\rangle =\int_\pP\D x\, \exp\left(\i \int_D  x^*(\omega)\right) f_1(r_1) f_2(r_2)\cdots f_k(r_k), \ee
manifestly depending only on $\omega$ and not $\frL$.

Deformation quantization also makes sense in a more general context of Poisson manifolds that are not necessarily symplectic.
A Poisson manifold by definition has Poisson brackets defined by a formula
$\{f,g\}=\sum_{ij}r^{ij}\partial_i f \partial_j g$, where $r^{ij}$ is an antisymmetric tensor that is such that the usual Jacobi identity is satisfied, but $r^{ij}$ may not be
 invertible (on a symplectic manifold, $r^{ij}$ is the inverse of the symplectic form and in particular is invertible).   A path integral description of deformation quantization in
this context is still possible, but one has to use
two-dimensional path integrals rather than one-dimensional ones \cite{Kon,CF,CF2}.   

\subsection{The $A$-Model}\label{amo}

In contrast to geometric quantization, which is based on choosing a polarization,
quantization via branes \cite{GW} involves a different choice of the additional structure of $M$ that
enables quantization. 

The starting point is to assume that the real symplectic manifold $M$ has a complexification $Y$ with some
favorable properties.    In particular, we assume that $Y$ is a complex manifold with
an antiholomorphic involution  $\tau:Y\to Y$, satisfying $\tau^2=1$ and with $M$ as a component of its fixed point set.  
Moreover, we assume that the real symplectic form $\omega$ of $M$ analytically continues
to a holomorphic symplectic form $\Omega$ on $Y$, making $Y$  a complex symplectic manifold.  
We take the assertion that $\Omega$ is the analytic continuation of $\omega$ to $Y$ to mean that $\Omega$ restricted
to $M$ equals $\omega$:
\be\label{circo}\left.\Omega\right|_M=\omega. \ee
A nondegenerate and holomorphic two-form  $\Omega$ satisfying that property
automatically exists in a neighborhood of $M$, but we ask that $\Omega$ exists globally on $Y$.
From these assumptions, it follows that 
\be\label{taupl}\tau^*(\Omega)=\bar\Omega,\ee
 so in particular
\begin{align}\label{tauaction} \tau^*(\Re\,\Omega)&=\Re\,\Omega \cr \tau^*(\Im\,\Omega)&=-\Im\,\Omega. \end{align}
Since $M$ is invariant under $\tau$ and $\Im\,\Omega$ is odd, we also have
\be\label{lagm} \left.\Im\,\Omega\right|_M=0. \ee

An important special case
 is that $M$ is   a real algebraic variety, and $Y$ is  the corresponding
complex algebraic variety.  All examples we consider are of this type,
as were the examples considered in \cite{GW}. 

Just like geometric quantization, brane quantization involves a choice, and the resulting quantum Hilbert space depends on this
choice in an essential way.  In particular, if $M$ has a complexification $Y$ with the properties needed for brane quantization,
there are always many possibilities for $Y$.   A symplectomorphism of $M$ that does not extend to a holomorphic symplectomorphism of
$Y$ will map $Y$ to another possible complexification $Y'$ of $M$, leading upon quantization to an inequivalent Hilbert space.  The
analogous statement in geometric quantization is that a symplectomorphism of $M$ that does not preserve a given polarization maps
the polarization to an inequivalent one, again leading to an inequivalent Hilbert space.

For quantization by branes, we want to view $Y$ as a real symplectic manifold and to study the corresponding $A$-model.
Any nontrivial real linear combination of $\Re\,\Omega$ and $\Im\,\Omega$ could be viewed as a real symplectic form on $Y$.
However, for quantization of $M$, we will want $M$ to be a Lagrangian submanifold.   For this, in view of eqn. (\ref{tauaction}),
we should take the symplectic
form of $Y$ to be a multiple of $\Im\,\Omega$.    Since we have not said anything that fixed the normalization of $\omega$ or $\Omega$,
we will simply take the symplectic form of $Y$ to be $\Im\,\Omega$.    When we speak of the $A$-model of $Y$, we always mean the
$A$-model with that symplectic form  (possibly generalized to include a $B$-field in the $A$-model).  

A nontrivial condition on $Y$ is needed so that the 
 $A$-model of $Y$ with the real symplectic form $\Im\,\Omega$ is well-defined.   For example, given any $Y$ that is a candidate complexification
 of some $M$, we could make another candidate $Y'$ by omitting from $Y$ a $\tau$-conjugate pair of points.  This seems like a rather unnatural example, and one expects to have to exclude it.   The general condition that should be placed
 on $Y$ is not very clear from a physical point of view.   However, there is a known sufficient condition that suffices 
for all examples that will be considered in the present article.   This is that $Y$ admits a complete
hyper-Kahler metric, which in a sense that will be explained momentarily extends the complex symplectic structure of $Y$.      Such a complete
hyper-Kahler metric on $Y$ ensures the existence of a quantum $\sigma$-model with target $Y$, and a standard twisting of this $\sigma$-model
is the desired $A$-model.

We pause for a moment to explain why it is important that the $\sigma$-model of $Y$ actually exists as a two-dimensional quantum
field theory.   In Section \ref{abranes}, we will review how the $A$-model of $Y$ is related to deformation quantization of the algebra of holomorphic 
functions on $Y$.  As a deformation over a formal power series ring $\C[[\hbar]]$, where $\hbar$ is the deformation parameter, this
construction is formal and
does not require any assumption that a quantum  $\sigma$-model of $Y$ actually exists.\footnote{The underlying $\sigma$-model
from which the $A$-model is derived may itself have been constructed by quantizing a classical system, with a quantum parameter usually called $\hbar$.
That underlying quantum parameter is not necessarily related in a simple way to the deformation parameter of the $A$-model,
which is really a $\sigma$-model modulus.
It will hopefully cause no confusion to refer to this $A$-model deformation parameter as $\hbar$.  In any event, we usually do not introduce
the deformation parameter explicitly; instead we write formulas with a real symplectic form $\omega$ or a complex symplectic form $\Omega$.
To introduce $\hbar$ in such formulas, one can pick a specific real or complex symplectic form $\omega_0$ or $\Omega_0$, and write
$\omega=\omega_0/\hbar$ or $\Omega=\Omega_0/\hbar$.  When we do write explicit factors of $\hbar$ in formulas, it is often at a preliminary stage
before learning that it makes sense to set $\hbar=1$.}   But for quantization (and not just
deformation quantization), we want to get a deformed algebra in which $\hbar$ can be viewed as a complex parameter, not just as the
argument of a formal power series.   This is what we get when the quantum $\sigma$-model actually exists; $\hbar$ is then one of the moduli
of the $\sigma$-model.   It may be that the existence of a full, utraviolet-complete $\sigma$-model is more than is  needed
for the constructions discussed in the present article, but existence of the quantum $\sigma$-model is the sufficient condition that we know.

A hyper-Kahler manifold
has a family of complex structures, associated to an action on its tangent space of the quaternion units $I,J,K$, obeying
the usual quaternion relations $I^2=J^2=K^2=-1$, $IJ=K$.  Associated to $I,J,K$ is a triple of Kahler forms $\omega_I$, $\omega_J$,
and $\omega_K$ (where $\omega_I$ is of type $(1,1)$ and positive with respect to complex structure $I$, and similarly for $\omega_J$
and $\omega_K$), along with
 holomorphic two-forms $\Omega_I=\omega_J+\i\omega_K$, and cyclic permutations of $I,J,K$. The complex structures
and symplectic forms obey 
\be\label{tono}I^t\omega_J=-\omega_K, ~ I^t\omega_K=\omega_J, \ee
and cyclic permutations thereof, where $I^t$ is the transpose of $I$, viewed as a linear transformation acting on 1-forms on $Y$.
The hyper-Kahler metric of $Y$ is
\be\label{bono} g=I^t \omega_I=J^t\omega_J= K^t\omega_K.\ee

When we say that the hyper-Kahler structure of $Y$ extends its complex symplectic structure as a complexification of $M$, 
we mean that $I$ is the 
 complex structure of $Y$  as a complexification of $M$, and the corresponding holomorphic
symplectic form is $\Omega=\Omega_I=\omega_J+\i \omega_K$.    We assume that $Y$ has an involution $\tau$ that satisfies $\tau^*(g)=g$
as well as  the conditions stated earlier, which now become $\tau^*(I)=-I$, $\tau^*(\omega_J)=\omega_J$, $\tau^*(\omega_K)=-\omega_K$.
From this and eqn. (\ref{bono}) we can further deduce:
\be\label{tabush}\tau^*\begin{pmatrix} I\cr J\cr K\end{pmatrix}=\begin{pmatrix} -I \cr J \cr -K\end{pmatrix},~~~~
    \tau^*\begin{pmatrix} \omega_I\cr \omega_J\cr \omega_K\end{pmatrix}=\begin{pmatrix} -\omega_I \cr \omega_J \cr -\omega_K\end{pmatrix} .\ee                                

In studying the $A$-model of $Y$, we
 will use a notation adapted to this hyper-Kahler case, and thus we will write $I$ for the complex structure of $Y$
as a complexification of $M$,  $\omega_J$ for $\Re\,\Omega$, and $\omega_K$ for $\Im\,\Omega$.

An important type of example is the case that 
 $M$ is a real affine variety, meaning that it is a submanifold of
some $\R^N$ defined by the vanishing of some real-valued polynomials in the coordinates $x_1,x_2,\cdots, x_N$ of $\R^N$. 
In that case, we can take
$Y$ to be  the corresponding complex
affine variety (the submanifold of $\C^N$ defined by the vanishing of the same polynomials in the $x_i$, now viewed as complex
variables).     

When we quantize $M$, it will turn out that the observables correspond to functions on $M$ that analytically continue to
holomorphic functions on $Y$.   But it is important to decide
what class of holomorphic functions we want to allow.   Do we want to consider only polynomial functions of the $x_i$, or do we
want to allow functions that grow exponentially at infinity?   Since the observables will correspond to physical string states of the
$A$-model, the question is somewhat analogous to asking what class of quantum mechanical wavefunctions we want to allow.  
We will allow only polynomial functions on $Y$, rather than functions that grow exponentially, since this leads to a much simpler theory.

Some simple examples of affine varieties, studied in \cite{GW}, are a coadjoint orbit  $M$ of $\SU(2)$, defined by
\be\label{cosu} x^2+y^2+z^2  = \j^2 \ee
and  a coadjoint orbit $M'$ of $\SL(2,\R)$, defined by
\be\label{osuc}-x^2-y^2+z^2=\j^2, \ee
in each case with real variables $x,y,z$ and symplectic form $\omega= \d x\d y/z$.    
In eqn. (\ref{cosu}), $\j^2>0$; in eqn. (\ref{osuc}), both signs of $\j^2$ are of interest.   
The corresponding complex variety $Y$, which is a coadjoint orbit of $\SL(2,\C)$,  is obtained by taking $x,y,z$ to be complex-valued.   After complexification, the signs
in the equations do not matter, so for definiteness we define $Y$ by  eqn. (\ref{cosu}).    $M$ is embedded in $Y$ as the locus
with $x,y,z$ real, and $M'$ is the locus with $x,y $ imaginary, and $z$ real.   $M$ is the fixed point set of the antiholomorphic involution
of $Y$ defined by $\tau:(x,y,z)\to (\bar x,\bar y,\bar z)$ (assuming that $\j^2>0$ so that this fixed point set exists), and 
$M'$ is the fixed point set of a different antiholomorphic involution
$\t\tau$ defined by $\t\tau:(x,y,z)\to (-\bar x, -\bar y, \bar z)$.  The brane quantization method applied to $Y$, with the involutions 
$\tau$ and $\t\tau$,  reproduces classical
results about  representation theory of $\SU(2)$ and $\SL(2,\R)$.   Some of this was described in \cite{GW}, and we will make
some further remarks starting in Section \ref{scbq}.

\subsection{The Canonical Coisotropic Brane And Deformation Quantization}\label{abranes}

\subsubsection{$A$-Branes}\label{aobranes}

Now let us discuss the branes of the $A$-model.   The most familiar $A$-branes are Lagrangian branes.  The support of a Lagrangian
$A$-brane is a Lagrangian submanifold $L$ of $Y$.   Such a submanifold is always middle-dimensional, so if $Y$ has dimension
$\dim\,Y=2n$, then $\dim\,L=n=\frac{1}{2}\dim\,Y$.   The definition of an $A$-brane with support $L$ depends on the choice of
a Chan-Paton  bundle over $L$, or for short a $\CP$ bundle.     Our basic examples will be Lagrangian $A$-branes
of rank 1.     

The $\CP$ bundle of a rank 1 $A$-brane supported on $L$ is usually described as a complex line bundle
$\L\to L$ with a unitary flat connection.  
Though this description is adequate  locally, and it is also adequate globally if $L$ is a spin
manifold with a chosen spin structure,  it is in general slightly oversimplified because of
an anomaly \cite{FW} that is part of the relation of branes to $K$-theory.\footnote{\label{anomdesc} The anomaly has nothing to do with any topological twisting and is a general
property of branes in $\sigma$-models.   A brief explanation  is as follows.   First of all, there exists a brane $\B_0$
supported on $Y$ with trivial $\CP$ bundle.   In the relation of branes to $K$-theory, this brane corresponds to a trivial complex line bundle over $Y$.
Now consider any other brane $\B_1$ with support on some submanifold $L\subset Y$ and $\CP$ bundle $\L\to L$. If $\L$ has rank 1, then naively it is a
complex line bundle over $L$.   However, a standard analysis shows that the low-lying states of the $(\B_0,\B_1)$ system, in the Ramond sector, correspond
to spinors with values in $\L$.   
For such states lto  exist, given that $L$ was not assumed to be spin, $\L$ must be a 
$\spinc$ structure rather than a complex line bundle.}   The correct global statement is that $\L$ is a flat $\spinc$ structure over
$L$.  The concept of a flat $\spinc$ structure was already described in Section \ref{gq}: if $\L$ is a $\spinc$ structure, then $\L^2$ is an ordinary complex
line bundle with connection,  and the curvature of $\L$ is defined as one-half the curvature of $\L^2$.

Apart from the familiar Lagrangian branes,
 the $A$-model of a symplectic manifold $Y$ may have additional branes, discovered in \cite{KO}, known as coisotropic branes.  The 
support of a coisotropic brane is a coisotropic submanifold\footnote{A coisotropic submanifold is one that
can be defined locally by the vanishing of some Poisson-commuting functions.   For example, $Y$ itself is coisotropic; in that case, no Poisson-commuting
functions are required.   At the other extreme, a Lagrangian submanifold $L$ is coisotropic.   Locally one can pick coordinates so that $\omega_Y=\sum_i \d p_i \d q^i$
and $L$ is defined by $p_1=p_2=\cdots=p_n=0$.   The $p_i$ are the Poisson-commuting functions that locally define $L$.   A Lagrangian submanifold is a coisotropic submanifold of minimum possible dimension. } of $Y$  whose dimension  exceeds $\frac{1}{2}\dim\,Y$.    In this article,
we will be primarily concerned with the simplest case, which is a rank 1 coisotropic brane whose support is all of $Y$.    Since a complex symplectic manifold
has a canonical spin structure, the anomaly involving spin
 is not relevant in such a case; the $\CP$ bundle of a brane with support $Y$ is simply a complex line bundle $\RR$.

In introducing coisotropic branes supported on $Y$, we will allow for the possible presence in the $\sigma$-model of a two-form field $\sB$, 
usually called the $B$-field.   In the present article, the $B$-field is always
flat and topologically trivial, meaning that $\sB$ is simply a closed two-form.  
   As explained in \cite{KO}, in general
in the $A$-model of a symplectic manifold $Y$ with
symplectic form $\omega_Y$ and two-form field $\sB$, the condition for a rank 1 brane with support $Y$ and $\CP$ curvature $\sF$
to be an $A$-brane is that $I=\omega_Y^{-1}(\sF+\sB)$ should be an integrable complex structure on $Y$.  The simplest examples arise in precisely
the situation of interest in the present article.   If $Y$ is a complex symplectic manifold with complex structure $I$ and holomorphic
two-form $\Omega=\omega_J+\i\omega_K$, then in the $A$-model with $\omega_Y=\omega_K$, we can solve $I=\omega_Y^{-1}(\sF+\sB)$
by setting $\sF+\sB=\omega_J$.   

We still have some freedom to choose $\sF$ and $\sB$ separately, obeying $\sF+\sB=\omega_J$. 
The different possibilities are equivalent in the sense that they are related by $B$-field gauge transformations.   In general, the $\sigma$-model
is invariant under tensoring the $\CP$ bundle of every brane with an arbitrary unitary line bundle $\mathcal S$, of curvature $\sG$ (the same $\mathcal S$ for all
branes).    This induces a shift
$\sF\to \sF+\sG$ for the $\CP$ curvature of every brane, and a compensating shift $\sB\to \sB-\sG$.   

In the present article, it will generally be most convenient to choose $\sF=0$, $\sB=\omega_J$.  The $\CP$ bundle of a brane with $\sF=0$ can be completely
trivial, with a trivial flat connection.     The coisotropic $A$-brane with trivial $\CP$ bundle, in an $A$-model with symplectic form $\omega_K$ and $B$-field
$\omega_J$, is what we will call the canonical coisotropic $A$-brane, denoted $\B_\cc$.   Branes of this type are important in the gauge theory
approach to the geometric Langlands correspondence \cite{KW}. 

If there exists a unitary line bundle $\RR $ with curvature $\omega_J$, then by a $B$-field gauge transformation, we can set $\sB=0$ and obtain
an alternative description in which $\B_\cc$ has $\CP$ bundle $\RR$.   In fact, that description has been used previously \cite{KW,GW}, and we will see one reason that it
can be useful in Section \ref{qcomp}.    However, in this article it will generally be more simple to take the $\CP$ bundle of $\B_\cc$ to be completely trivial
and set $\sB=\omega_J$.

In general, in the $A$-model with symplectic form $\omega_Y$ and a given $B$-field, the sum $\omega_\C=\omega_Y-\i \sB$ is called the complexified symplectic form.
In the present context, this is
\be\label{zoffo}\omega_\C=\omega_Y-\i \sB=\omega_K-\i \omega_J = -\i \Omega. \ee

\begin{figure}
 \begin{center}
   \includegraphics[width=5in]{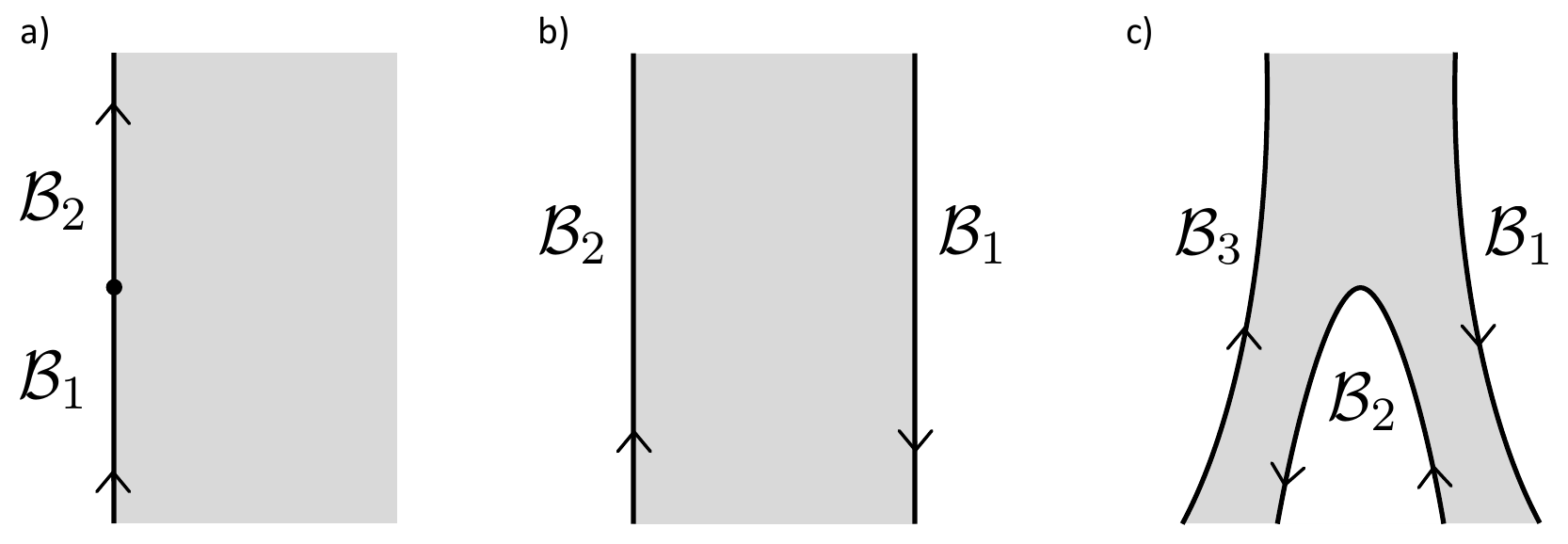}\caption{\small\small{
(a)   In the case of a left boundary of a two-manifold $\Sigma$, we consider ``time'' to flow upward, in the direction of the arrow,
so a junction of boundary conditions with $\B_1$ below $\B_2$ represents a transition from $\B_1$ to $\B_2$, described algebraically as
an element of $\Hom(\B_1,\B_2)$.     (b) By a conformal
transformation, one can map this to a strip with $\B_2$ on the left and $\B_1$ on the right.   The drawing represents an element of $\Hom(\B_1,\B_2)$
propagating from the bottom of the picture to the top.   Alternatively, viewed from top to bottom, the same drawing represents downwards propagation of an element
of the dual space $\Hom(\B_2,\B_1)$.   (One can exchange the two views by a $\pi$ rotation of the plane, preserving
its orientation and exchanging $\B_1$ and $\B_2$. 
A reflection would reverse the orientation of the plane and is not allowed.)  The upward and downward pointing arrows indicate that $\Hom(\B_1,\B_2)$ depends covariantly on $\B_2$
and contravariantly on $\B_1$. (c) The natural composition of morphisms
$\Hom(\B_2,\B_3)\otimes \Hom(\B_1,\B_2)\to \Hom(\B_1,\B_3)$ (where $\B_1$ is mapped to $\B_2$ which is then mapped to $\B_3$)
corresponds to this picture. }}\label{explained}
 \end{center}
\end{figure}

 In general, if $\B_1$ is an $A$-brane, then the space of physical states of $(\B_1,\B_1)$ strings in the $A$-model  is an associative algebra $\A$.   If $\B_2$ is
 another $A$-brane, then 
the analogous spaces of $(\B_1,\B_2)$  or $(\B_2,\B_1)$ physical states are $\A$-modules (right and left $\A$-modules, to be precise).   
The connection of the $A$-model with quantization  arises upon implementing this idea for the case that $\B_1=\B_\cc$, and $\B_2$ 
is a Lagrangian brane.   As an abbreviation, we will use a common mathematical terminology, and refer to the space of $(\B_1,\B_2)$
physical states in the $A$-model (for any $\B_1$ and $\B_2$) as $\Hom(\B_1,\B_2)$. 
The basic idea is that the joining of a $(\B_1,\B_2)$ string to a $(\B_2,\B_3)$ string to make a $(\B_1,\B_3)$ string is described as a
 composition of morphisms $\Hom(\B_2,\B_3)\otimes \Hom(\B_1,\B_2)\to \Hom(\B_1,\B_3)$, where $\B_1$ is mapped
to $\B_2$ and then $\B_2$ is mapped to $\B_3$.  In general, $\Hom(\B_1,\B_2)$ depends covariantly on $\B_2$ and contravariantly on $\B_1$.
Related to this, if $\B_1$ has
support $T_1$ and $\CP$ bundle $E_1$, while $\B_2$ has support $T_2$ and $\CP$ bundle $E_2$, then very roughly $\Hom(\B_1,\B_2)$ is some
sort of space of sections over $T_1\cap T_2$ of $E_1^\vee \otimes E_2$ (here $E_1^\vee$ is the dual of $E_1$).   Schematically
\be\label{kogo}\Hom(\B_1,\B_2)\sim \Gamma(T_1\cap T_2, E_1^\vee\otimes E_2). \ee
This is a very schematic statement, because we have not said what kind of sections are allowed (this depends on details of the type of brane
considered), and in general the full story is much more complicated.  The point of eqn. (\ref{kogo}) is only to stress that $E_1$ is dualized and $E_2$ is not.  
Some
 conventions for the relations between strings and morphisms are explained in fig. \ref{explained}.

Additively, at least for small $\hbar$,  $\Hom(\B_\cc,\B_\cc)$ is a direct sum of cohomology groups of $Y$,
\be\label{poklo} \Hom(\B_\cc,\B_\cc)=\oplus_{r=0}^{\dim_\C Y} H^r(Y,\O)\ee
(where $\O$ is a trivial complex line bundle over $Y$).    The integer $r$ labels the fermion number or ``ghost number'' of
a cohomology class.    The derivation of this result is sketched in  Appendix \ref{details}, following Section 2.2 of \cite{GW}.  
Since the fermion number is additive when string states are joined (or multiplied), the states with $r=0$
form a ring in their own right.   This part of the cohomology is therefore particularly important.   The cohomology with $r>0$ actually
vanishes in many interesting examples.

The ring $\Hom(\B_\cc,\B_\cc)$ is particularly interesting in the case that $Y$ is an affine algebraic variety (defined by vanishing of some
polynomial functions on $\C^N$), so we continue the discussion
for that case.    Since the sheaf cohomology of an affine variety vanishes for $r>0$, we only have to consider the $r=0$ cohomology, which as a vector
space is the space of holomorphic functions on $Y$.   As explained in Section \ref{amo},
we consider only functions that are polynomials in the coordinates. 
Under ordinary multiplication, the holomorphic functions on $Y$ form a commutative, associative ring $\A_0$. 
But that is not the ring structure of $\A=\Hom(\B_\cc,\B_\cc)$.   Rather, as shown  in Section 11.1 of \cite{KW}, following a number
of prior developments \cite{KO,BrS,Ka,Pe,Gu,AZ}, and as we explain next,
$\A$ is an associative but noncommutative deformation of $\A_0$.   

\subsubsection{The Relation To Deformation Quantization}\label{defrel}

 The basic idea of this derivation is as follows.   In general, the $A$-model
is constructed from an underlying supersymmetric $\sigma$-model by choosing one of its supercharges $Q$, chosen to satisfy $Q^2=0$,
and restricting to $Q$-invariant observables.   Since $Q^2=0$, $Q$-invariant observables are not affected by $Q$-exact 
terms $\{Q,\cdots\}$ in the action.  On a two-manifold $\Sigma$, consider a $\sigma$-model with target $Y$ and $\sigma$-model map 
$X:\Sigma\to Y$.   Let $\omega_\C=\omega-\i B$  be the complexified symplectic form of the $A$-model. If $\Sigma$ has no boundary, then the action takes the
 simple form
\be\label{simpform}\sI = \int_\Sigma X^*(\omega_\C) +\int_\Sigma \{Q,\cdots\}. \ee
In other words, everything is $Q$-exact except the pullback of $\omega_\C$.   In our setup, $\omega_\C=-\i\Omega$.
So for example if $\Sigma$ is a disc $D$ with boundary $S^1$, then the action is
\be\label{toldo} \sI=
-\i\int_D X^*(
\Omega)+\int_D \{Q,\cdots\}. \ee
The integrand of the Feynman integral in Euclidean signature is $\exp(-\sI)$, so this integral is schematically
\be\label{oldo}\int DX \cdots \exp\left(\i\int_D \Omega\right) \exp\left(-\int_D \{Q,\cdots\}\right). \ee   
(Here $DX \cdots $ refers to an integral over $X$ and other fields.) 
More generally, we can include operator
insertions on the boundary of the disc.   Such operators represent $(\B_\cc,\B_\cc)$ strings, so they
correspond to holomorphic functions $f_1,\cdots, f_k$ on $Y$ (or cohomology classes of higher degree, as discussed later).   As in Section \ref{uses}, we pick
points $r_1,r_2,\cdots,r_k$ in the boundary of the disc
and write $f_i(r_i)$ for $f_i(X(r_i))$.    Then the Feynman integral that represents an $A$-model calculation on a disc whose boundary
is labeled by $\B_\cc$ with operator insertions on the boundary is
\be\label{noldo}\int DX \cdots \exp\left(\i\int_D \Omega\right) \exp\left(-\int_D \{Q,\cdots\}\right) f_1(r_1)f_2(r_2)\cdots f_k(r_k). \ee 
    
This integral is related to deformation quantization of the ring $\A_0$ of holomorphic functions on $Y$.   Let us first describe that notion.
Formally, deformation quantization of the ring of holomorphic functions on a complex manifold $Y$ is rather analogous to
the deformation 
quantization of real symplectic (or Poisson) manifolds that was  briefly described in Section \ref{uses},
except that what is deformed is the ring of holomorphic functions.     If $\Omega=\frac{1}{2}\sum_{i,j}\Omega_{ij}\d y^i
\d y^j$ (where the $y^i$ are local holomorphic coordinates on $Y$), and $\Omega^{ij}$ is the inverse matrix to $\Omega_{ij}$,
then the Poisson bracket of holomorphic functions is defined in the usual way by
\be\label{pb}\{f,g\}=\sum_{i,j}\Omega^{ij}\partial_i f \,\partial_j g. \ee
Then in the deformed ring $\A$,  the algebra elements $\h f$, $\h g$ corresponding to holomorphic functions $f,g$ satisfy
\be\label{hb} \h f \h g - \h g \h f =-\i\hbar\h{\{f,g\}}_0 +\O(\hbar^2),\ee
where we have introduced a deformation parameter by writing  
 $\Omega =\frac{1}{\hbar}\Omega_0$, with some fixed $\Omega_0$, and $\{~,~\}_0$ is the Poisson bracket defined using $\Omega_0$.  In the expansion of $\h f \h g$ in powers of $\hbar$, the term of order $\hbar^n$
 is supposed to be a sum of terms schematically of the form  $\partial^k f \partial^{2n-k}g$.

The most precise explanation of how the integral (\ref{noldo})
  is related to deformation quantization was given in \cite{W}.   First recall from Section \ref{uses} that deformation quantization of the real symplectic manifold
  $M$  can be described by integration over the space $\pP$ of maps $x:S^1\to M$:
  \be\label{pyccl}\bigl\langle f_1 f_2 \cdots f_k\bigr\rangle =\int_\pP\D x \,\exp\left(\i \int_D  x^*(\omega)
  \right) \, f_1(r_1) f_2(r_2)\cdots f_k(r_k). \ee
  Here the map $x:S^1\to M$, assumed to be nearly constant, has been extended to $ x:D\to M$, where $D$ is  a disc with boundary $S^1$, by shrinking
  its image to a point in $M$.   
   $M$ has $Y$ as a complexification, and $\omega=\omega_J$ analytically continues to $\Omega$.
    The complexification of $\pP$ is the space $\h \pP$ of maps $X:S^1\to Y$.
   In deformation quantization, $X:S^1\to Y$ can be assumed to be nearly constant, again leading to a homotopy class of extensions
   to $ X:D\to Y$.  The function $\exp(\i\int_D  x^*(\omega))$ analytically continues to a holomorphic function on $\h \pP$, namely
   $\exp(\i\int_D X^*(\Omega))$.   We note that precisely this function appears as a factor in eqn. (\ref{oldo}), the rest of the action being $Q$-exact.
   Suitable smooth functions $f_1,f_2,\cdots, f_k$ on $M$ can be analytically continued to holomorphic functions on $Y$, which we also call $f_1,f_2,\cdots ,f_k$.
   Formally, the expression $\D x \exp(\i \int_D  x^*(\omega)) $ in eqn. (\ref{pyccl}), where $\D x$ is the measure of the Feynman path integral,
    is a top degree differential form on the infinite-dimensional manifold $\pP$.
   It analytically continues to a holomorphic form    $\D X \,\exp(\i\int_D X^*(\Omega))$ of top degree on $\h \pP$.   To integrate a top degree
   holomorphic form on a complex manifold,
   in this case the infinite-dimensional manifold $\h \pP$,
   one needs a middle-dimensional integration cycle.   So for example, if $\Gamma$ is a middle-dimensional cycle in $\h \pP$ of an appropriate kind,
   we might hope to make sense of the integral 
     \be\label{pyclo}\int_\Gamma\D X \exp\left(\i \int_D  X^*(\Omega)\right) f_1(r_1) f_2(r_2)\cdots f_k(r_k). \ee
     
One example of a middle-dimensional cycle in $\h \pP$ is the original $\pP$ consisting of maps to $M\subset Y$.   With this choice of $\Gamma$,
     the integral (\ref{pyclo}) reduces back to (\ref{pyccl}), and provides a framework to understand deformation quantization.   
     What is shown in \cite{W} is that the integral (\ref{noldo}) on a disc with $\B_\cc$ boundary and operator insertions on the boundary can be interpreted
     as the integral (\ref{pyclo}) with a different integration cycle, namely the cycle\footnote{As we discuss shortly, a correction is
     needed because this cycle is not quite middle-dimensional.}  consisting of maps $X:S^1\to Y$ that extend
     to maps $X:D\to Y$ that are holomorphic in complex structure $K$.   The choice of $\Gamma$, within a class of almost middle-dimensional cycles such that
     the integral (\ref{pyclo}) makes sense, does not affect the arguments relating this integral to deformation quantization of $Y$.   A finite-dimensional
     analog of this statement is that if $Z$ is a complex manifold with a holomorphic form $\Omega$ of top degree, then the differential equations
     obeyed by a period integral $\int_\Gamma\Omega$ do not depend on the choice of the cycle $\Gamma\subset Z$.

We have omitted much in this short summary, but one  further detail requires explanation, though it will not play any direct role in the present article.
As explained in the discussion of eqn. (\ref{cycl}), if we can make sense of the integral
     (\ref{pyclo}), we will get an associative algebra $\A$ that deforms $\A_0$, and a trace on $\A$.  A trace is
      a complex-valued linear function $\h f\to \la\, \h f\,\ra$ obeying $\la\,\h f
     \h g\,\ra = \la \,\h g \h f\,\ra$.   In the case of a smooth manifold $M$, the classical limit of the trace is integration with the symplectic measure: $f\to \int_M  \frac {\omega^n}{n!} f$, where $n=\frac{1}{2}\dim M$.   This does not have such a direct analog for a complex symplectic manifold $Y$ with a holomorphic function $f$.
     The closest analog of $\frac{\omega^n}{n!}f$ is 
  $\frac{\Omega^n}{n!} f$, which 
     is a holomorphic differential form of top degree; there is no way to integrate   $\frac{\Omega^n}{n!} f$ over $Y$.   If we use the integration cycle $\pP$ that involves
     restriction to $M$, this presents no difficulty, but if we use the integration cycle that is related to $\B_\cc$, we need to modify the definition of one of the
     operators and as a result we will not get a trace.
        We are primarily interested in noncompact $Y$, since if $Y$ is compact, there are no nonconstant holomorphic functions to be quantized.   If $Y$ is
     noncompact, duality is a pairing between the ordinary $\bar\partial$ cohomology and $\bar\partial$  cohomology with compact support.   If $f_1,\cdots,
     f_{k-1}$ are holomorphic functions on $Y$, then we should take $f_k$ to be an element of $H^{\dim_\C Y}_{\mathrm{cpct}}(Y,\O)$, a top degree $\bar\partial$
     cohomology class with compact support.   The classical limit of the integral in eqn. (\ref{pyclo}) is then $\int_Y f_1 f_2\cdots f_k \frac{\Omega^n}{n!}$. This is
     a well-defined integral, 
     and  the quantum corrections are also well-defined.   From
     the correlation functions of eqn. (\ref{pyclo}), by using the operator product expansion as in eqn.
     (\ref{woppo}),  we can construct an associative algebra $\A$ that reduces to the commutative algebra $\A_0$ of holomorphic functions for $\hbar\to 0$.
     But this algebra does not come with a trace, because the cyclic symmetry is spoiled by taking one of the operators to represent a cohomology class with
     compact support.
          
The explanation of this point  given in \cite{W} is that because of an index theorem, the cycle $\Gamma$ associated to $\B_\cc$ is actually above the middle
dimension by an amount $2n=\dim_\C\, Y$, so it is not quite suitable for integration of a holomorphic form.
  One compensates for this by replacing one of the fiunctions $f_i$ by a $(0,2n$)-form.

  \subsubsection{Actual Deformations}\label{actual}

For a general $Y$, this construction will produce a deformation quantization over a ring of formal power series in $\hbar$.   That is not
good enough for the application of the $A$-model to quantization; for this, one needs to be able to treat $\hbar$ as a complex 
number.  Equivalently, one needs to be able to take $\Omega$ to be a definite holomorphic symplectic form on $Y$.  When this is the case, 
we will say that the $A$-model produces an actual deformed algebra $\A$, not just a formal one.
A sufficient condition for this, as remarked in Section \ref{amo}, is that the underlying quantum $\sigma$-model with target $Y$ (of which
the $A$-model is a twisted version) should actually exist as a full-fledged quantum field theory.   In particular, this is so if $Y$ admits a complete hyper-Kahler metric.

Discussing the same question concerning deformation quantization of the ring of holomorphic functions, 
Kontsevich \cite{K2} proposed that an actual deformation will exist if, roughly,
$Y$ can be compactified by adding a divisor at infinity along which $\Omega$ has a pole.   Kontsevich's criterion is somewhat similar to the conditions
under which a hyper-Kahler version of the theorem of Tian and Yau on complete Calabi-Yau metrics \cite{TY} would predict that $Y$ has a complete
hyper-Kahler metric.  (Such a theorem does not seem to be available except in complex dimension two, where Calabi-Yau metrics are automatically
hyper-Kahler.) 

In some important cases discussed in \cite{KW,GW}, one can show
directly (without knowledge about  the quantum $\sigma$-model) that $\sigma$-model perturbation theory terminates 
after finitely many steps and therefore that the $A$-model provides an actual
deformation.   This happens if $Y$ has a scaling symmetry, or at least an asymptotic scaling symmetry, under which $\hbar$ scales
with positive degree.    

For example, consider the complex variety $x^2+y^2+z^2=\j^2$ (eqn. \ref{osuc}).    This manifold is asymptotic at infinity to the cone
$x^2+y^2+z^2=0$.   It has an asymptotic scaling symmetry $(x,y,z)\to \lambda(x,y,z)$, under which $\j$ scales with degree 1.
We consider the holomorphic symplectic form $\Omega=\d x\,\d y/z\hbar$, where $\hbar$ also scales with degree 1.  The scaling
symmetry means that nothing is lost by setting $\hbar=1$, after learning that it makes sense to do so.
The classical ring $\A_0$ is generated by $x,y,z$, all of degree 1, with commutativity  relations such as $xy=yx$ and the additional relation
$x^2+y^2+z^2=\j^2$.   The deformed ring will be invariant under joint scaling of $x,y,z,\j,\hbar$, and also has a manifest $\SO(3)$ symmetry
rotating $x,y,z$,   Taking these facts into account, one finds \cite{GW} that the deformed ring is given by $\h x \h y-\h y \h x =\hbar \h z$,
and cyclic permutations of this statement, along with $\h x^2+\h y^2+\h z^2=\j^2-\frac{1}{4}\hbar^2$, where only the numerical coefficient
$\frac{1}{4}$ in the last term requires some analysis.   This example also illustrates the other criteria; the space $Y$ admits a complete hyper-Kahler
metric (the Eguchi-Hansen metric), and it has a compactification with the properties required by Kontsevich (namely the 
quadric  $x^2+y^2+z^2=w^2$ in $\bCP^3$).    One can similarly consider an affine variety defined by
a polynomial of higher degree. For the complex symplectic manifold $x^n+y^n+z^n=b$
(where $b$ is  a complex constant),
with symplectic form $\d x \d y/z^{n-1}$, one finds that both criteria are satisfied for $n=3$ and neither for $n>3$.

Complex cotangent bundles, which we will discuss at length in Section \ref{cotangent}, have an exact scaling symmetry, by virtue of which they
are examples with an actual deformation,    
even though in general they are believed not to admit complete hyper-Kahler metrics.   They do satisfy the criterion of Kontsevich.

\subsection{Lagrangian $A$-Branes And Quantization}\label{abq}

\subsubsection{Definition Of The Hilbert Space}\label{hilbdef}

Now we want to incorporate a Lagrangian $A$-brane and explain the relation to quantization.   We follow the explanation in
\cite{GW}, Section 2.3.   Some important illustrative examples had been analyzed earlier \cite{AZ}.

To quantize the  real symplectic manifold $M$,
we pick a rank 1 Lagrangian $A$-brane $\B$ with support $M$ and some $\CP$ bundle.    What characterizes a rank 1 Lagrangian $A$-brane is that
the curvature $\sF$ of its $\CP$ bundle satisfies
$\sF+\sB=0$.   As we have taken $\sB=\omega_J$, the curvature must be $\sF=-\omega_J$.   Our goal, as described in Section \ref{amo},
is to quantize $M$ with the symplectic structure $\omega_J$.   We recall as well that in geometric quantization, the first step is to introduce
a prequantum line bundle $\frL$ with curvature equal to the symplectic form, in this case $\omega_J$.   As this is the negative of the curvature of $\frL$,   
we can interpret the $\CP$ bundle of $\B$
 as $\frL^{-1}$.   One sign that this identification makes sense is that the anomalies match.  What is loosely
called the prequantum line bundle in geometric quantization is more accurately a $\spinc$ structure, as we discussed in Section \ref{gq}.
And similarly, the $\CP$ bundle of a rank 1 brane is really a $\spinc$ structure rather than a complex line bundle, as sketched in footnote \ref{anomdesc}.

The Hilbert space for brane quantization of $M$ with the prequantum line bundle $\frL$ is then defined to be
$\H=\Hom(\B,\B_\cc)$, the space of physical states of the $(\B,\B_\cc)$ system.   The dual space is $\H'=\Hom(\B_\cc,\B)$.
The reason that we identify $\H$, not $\H'$, as the Hilbert space is that $\A=\Hom(\B_\cc,\B_\cc)$ acts on $\H=\Hom(\B,\B_\cc)$ on the left,
while it acts on $\H'=\Hom(\B_\cc,\B)$ on the right.  To the extent that one prefers to work with left actions of a ring if possible, it is natural
to define $\H$ as we have.   In referring to $\H$ as a Hilbert space, not just a vector space, we are referring to the
fact that $\H$ has a natural hermitian inner product; this is described in Section \ref{hermitian}.

To understand the motivation for this definition of the quantum Hilbert space, we construct the effective action
 for low-energy states of an open string whose left
and right boundaries end on $\B_\cc$ and $\B$, respectively.  Let the string worldsheet be $\R\times I$, where $\R$ is parametrized
by $\tau$ and $I$ is an interval.    The boundary conditions at the two ends of $I$ are
such that there are no fermion zero-modes.  See Appendix \ref{details} for an explanation of
this statement.     The boson zero-modes are constant along $I$ and so describe a map $X:\R\to M$, which we can
describe by functions $x^i(\tau)$, where $x^i$ are local coordinates for $M$.   What is the effective action for $x^i(\tau)$? The bulk effective
action contributes various terms that are quadratic in $\d x^i/\d \tau$, but the most relevant terms are linear in $\d x^i/\d \tau$ and
come from the boundary couplings to the $\CP$ bundles at the two ends of the string.   Referring back to eqn. (\ref{kogo}), we see that in the definition of
 $\Hom(\B,\B_\cc)$, we have the $\CP$ bundle of $\B_\cc$ at one end of the string and the dual, or inverse, of the $\CP$ bundle
of $\B$ at the other end.   The $\CP$ bundle of $\B_\cc$ is completely trivial, and the $\CP$ bundle of $\B$ is $\frL^{-1}$, so its dual
is $\frL$, the prequantum line bundle.     If $\sA$ is the connection on $\frL$, then the contribution of the coupling of the string endpoint to $\frL$ to the low
energy effective action is
just
\be\label{kixx}\sI=\int_\R x^*(\sA). \ee
This is the action that was introduced Section \ref{uses}, where the connection on the prequantum line bundle is called $\sT/\hbar$.   
As explained there, this is formally the action associated to quantization.

  As we have discussed in Section \ref{uses}, the actions of eqns. (\ref{iff}) or 
 (\ref{kixx}) actually
cannot be quantized without more information.   The correct statement is not that the $A$-model computation of $\Hom(\B,\B_\cc)$
reduces to quantization of  the action (\ref{kixx}).   Rather, the  fact that the $A$-model is well-defined (for suitable $Y$) while quantization of eqn.
(\ref{kixx}) is ambiguous means that some information has been lost in the reduction to (\ref{kixx}).  The right way
to state the conclusion is that $\H=\Hom(\B,\B_\cc)$ is the answer for quantization of $M$ with prequantum line bundle $\frL$ that
is provided by the $A$-model.

\subsubsection{Some Questions}\label{ques}

The claim that the $A$-model is providing this answer for the quantization of $M$ raises a number of questions.  Here are some obvious ones:

(1)  The space of physical states of the $A$-model is not usually a Hilbert space.   Usually it does not have a 
hermitian inner product of any kind (even an indefinite
one).  How does the $A$-model generate a hermitian inner product in the present context?

(2) What are the observables that act on the quantum Hilbert space $\H$?

(3) Is this notion of quantization compatible with standard methods of quantization when those are available?

The first question will be answered in Section \ref{hermitian}.

Concerning the second question, the immediate answer that the $A$-model provides is that $\A=\Hom(\B_\cc,\B_\cc)$
acts on $\H=\Hom(\B,\B_\cc)$.    This answer is most satisfactory if $M$ is  a real affine variety and $Y$ is its complexification.
In this case, as we learned in Section \ref{abranes}, $\A$ is an associative noncommutative deformation of the 
 ring $\A_0$ of polynomial functions on $Y$.   The ring
of polynomial functions  on $Y$ is the same as the ring of polynomial functions on $M$  (a polynomial function on $M$ can be
analytically continued to a polynomial function on $Y$, and 
a nonzero polynomial on $Y$ has a nonzero restriction to $M$).  So the functions on $M$
that are quantized are the polynomial functions.   That is a fairly satisfactory answer.   For example, if $M=\R^{2n}$, the functions
that are being quantized are simply the polynomials in the coordinates and momenta.
In Section \ref{hermitian}, we will learn that the elements of $\A$  that correspond to hermitian operators on $\H$ are the ones that are $\tau$-invariant.

However, quantization  of a symplectic manifold $M$ via branes makes sense for many examples other than real affine varieties.
In many other instances,  $\A_0$ is much smaller, and then we do not get such a rich set of functions
on $M$ that are quantized to operators on $\H$.     It is then less obvious, in general, what
are interesting observables.    However, there is another way to define operators that act on $\H$, namely operators associated to Lagrangian
correspondences.   This will be explained in Section \ref{symq}.  
 For example, in a case relevant to geometric Langlands that will be studied elsewhere \cite{GaW},
$\A_0$ is ``small'' and $\A$ is actually abelian and isomorphic to $\A_0$.   In that particular example,  it is important to consider
operators associated to Lagrangian correspondences (Hecke correspondences, to be precise, which physically are associated to 't Hooft operators).   

The last question of comparing quantization by branes to standard methods of quantization is the subject of Section \ref{comparing}.

\subsubsection{Generalizations}\label{genera}

So far, we have considered a Lagrangian brane that is supported on  $M$, a component of the fixed point set of 
an antiholomorphic involution $\tau:Y\to Y$.   
As we will discuss in Section \ref{hermitian}, this condition is important in defining a hermitian inner product
on the space $\H$.   But other types of Lagrangian brane are also important.

  Since we consider $Y$ as a symplectic manifold with symplectic structure
$\omega_K$, by definition to say that $L$ is a Lagrangian submanifold means that the restriction of $\omega_K$ to $L$ vanishes.
But there is no general constraint on the restriction of $\omega_J$ to $L$.   If that restriction vanishes as well, then $L$ is actually
Lagrangian for the complex symplectic structure $\Omega_I=\omega_J+\i\omega_K$ of $Y$.  Such an $L$ is a complex
submanifold\footnote{If $\omega_J$ and $\omega_K$ vanish when restricted to $L$, this means that, along $L$, each is a pairing
between the tangent bundle and normal bundle to $L$ in $Y$.   It follows that $I=\omega_K{}^{-1}\omega_J$ maps the tangent
bundle of $L$ to itself, so $L$ is a complex submanifold.} of $Y$, so it is actually a complex Lagrangian submanifold. Since $\Hom(\B,\B_\cc)$ is a module for $\A=\Hom(\B_\cc,\B_\cc)$
for any $A$-brane $\B$, in particular this is so if $\B$ is a Lagrangian brane supported on a 
complex symplectic manifold $L$.  Such branes are of type $(B,A,A)$ in the sense that they are $B$-branes in complex structure $I$
(since $L$ is a complex submanifold, and the $\CP$ bundle of $\B$, being flat, is holomorphic in complex structure $I$), and $A$-branes for symplectic
structures $\omega_J$ and $\omega_K$ (since $L$ is Lagrangian in each of these structures and its $\CP$ bundle is flat).   
   Modules associated to branes of type $(B,A,A)$ are interesting and will be important tools 
in Section \ref{comparing}, but they are not closely related to quantization of $L$.  Indeed, since both $\omega_J$ and $\omega_K$ vanish
when restricted to $L$, nothing in this discussion has provided a symplectic
structure on $L$, so we do not have a natural problem of quantizing $L$.   (If $Y$ is hyper-Kahler, it has a third symplectic structure, namely 
$\omega_I$, and a complex Lagrangian submanifold $L$ is symplectic with respect to $\omega_I$.   But  the $A$-model that we are studying and the branes $\B_\cc$ and $\B$
 are not  sensitive to $\omega_I$.)

The case that is related to quantization of $L$  is the opposite case that $\omega_J$ is nondegenerate when restricted to $L$.     If $\B$ is
a rank one $A$-brane supported on any such $\L$, then  
 $\Hom(\B,\B_\cc)$  has most of the properties that one would expect for a quantization of $L$ with symplectic
 structure $\omega=\omega_J|_L$.   Only one step in the  construction fails, namely the definition of a hermitian inner
 product.   As we will explain in Section \ref{hermitian}, 
 the definition of this inner product requires that $L$ should be the
 fixed point set of an antiholomorphic involution of $Y$.  Intuitively, one should expect this. If $L$ is not the fixed point set of an antiholomorphic
 symmetry of $Y$, then it is not natural for a holomorphic function on $Y$ to be real when restricted to $L$, so there is no evident
 way to decide  what holomorphic functions on $Y$ should become hermitian operators in a quantization of $L$.  Hence
 it should not be possible to define a hermitian inner product on $\Hom(\B,\B_\cc)$.
 
 In Section 3.7 of \cite{GW}, examples are described of Lagrangian
 submanifolds $L$ of a coadjoint orbit of $\SL(2,\C)$ (the affine variety  $Y$ 
 defined by eqn. (\ref{cosu})), such that $\omega_J$ is nondegenerate when
 restricted to $L$, but $L$ is not the fixed point set of any antiholomorphic involution of $Y$.   Such $L$'s were related, for example, to
various kinds of interesting but not unitary representations of the group $\SL(2,\R)$ or its Lie algebra. 
 
 Another generalization is to consider the case that $\B$ is a Lagrangian brane supported on $M$ (or on a more general Lagrangian
 submanifold $L$)
 of rank $m>1$.   In discussing
 this case, let us assume for simplicity that $M$ is a spin manifold, so that we do not have to discuss the usual anomaly involving spin.
 The $\CP$ bundle of $M$ is then a flat vector bundle $\U\to M$ of rank $m$, with structure group $\UU(m)$.   In this situation,
 $\Hom(\B,\B_\cc)$ will  be a sort of rank $m$ generalization of the
 notion of quantization.  The existence of this generalization of the notion of quantization may come as a slight surprise.   However, it is not hard to see
 that such a generalization does exist in the cases in which geometric quantization is available.   For example, suppose that $M=T^*N$
 for some $N$.   Then in the spirit of geometric quantization, one can consider the Hilbert space consisting of $\ltwo$ sections of
 $\K^{1/2}\otimes \U\to N$.   Alternatively, suppose that $M$ has a Kahler polarization.   Then one can consider the Hilbert space
 $H^0(M,K^{1/2}\otimes \U)$.

\subsection{Hermitian Inner Product}\label{hermitian}

Here, following Section 2.4 of \cite{GW}, we will discuss what is required to define a hermitian inner product in the $A$-model.
Actually the two cases of the topological $A$-model and $B$-model can be treated in parallel, and 
 the $B$-model case is also interesting.   So we consider the two cases together: we consider
any two-dimensional topological field theory that is obtained by  twisting
a unitary, supersymmetric, two-dimensional $\sigma$-model.   The twisting is made by selecting a linear combination $Q$ of the
supercharges that satisfies $Q^2=0$, and treating it as a differential.

The usual inner product in such a topological field theory is a bilinear form (linear in each variable), not
a hermitian or sesquilinear form (linear in one variable, antilinear in the other).   Thus, if $\B_1$ and $\B_2$ are two branes,
there is always a nondegenerate bilinear pairing between $\Hom(\B_1,\B_2)$ and $\Hom(\B_2,\B_1)$.  It can be computed
by a two-point function on a disc.   So finding
a sesquilinear pairing of $\Hom(\B_1,\B_2)$ with itself is equivalent to finding an antilinear map from $\Hom(\B_1,\B_2)$
to $\Hom(\B_2,\B_1)$.

The underlying $\sigma$-model always has an antilinear symmetry, the $\CPT$ symmetry that
we will call $\Theta$.   It reverses the orientation of a string, and so maps $\Hom(\B_1,\B_2)$ to $\Hom(\B_2,\B_1)$.    However, $\Theta$
does not do the job we want, because it is not a symmetry of the topological field theory defined using a given $Q$.   Such a $Q$ is
never hermitian (since it obeys $Q^2=0$), and $\Theta$ conjugates $Q$ to $Q^\dagger$.  In other words, $\Theta$ maps an $A$- or
$B$-model with differential $Q$ to a conjugate $A$- or $B$-model with differential $Q^\dagger$.

To find an antilinear symmetry of the topological field theory associated to a given $Q$, we need to combine $\Theta$ with a linear
symmetry $\tau$ of the $\sigma$-model 
that conjugates $Q^\dagger$ back to $Q$.   Such a symmetry can exist in either an $A$-model or a $B$-model.
In an $A$-model, $\tau$ should reverse the sign of the symplectic form $\omega$ that is used in defining the $A$-model.\footnote{A diffeomorphism
of a real symplectic manifold that reverses the sign of the symplectic structure is said to be ``antisymplectic,'' by analogy with the use of the term
``antiholomorphic'' to describe a diffeomorphism that reverses the sign of the complex structure.}       
In a $B$-model, $\tau$ should reverse the sign of the complex structure $J$ that is used to define the $B$-model.     In other
words, we need $\tau^*(\omega)=-\omega$ in the $A$-model, and $\tau^*(J)=-J$ in the $B$-model. $\Theta_\tau=\tau\Theta$ will
then be an antilinear map from the $A$-model or $B$-model to itself.

More generally, in the presence of a $B$-field, $\Theta$ maps an $A$-model with complexified symplectic form $\omega_\C=\omega-\i \sB$ to
a conjugate $A$-model with $\omega_\C$ replaced by $-\bar\omega_\C$.
To compensate for this, while reversing the sign of $\omega$, $\tau$ should satisfy $\tau^*(\sB)=\sB$.

To get the best properties, 
$\tau$ should  be self-adjoint, satisfy $\tau^2=1$, and commute with $\Theta$.   We will assume
these conditions, which imply that   $\Theta_\tau^2=1$ and $\Theta_\tau$ is self-adjoint.

In general, $\tau$ will map branes $\B_1$ and $\B_2$ to some other branes $\bar \B_1$ and $\bar \B_2$, and $\Theta_\tau$
will then map $\Hom(\B_1,\B_2)$ to $\Hom(\bar\B_2,\bar\B_1)$.   
 So $\Theta_\tau$, composed with the natural bilinear pairing $(~,~)$ of the $A$- or $B$-model, will give a
sesquilinear pairing between  $\Hom(\B_1,\B_2)$ and $\Hom(\bar\B_1,\bar\B_2)$, defined by
\be\label{wofo} \langle \psi,\psi'\rangle =(\Theta_\tau \psi,\psi'). \ee
To get a sesquilinear pairing of $\Hom(\B_1,\B_2)$
with itself, we need $\bar\B_1=\B_1$, $\bar\B_2=\B_2$.   In other words, the branes considered must be $\tau$-invariant.    When that
is the case, it makes sense to ask if the sesquilinear form is hermitian, which means that $\bar{\langle \psi,\psi'\rangle}=\langle \psi',\psi\rangle$.
This condition is actually satisfied, because $(~,~)$ is symmetric and $\Theta_\tau$ is self-adjoint and antilinear.

For $\tau$-invariant branes $\B_1,$ $\B_2$, let us work out a general description of the adjoint of an operator $\O:\Hom(\B_1,\B_2)\to \Hom(\B_1,\B_2)$.
First of all, in general there is a dual or transpose operator $\O^*$ from the dual space $\Hom(\B_2,\B_1)$ to itself.    $\O^*$ is characterized by
the condition that for $\psi\in \Hom(\B_1,\B_2)$,
$\chi\in\Hom(\B_2,\B_1)$, 
\be\label{chardual} (\O^*\chi,\psi) =(\chi,\O\psi). \ee
On the other hand, for $\psi,\psi'\in \Hom(\B_1,\B_2)$, the adjoint operator $\O^\dagger$ should obey
\be\label{addual}\la \O^\dagger\psi',\psi\ra =\la\psi',\O\psi\ra. \ee
Combining these conditions with the definition of $\la~,~\ra$, we get $(\Theta_\tau\O^\dagger\psi',\psi)=(\Theta_\tau\psi',\O\psi)=(\O^*\Theta_\tau\psi',\psi)$.  
So $\Theta_\tau \O^\dagger=\O^*\Theta_\tau$, or, as $\Theta_\tau^2=1$,
\be\label{adjform} \O^\dagger=\Theta_\tau \O^*\Theta_\tau.\ee

Let us examine this construction of a hermitian product for the case that $Y$ is the complexification of a real symplectic manifold $M$.   $M$ is then a fixed
point set of an antiholomorphic involution $\tau:Y\to Y$, as discussed in Section \ref{amo}.   Therefore, a Lagrangian brane $\B$ supported
on $M$ is $\tau$-invariant.   Since we consider the $A$-model with $\omega=\omega_K$ and $\sB=\omega_J$,
 we want $\omega_K$ to be odd under $\tau$ (to satisfy $\tau^*(\omega)=-\omega$)
and $\omega_J$ to be even (so that $\sB$ is $\tau$-invariant\footnote{Alternatively, we could make a $B$-field gauge
transformation to set $\sB=0$ and give $\B_\cc$ a $\CP$ curvature $\omega_J$.   Then we would want $\tau^*(\omega_J)=\omega_J$ to make $\B_\cc$
$\tau$-invariant.}).   These conditions are satisfied (eqn. (\ref{tabush})).  The brane $\B_\cc$ is $\tau$-invariant, since its $\CP$ bundle is trivial.  
 So we do get
a hermitian structure on $\H=\Hom(\B,\B_\cc)$.  The same construction gives a hermitian structure on the dual space $\H'=\Hom(\B_\cc,\B)$.
These structures are equivalent under the antilinear isomorphism $\Theta_\tau:\H\cong \H'$.

Since $\A=\Hom(\B_\cc,\B_\cc)$ acts linearly on $\H$, 
we can now ask what elements $x\in \A$ correspond to hermitian operators on $\H$.   With the above definition of the hermitian
structure, the answer is that $x$ corresponds to a hermitian operator if and only if it commutes with $\Theta_\tau$.  Since $\Theta_\tau$
is antilinear and $\Theta_\tau^2=1$, every $x$ can be written $x=x_1+\i x_2$, where $x_1$ and $x_2$ commute with $\Theta_\tau$ and become
hermitian after quantization.

 In general, for $\tau$-invariant branes $\B_1,\B_2$,
the hermitian form that we have defined on $\Hom(\B_1,\B_2)$ is not positive-definite.   It is always nondegenerate, since $\Theta_\tau$ is an isomorphism
and the bilinear pairing $(~,~)$ of the $A$- or $B$-model is nondegenerate.   

In the more specific context of quantization,  the best we can say is that sufficiently close to a classical limit,
this hermitian form is positive-definite.  For example, in the case of a Kahler polarization, one is sufficiently close to a classical limit if the periods of the symplectic
form are large enough; in quantization of a cotangent bundle, one always expects positivity.

\subsection{Smooth Functions And Distributions}\label{smooth}

\begin{figure}
 \begin{center}
   \includegraphics[width=4in]{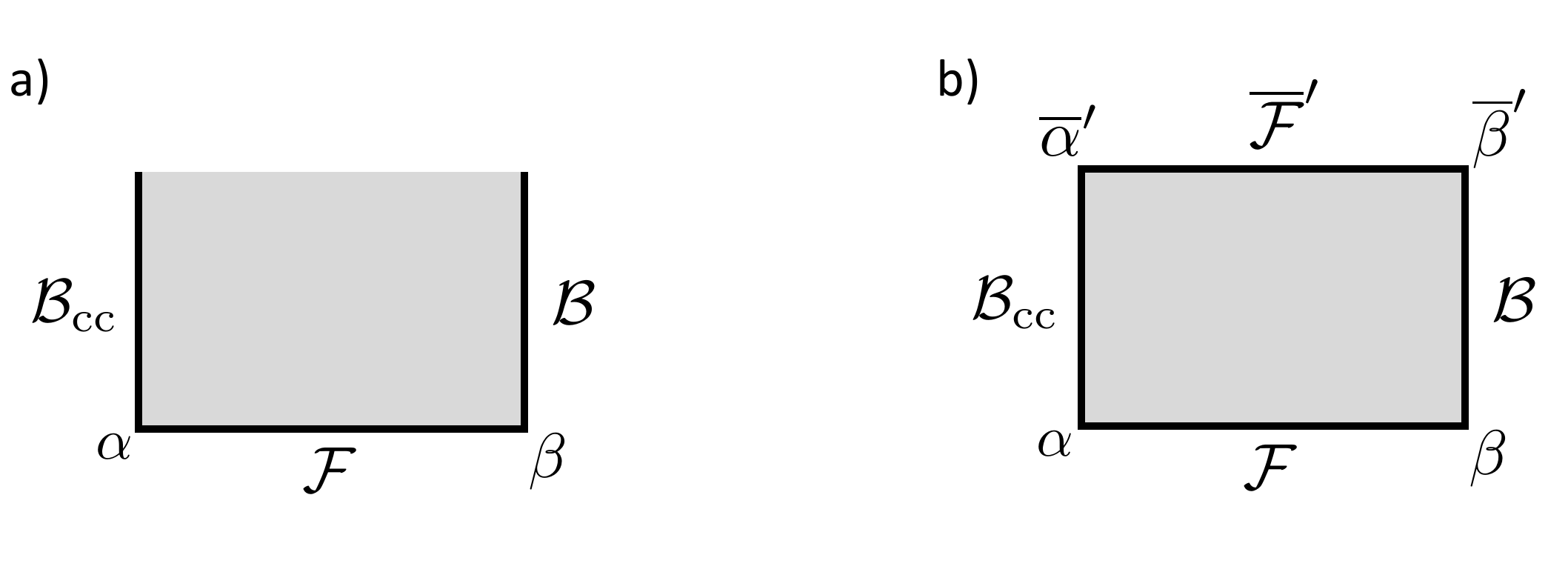}\caption{\small{(a) An auxiliary brane $\F$, with certain additional data, can be used
   to make a state in $\Hom(\B,\B_\cc)$.   In general this is a distributional state.  (b) Inner products between states constructed using branes $\F$, $\F'$
   can be computed by a path integral on a rectangle with $\F$ at the bottom and $\bar\F'$ -- the $\tau$ conjugate of $\F'$ -- at the top.  }  \label{outline}}
 \end{center}
\end{figure}

We will conclude this introduction by sketching a useful way to exhibit elements of $\H=\Hom(\B,\B_\cc)$.
More detail will be presented in Section \ref{comparing}.
The main ingredient is  another $A$-brane $\F$.   In practice, we will generally choose $\F$ to be a rank 1 brane supported on some
Lagrangian submanifold $L\subset Y$.   $L$ will typically {\it not} coincide with the phase space $M$ whose quantization we are studying.   
Then picking elements $\alpha\in \Hom(\F,\B_\cc)$ and $\beta\in \Hom(\B,\F)$, we compose them to make $\psi=\alpha\circ\beta\in \Hom(\B,\B_\cc)$,
as sketched in  fig. \ref{outline}(a). Similarly if $\F'$ is yet another $A$-brane, also equipped with $\alpha'\in \Hom(\F',\B_\cc)$, $\beta'\in \Hom(\B,\B')$, we
can compose them to make $\psi'=\alpha'\circ\beta'\in \Hom(\B,\B_\cc)$.     The inner product $\la\psi',\psi\ra$, defined as in Section \ref{hermitian},
can be computed by a path integral on a rectangle; see fig. \ref{outline}(b).  

However, we should ask what kind of states are $\psi$ and $\psi'$.    Here we have in mind questions of analysis like the distinction, in the quantum mechanics
of a particle moving in ordinary space $\R^3$, between different classes of wavefunction.   The arena of quantum mechanics of a particle on $\R^3$ is usually understood
to be the Hilbert space of $\ltwo$ functions.  But in a well-motivated physics problem, the wavefunction of a particle is usually much nicer than that -- a smooth function
with rapid decay at infinity.   Mathematically, such functions make up what is known as a Schwartz space.   On the other hand, 
 in developing quantum mechanics, it is convenient to also consider unnormalizable states, such as a wavefunction $\delta^3(\vec x-\vec x_0)$ with
delta function support at a point $\vec x_0\in \R^3$, or a plane wave $\exp(\i \vec p\cdot \vec x)$, which corresponds to a wavefunction with delta
function support in momentum space.     Mathematically, these unnormalizable states are called distributions; a distribution, by definition, has a well-defined
inner product with any vector in the Schwartz space.

In the case of quantization of a compact phase space $M$, the Hilbert space $\H$ is finite-dimensional and no such questions of analysis will arise.
However, if $M$ is not compact, we should ask what sort of wavefunctions are produced by the construction that was just sketched.  A simple observation is that if either
$\F$ or $\F'$ has compact support, one can expect the amplitude associated with the rectangle of fig. \ref{outline}(b) to be well-defined and convergent.
If neither $\F$ nor $\F'$ has compact support, in general this amplitude may not be well-defined.   This motivates the following tentative interpretation.  A completely
general $A$-brane $\F$ produces, in this construction, a distributional state that is not normalizable.    But an $A$-brane $\F$ with compact support produces
the analog of a vector in Schwartz space -- a vector that can be paired with an arbitrary distribution.

\section{Comparing Quantization by Branes to Geometric Quantization}\label{comparing}

\subsection{Complex Cotangent Bundles and Deformation Quantization}\label{cotangent}

Geometric quantization and quantization by branes are two different ways to introduce additional structure that is needed
in order to quantize a symplectic manifold $M$.    Are there  circumstances in which these two methods of quantization
agree?    The best answer we can hope for is to find conditions on a complexification $Y$  of $M$ and a polarization $\P$ of $M$
such that brane quantization based on $Y$ is equivalent to geometric quantization based on $\P$.   In the present section,
we will explain a criterion for such equivalence  for the two basic types of polarization that were introduced in Section \ref{gq}:
real polarizations of $M$ and complex  polarizations.  

We begin by describing the most obvious situation in which one might hope to compare brane quantization to geometric quantization.
Suppose that $M=T^*N$ is a cotangent bundle, with the natural symplectic structure $\omega=\sum_i \d p_i \d q^i$, where $q^i$ are local coordinates
on $N$ and $p_i$ are linear functions on the fibers of the cotangent bundle.   Suppose further that $W$ is a complexification of $N$ in the sense
of Section \ref{amo}.   Then  $Y=T^*W$ is
a complexification of $M$.  
The natural holomorphic symplectic form of $Y$, which can again be written as
 $\Omega=\sum_i \d p_i \d q^i$ (where now the $q^i$ and $p_i$ are local holomorphic coordinates on $W$ and on the fibers of $T^*W\to W$), is
 an analytic continuation of $\omega$.   If the $A$-model of $Y$ exists, it can be used to quantize $M$, and the result can be compared to geometric
 quantization using the description of $M$ as a cotangent bundle, $M=T^*N$.    
 
 First we discuss deformation quantization of $Y$, in other words, we discuss $\Hom(\B_\cc,\B_\cc)$.    The cotangent bundle has a scaling symmetry
 that is highly constraining.   To exploit this symmetry, it is convenient to generalize slightly to the case $\Omega=\sum_i \d p_i\d q^i/\hbar$ (where the parameter
 $\hbar$ can be set to 1 once one knows it makes sense to do so).   
 There is a scaling symmetry under which the $q^i$ are invariant while $p_j$ and $\hbar$ have degree 1.

 The scaling symmetry is an action of $\C^*$ on these variables.  We will assume that deformation quantization should be carried out in a $\C^*$ invariant way.
 This is so if a $\UU(1)$ subgroup of $\C^*$ is a symmetry of an underlying $\sigma$-model,
 from which the $A$-model of interest is deduced.  (This is true in an application to geometric Langlands that will be discussed elsewhere \cite{GaW}.)
Holomorphy will promote such a $\UU(1)$ symmetry to $\C^*$.   In the absence of an underlying $\UU(1)$ symmetry,  deformation quantization via $\B_\cc$
might not be $\C^*$-invariant.
 
 Suppose for example that $W=\C^n$, $Y=\C^{2n}$, so that the coordinates $q^i$, $p_j$ can actually  be defined globally.  
  The Poisson brackets
 are $\{p_i,q^j\}=\delta_i^j$, $\{q^i,q^j\}=\{p_i,p_j\}=0$.   Requiring that quantum commutators agree with Poisson brackets up to corrections of order $\hbar^2$
 and imposing the scaling symmetry, the most general possible form of the quantum commutators is
 \begin{align}\label{genform} [\h p_i,\h q^j] & = -i\hbar \delta^j_i   \cr
                                               [\h p_i,\h p_j ]   &    =    \hbar^2 f_{ij}(\h q^k)    \cr
                                              [\h q^i,\h q^j]& = 0 . \end{align}
   The Jacobi identity $[\h p_i,[\h p_j,\h p_k]]+\cdots =0$ implies that $\partial_i f_{jk}+\partial _j f_{ki}+\partial_k f_{ij}=0$, so there are some functions
   $g_k$ with $f_{ij}=\partial_i g_j-\partial_j g_i$.   Then upon replacing $\h p_i$ by $\h p_i' =\h p_i -\i \hbar g_i(q^k)$ -- a substitution that is consistent
   with the scaling relation -- we reduce eqn. (\ref{genform}) to
   \begin{align}\label{genformtwo} [\h p'_i,\h q^j] & = -i\hbar \delta^j_i  \cr 
   [\h p'_i,\h p'_j]&=0 
   \cr 
    [\h q^i,\h q^j]& = 0 . \end{align}     
    The quantum-deformed algebra of polynomial functions of $p_i$ and $q^j$ -- which are the only functions we allow, as discussed in Section \ref{abranes} --
    is therefore the algebra we get if we represent $q^j$ as a multiplication operator, and $p_i$ as $-\i\hbar\frac{\partial}{\partial q^i}$.   Of course
    this last formula is familiar from elementary quantum mechanics.
With this characterization of $\h q^i$ and $\h p_j$, 
   the associative algebra $\Hom(\B_\cc,\B_\cc)$  can be characterized as the algebra
  of polynomial holomorphic differential operators on $\C^n$.
      In making this relation to differential operators, it is important to only allow functions whose dependence on the $p$'s
  is polynomial, and we will always impose this condition in what follows.  Similarly we only allow a polynomial dependence on $q$.  Thus a function of $\h p$ and $\h q$
  corresponds to a polynomial holomorphic differential operator.
  
  Now let us consider a more general $W$.   Locally, we can reduce to the previous case by replacing $W$ with an open set $\U\subset W$, chosen so
  that $\U$ is isomorphic to a region in $\C^n$, and replacing $Y$ with
  $T^*\U$.  Replacing the target space $Y=T^*W$ by an open set, in this case $T^*\U$, is in general not a good operation in a $\sigma$-model.   However, 
  since $\sigma$-model perturbation theory involves small fluctuations around constant maps to the target space, there is no problem in $\sigma$-model perturbation
  theory to restrict to an open set.   Usually this would give only perturbative information, but we are in an unusual situation in which this is enough,
  since the scaling symmetry implies that perturbation theory up to order $\hbar^2$ gives the full answer.
  We have chosen an open set of the form $T^*\U$ because such a set is invariant under scaling, so that the scaling
  argument  applies.   Therefore, it is possible to think of $\Hom(\B_\cc,\B_\cc)$ as a sheaf over $W$: to any open set $\U\subset W$, we associate
  $\Hom(\B_\cc,\B_\cc)$ over $T^*\U$.
  We should acknowledge that while this step makes sense and seems very natural mathematically, it is not clear that it is entirely natural physically.
  It would be desirable to find a more natural physical interpretation of this step, or possibly a way to avoid it in the following arguments.
  
  If we choose $\U$ small enough as to be isomorphic to an  open ball in $\C^n$, then matters are simple.
  Once we do restrict to $T^*\U$ for such a $\U$,
the coordinates $q_i$, $p^j$ can be defined globally, and promoted as before to operators that represent
  $(\B_\cc,\B_\cc)$ strings.  Therefore the above analysis applies in any such set.
  
  Now we cover $W$ with small open sets $\U_\alpha$ that are each isomorphic to an open in $\C^n$, and
  have topologically trivial intersections.   $Y$ is then covered by the open sets $T^*\U_\alpha$.
 For each $\alpha$, we pick local coordinates $q^i_\alpha$ for $\U_\alpha$, and fiber coordinates $p_{j,\alpha}$ for the cotangent bundle.
    In intersection  regions $T^*(\U_\alpha\cap \U_\beta)$, the relations between the coordinates are 
 \begin{align}\label{relco} q^i_\alpha & = h^i_{\alpha\beta}(q^k_\beta) \cr
                                 p_{i\,\alpha}& = M_{\alpha\beta}(q)^k_i p_{k,\beta}, \end{align} 
                                 where $h^i_{\alpha\beta}$ are some holomorphic functions and     $M_{\alpha\beta}{}^i_j$ is the inverse matrix to $\partial_i h^j_{\alpha\beta}$.   
                                 
 At the quantum level, the scaling symmetry does not allow any change in the transition formula for the $q$'s, so we must have $\h q^i_\beta =h^i_{\alpha\beta}(\h q^k)$,
 with the same functions of the $q$'s as before.   However, for the $p$'s, the scaling symmetry allows a correction of order $\hbar$, so that the formula
 would become $\h p_{i\,\alpha}     = M_{\alpha\beta}(\h q)^k_i \h p_{k,\beta}+\i \hbar r_{i\,\alpha\beta}(\hat q)$ for some holomorphic 
 functions $r_{i\,\alpha\beta}$.   However, to satisfy 
 $[\h p_{i\,\alpha},\h p_{j\,\beta}]=0$, we need $\partial_i r_{j\,\alpha\beta}-\partial_j r_{i\,\alpha\beta}=0$, and therefore (since  $\U_\alpha\cap \U_\beta$ is topologically trivial)  
 $ r_{j\,\alpha\beta}=\partial_j w_{\alpha\beta}(\h q)$ for some  holomorphic functions 
 $w_{\alpha\beta}$.  The transformation of the $\h p$'s is then 
 \be\label{transp} \h p_{i\,\alpha}     = M_{\alpha\beta}(\h q)^k_i \h p_{k,\beta}+\i \hbar \partial_i w_{\,\alpha\beta}(\hat q).\ee
 We can make this formula look much simpler if we use a single set of coordinates $q^i$ in the region $\U_\alpha\cap\U_\beta$ (rather than
 two sets of coordinates $q^i_\alpha$ and $q^i_\beta$).   Then the matrix $M$ is replaced by 1 and the formula becomes simply
 \be\label{transpo} \h p_{i\,\alpha}=\h p_{i\,\beta}+\i\hbar \partial_i w_{\alpha\beta}.\ee
 This isolates the nonclassical part of the transformation of $\h p$.
 
   It is immediate that $w_{\alpha\beta}=-w_{\beta\alpha}$, 
 and  consistency in triple overlaps tells us that $\partial_k(w_{\alpha\beta}+w_{\beta\gamma}+w_{\gamma\alpha})=0$,
 and therefore that $\phi_{\alpha\beta\gamma}= w_{\alpha\beta}+w_{\beta\gamma}+w_{\gamma\alpha} $    is a complex constant.
 In quadruple intersections, these constants automatically satisfy 
 \be\label{cocrel}\phi_{\alpha\beta\gamma}-\phi_{\beta\gamma\delta }+\phi_{\gamma\delta\alpha}-\phi_{\delta\alpha\beta}=0,\ee
 and therefore define an element $[\phi]$ of $H^2(W,\C)$.
 
 If $[\phi]$ is actually valued in $2\pi\i\cdot H^2(W,\Z)$, meaning that, after possibly shifting the $w_{\alpha\beta}$ by constants,  
 each $\phi_{\alpha\beta\gamma}$ is an integer multiple of $2\pi \i$,
 then the quantities $S_{\alpha\beta}=\exp(w_{\alpha\beta})$   satisfy $S_{\alpha\beta}S_{\beta\gamma}S_{\gamma\alpha}=1$ and can be regarded
 as the transition functions of a complex line bundle $\L\to W$.    Explicitly, $\L$ is as follows.  In each $\U_\alpha$, a section of $\L$
 is simply a function $\psi_\alpha$.   But the functions $\psi_\alpha$ and $\psi_\beta$ in different open sets $\U_\alpha$ and $\U_\beta$ are related
 by $\psi_\alpha=S_{\alpha\beta}\psi_\beta$.
  In this situation, the above formulas can be interpreted to mean that the quantization of
 a function of $q$ and $p$    gives a holomorphic differential operator acting on sections of $\L$.        In other words, suppose that $D_\alpha$ and $D_\beta$
 are holomorphic differential operators acting on functions in the respective open sets $\U_\alpha$ and $\U_\beta$.   In order to be able to interpret
 these as the restrictions to the given regions of a globally defined holomorphic differential operator on $W$, we need $D_\alpha (S_{\alpha\beta}\psi)
 =S_{\alpha\beta} D_\beta\psi$ or  $D_\alpha=S_{\alpha\beta} D_\beta S^{-1}_{\alpha\beta}$.   In particular, the operator that in region $\beta$ is
  $\h p_i=-\i\hbar\frac{\partial}{\partial q^i}$
 will in region $\alpha$ be $\h p_i+\i\hbar \partial_i \log S_{\alpha\beta}=\h p_i+\i\hbar \partial_i w_{\alpha\beta}$, matching (\ref{transp}).
 
This has the following generalization.\footnote{\label{zany} The situation that we will describe arises in important examples of quantization, but it is not
completely general.   For example, if $W$ is compact and the Hodge decomposition of $[\phi]$ has a piece of type $(2,0)$, then what we are about
to describe is not applicable.}  Let $\L\to W$ be a holomorphic line bundle, and
suppose that after picking trivializations of $\L$ in the open sets
$\U_\alpha$, $\L$ can be described by transition functions $e^{x_{\alpha\beta}}$ in $\U_\alpha\cap \U_\beta$.
Suppose that $w_{\alpha\beta}=\lambda x_{\alpha\beta}$ for some complex $\lambda$.   If $\lambda$ is an integer, then the functions $e^{w_{\alpha\beta}}=
e^{\lambda x_{\alpha\beta}}$ are transitions functions for a holomorphic line bundle, namely $\L^\lambda$.   In that case, the discussion in the last paragraph
applies and the quantization of a function of $q$ and $p$ is a holomorphic differential operator that acts on sections of $\L^\lambda$.   For generic
complex $\lambda$, a line bundle $\L^\lambda$ does not exist, but only because in triple overlaps $\U_\alpha\cap\U_\beta\cap \U_\gamma$,
the product of transition functions $e^{\lambda(x_{\alpha\beta}+x_{\beta\gamma}+x_{\gamma\alpha})}$ is a $c$-number, not equal to 1.   Since differential
operators commute with $c$-numbers, the notion of differential operators acting on $\L^\lambda$ makes sense whether $\L^\lambda$ exists as a line bundle or not.
And the quantized functions of $q$ and $p$ are naturally understood as differential operators acting on $\L^\lambda$.

More generally, let $\L_i\to W$ be holomorphic line bundles, described by transition functions $e^{x_{i\,\alpha\beta}}$, with some holomorphic functions
$x_{i\,\alpha\beta}$.  If $w_{\alpha\beta}=\sum_i\lambda_i x_{i\,\alpha\beta}$ for some integers $\lambda_i$, then  $\otimes_i \L_i^{\lambda_i}$ makes sense as a line bundle and the quantized functions of $p$ and $q$ are the differential operators acting on 
$\otimes_i \L_i^{\lambda_i}$.   If  
 $w_{\alpha\beta}=\sum_i\lambda_i x_{i\,\alpha\beta}$ with more general complex coefficients $\lambda_i$, then
$\otimes_i \L_i^{\lambda_i}$ does not make sense as a line bundle but differential operators acting on 
$\otimes_i \L_i^{\lambda_i}$ do make sense, and in a natural way the quantized
functions of $p$ and $q$ are holomorphic differential operators acting on $\otimes_i \L_i^{\lambda_i}$.

We will loosely refer to $\otimes_i \L_i^{\lambda_i}$ as a complex power of a line bundle (though it is more accurately a tensor product of such complex powers).
The situation that we have described is important for quantization, but it is not the most general possibility, as noted in footnote \ref{zany}.

An important detail is that if $\RR$ is a holomorphic line bundle that admits a flat connection,
 then the holomorphic differential operators acting on 
$\otimes_i \L_i^{\lambda_i}$ are the same as the holomorphic differential operators acting on $\RR\otimes_i \L_i^{\lambda_i}$.   That is because a flat
line bundle can be characterized by its $c$-number global holonomies.   Since a differential operator commutes with $c$-numbers, holomorphic
differential operators are not sensitive to the difference between $\otimes_i \L_i^{\lambda_i}$ and $\RR\otimes_i \L_i^{\lambda_i}$.

It turns out that in the case of $Y=T^*W$, $\Hom(\B_\cc,\B_\cc)$ is the sheaf of differential operators acting on a complex
power of a line bundle. 
 A quick way to show this is to make a convenient choice of $A$-brane $\B_0$ and use the fact that $\Hom(\B_\cc,\B_\cc)$ can act on $\Hom(\B_0,\B_\cc)$.   To do so, we will make use of Lagrangian branes of type $(B,A,A)$.

\subsection{Branes Of Type $(B,A,A)$ As Tools}\label{lagbr}

Branes of type $(B,A,A)$ are $B$-branes in complex structure $I$ and $A$-branes with respect to $\omega_J$ and $\omega_K$.  They were introduced
in Section \ref{genera}.    We will
be particularly interested in rank 1 Lagrangian branes of type $(B,A,A)$.   The support of such a brane  is a complex Lagrangian submanifold $L$.
  We have taken the $B$-field to be $\sB=\omega_J$, which vanishes when restricted to a complex Lagrangian submanifold $L$.   Therefore the condition
 $\sF+\sB=0$ for the $\CP$ curvature of a rank one $A$-brane $\B_0$ supported on $L$ reduces to $\sF=0$.   So the $\CP$ bundle of such a brane is a flat $\spinc$ structure.
 As explained in Section \ref{genera}, if $\B_0$ is such a brane, then $\Hom(\B_0,\B_\cc)$ is a module for the quantum deformed algebra of functions on $Y$,
though this module is not naturally interpreted in terms of quantization of $L$. 
 
 We can use such modules as a tool in understanding $\Hom(\B_\cc,\B_\cc)$.   As a probe, we consider a convenient brane 
of type $(B,A,A)$, namely a brane
 supported on the zero-section of the cotangent bundle, namely $W\subset T^*W$. 
 We assume that $W$ admits a flat $\spinc$ structure $\L$, and therefore
 is the support of a rank 1 Lagrangian brane $\B_0$.    As the considerations will be local along $W$, it is not really important if $\L$ exists globally, and likewise
 we will see in a moment that the choice of $\L$ is irrelevant.  
 
The zero-section of the cotangent bundle is invariant under scaling of the fibers.     So $\Hom(\B_0,\B_\cc)$ can be evaluated in the small $\hbar$ limit.
Taking into account the fact that the $\CP$ bundle of $\B_\cc$ is trivial, the result, following a general analysis in Appendix \ref{details},
is that  $\Hom(\B_0,\B_\cc)\cong \oplus_{q=0}^{\dim_\C\,W} H^q(W,K_W^{1/2}\otimes \L^{-1})$.
In particular, just setting $q=0$, $\Hom(\B_\cc,\B_\cc)$ must be able to act on
$H^0(W,K_W^{1/2}\otimes \L^{-1})$, the space of holomorphic sections of $K_W^{1/2}\otimes \L^{-1}$.   
Since $\L^{-1}$ is flat, it plays no role here; holomorphic differential operators acting on sections of $K_W^{1/2}\otimes \L^{-1}$ are the same
as holomorphic differential operators acting on sections of $K_W^{1/2}$.    
 Thus, $\Hom(\B_\cc,\B_\cc)$ must be the sheaf of differential operators acting on
sections of $K_W^{1/2}$, not on some other line bundle.

In view of what was explained earlier, the
differential operators acting on $K_W^{1/2}$ make sense regardless of whether $K_W^{1/2}$ exists globally as a line bundle.   But perhaps it is worth spelling
out that similarly, if $K_W^{1/2}$ does exist as a line bundle, the choice of this line bundle does not matter.
Different choices of $K_W^{1/2}$ differ by tensoring with a line bundle $\RR$ such
that $\RR^2$ is trivial.  Such an $\RR$ admits a flat connection, so tensoring by $\RR$ has no effect on the holomorphic differential operators acting on $K_W^{1/2}$.

In Appendix \ref{opposite}, we explain
another (and mathematically more standard) way to show  that in the case of a cotangent bundle, $\Hom(\B_\cc,\B_\cc)$ is the sheaf of differential operators acting on sections of $K_W^{1/2}$
and not some other complex power of a line bundle.

\subsection{Complex Cotangent Bundles and Quantization}\label{ccb}

Now we return to the real symplectic manifold $M=T^*N$, and ask how the $A$-model of $Y=T^*W$, where $W$ is a complexification of $M$,
can be used to study the quantization of $M$.

We will make use of another Lagrangian brane $\cF$ of type $(B,A,A)$.    Before getting into details, we explain the strategy (which was briefly
described in Section \ref{smooth}).   
  We will describe fairly natural elements $\alpha\in \Hom(\B_\cc,\cF)$ and $\beta\in \Hom(\cF,\B)$,
where as before $\B_\cc$ is a canonical coisotropic $A$-brane and $\B$ is a rank 1 Lagrangian $A$-brane supported on $M$.
Once $\alpha$ and $\beta$ are defined, it is straightforward to use the triple $\cF,\alpha,\beta$ to define an 
element  $\psi'$ in $\H'=\Hom(\B_\cc,\B)$, which is the dual of the Hilbert space $\H=\Hom(\B,\B_\cc)$.   
(The natural antilinear map $\Theta_\tau$, defined in Section \ref{hermitian},  establishes an isomorphism between $\H$ and $\H'$,
but when we do not invoke it, we distinguish $\H$ and $\H'$.)   Actually, for the choice of $\cF$ that we will make, $\psi'$ will be a distributional
wavefunction.  This is consistent  with the discussion in Section \ref{smooth}, since $\cF$ will not have compact support.

\begin{figure}
 \begin{center}
   \includegraphics[width=1.5in]{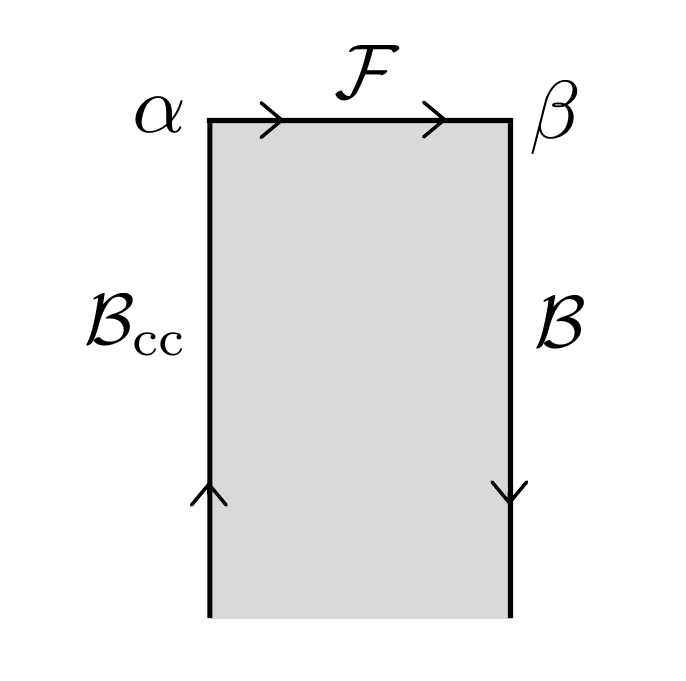}\caption{\small{A strip with $\B_\cc$ on the left, $\F$ at the top, and $\B$ on the right, with appropriate ``corners''
   $\alpha,\beta$.    With conventions as in fig. \ref{explained}(a), $\alpha$ represents an element of $\Hom(\B_\cc,\F)$ and $\beta$ represents
   an element of $\Hom(\F,\B)$.   Their composition is an element of $\H'=\Hom(\B_\cc,\B)$.   In the picture, this element propagates downward from
   the top of the strip, where is is created. }  \label{strip}}
 \end{center}
\end{figure}

Algebraically, we compose $\alpha\in\Hom(\B_\cc,\cF)$ with $\beta\in\Hom(\cF,\B)$ to make $\psi' = \beta\circ\alpha\in \H'=\Hom(\B_\cc,\B)$.  
In terms of pictures, we simply consider a strip, as in fig. \ref{strip}, with $\B_\cc$ on the left,
$\cF$ on the top,  $\B$ on the right, and $\alpha,\beta$ at the corners.   Motivated by such pictures, we will refer to objects
such as $\alpha$ and $\beta$ as corners.\footnote{In two dimensions, calling these objects ``corners'' may seem ill-motivated,
as these corners can be straightened out by a conformal mapping.   The terminology is motivated by
analogous constructions above two dimensions in which such corners, which represent junctions between two different boundary
conditions, cannot be straightened out conformally.   See \cite{GR} for a systematic study of such junctions in certain four-dimensional
supersymmetric theories.   Such four-dimensional constructions can be reduced to two dimensions by compactification on a Riemann
surface, and this  is relevant to geometric Langlands.}  
The path integral on the strip naturally produces a vector $\psi'\in\H'=\Hom(\B_\cc,\B)$.  Indeed,
 if we supply a state $\psi\in\Hom(\B,\B_\cc)$ to provide a boundary condition at the bottom of the strip,
then the path integral on the strip will simply produce a number. 
  A picture similar to  fig. \ref{strip} but with $\F$ at the bottom  produces a vector in $\H$, as already explained in Section \ref{smooth}.  

For $\cF$, we take a brane that is  supported on a fiber  $F$ of the cotangent bundle $T^*W$, namely the fiber that contains some point $w\in W$.   
Eventually we will specialize to the case $w\in N$, but for the moment we allow arbitrary $w$.    We will denote local coordinates on $W$ in a neighborhood of
$w$ as $q^i$, and write $p_i$ for conjugate functions on the fibers of the cotangent bundle.

Since $F\cong \C^n$ has a unique spin structure, 
the $\CP$ bundle of $\cF$ is simply a flat line bundle; since $F$ is simply-connected, this flat line bundle is trivial.
$F$ is invariant under scaling of the cotangent bundle (though not pointwise), so $\cF$ is likewise invariant under this scaling.
Hence as in Section \ref{lagbr},  $\Hom(\B_\cc,\cF)$ can be evaluated in the $\hbar\to 0$ limit.     The result,
again from Appendix \ref{details}, is that $\Hom(\B_\cc,\cF)= H^0(F,K_{F}^{1/2})$, where we use the fact
that for $F\cong\C^n$, the higher cohomology vanishes.  
Moreover, $K_F^{1/2}$ is trivial, so 
 $H^0(F,K_F^{1/2})$ is equivalent to the space of holomorphic functions on $F$.
Since we do not allow wavefunctions that grow exponentially at infinity,
$H^0(F,K_F^{1/2})$ can be identified with
the space of polynomial functions in the coordinates $p_i$ that parametrize $F$.    We want to pick\footnote{We will not distinguish in the notation
between $\alpha\in H^0(F, K_F^{1/2})$ and the corresponding element of $\Hom(\B_\cc,\F)$, and similarly later for $\beta$.}
$\alpha\in H^0(F,K_F^{1/2})$ to be everywhere nonzero, meaning that it provides a trivialization of the trivial line bundle $K_F^{1/2}$.
Such an $\alpha$ is uniquely determined once it is fixed at a particular point, which we take to be the unique point $w\in F\cap W$.   
We choose\footnote{\label{funnysign}  The following description does not determine the sign of $\alpha$.   For the possible significance of this sign, see footnote \ref{maybe}. }   $\alpha|_w=(\d p_1\d p_2\cdots\d p_n)^{1/2}$, by which we mean a vector in $K_F^{1/2}$ 
 that maps to
 $\d p_1\d p_2 \cdots \d p_n$
under the isomorphism $(K_F^{1/2})^2=K_F$.   We abbreviate this as $\alpha=(\d\vec p)^{1/2}$.

We also need to define an element $\beta\in \Hom(\cF,\B)$.   Since $\B$ and $\cF$ are both Lagrangian branes -- supported respectively on $M$ and $F$ --
this is a standard ingredient in the $A$-model.   The leading approximation to $\Hom(\cF,\B)$ is the cohomology of the intersection $M\cap F$ with values 
in the appropriate tensor product of $\CP$ bundles.    Under the projection $T^*W\to W$, $M$ projects to $N$ and $F$ projects to $w$, so the
intersection $M\cap F$ is empty unless $w\in N$.   So if $w\notin N$, then $\Hom(\cF,\B)=0$.   We therefore restrict to the case that $w\in N$.   In this
case, $M\cap F$ is simply the fiber at $w$ of the real cotangent bundle $T^*N$.    Thus it is a copy of $\R^n$; we call this $F_M$.    Since the $\CP$ bundle of
$\cF$ is trivial, the relevant
tensor product of $\CP$ bundles is just $\L$, the $\CP$ bundle of $\B$.
So the leading approximation to $\Hom(\cF,\B)$ is $\oplus_{r=0}^n H_{\mathrm{dR}}^r(F_M,\L)$, where here the cohomology in question is the de Rham cohomology of $F_M$
with values in the flat line bundle $\L|_{F_M}$.   Since $F_M\cong \R^n$, this cohomology vanishes except for $r=0$.   In general, for Lagrangian $A$-branes $\B,\cF$
the cohomology of the intersection
is only an initial approximation to $\Hom(\cF,\B)$, but in this case, since the cohomology vanishes except for one value of $r$, this is the full answer:
$\Hom(\cF,\B)=H_{\mathrm{dR}}^0(F_M,\L)$.   This cohomology group consists of covariantly constant sections of $\L\to F_M$.   So for $\beta$ we can
take any covariantly constant section of $\L$ over $F_M$.   It is convenient to identify a covariantly constant section with its value at the point $w\in F_M$,
so we can think of $\beta$ as a vector in $\L_w$, the fiber of $\L$ at $w$.

\begin{figure}
 \begin{center}
   \includegraphics[width=1.5in]{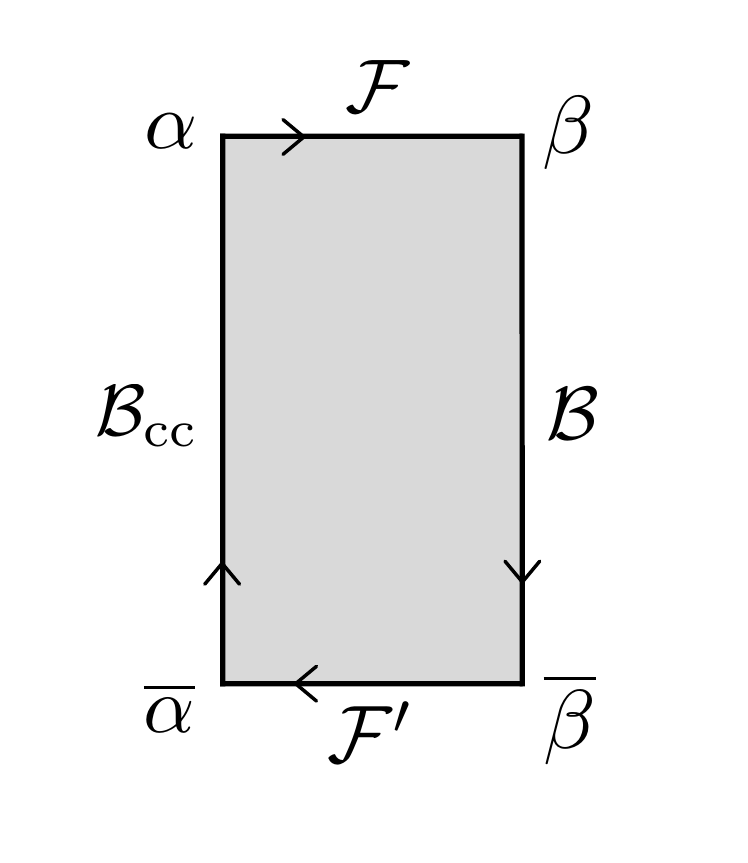}
 \end{center}
\caption{\small  The inner product between states produced by $\F$ and by $\F'$ can be computed via a path integral on this rectangle.
\label{rectangle}}
\end{figure}

For any $w\in N$ and for $\alpha,\beta$ chosen as above, this construction determines a state $\psi'(w;\alpha,\beta)\in \H'$.
To get an idea of the meaning of these states, let us compute the inner product of $\psi'(w;\alpha,\beta)$ with some other state $\psi'(w';\alpha',\beta')$,
defined similarly starting with a brane $\cF'$ supported on the fiber $F'$ of the cotangent bundle at 
another point $w'\in N$, and with $\alpha',\beta'$ defined as above.    To define this inner product, we use the Hilbert space
structure of $\H'$, as introduced in Section \ref{hermitian}.    The main step in the definition is to apply to one of the states
the product $\Theta_\tau=\Theta\tau$, where $\Theta$ is $\CPT$ and $\tau$ is the antiholomorphic involution of $Y$ that keeps $M$ fixed,   Let us, for example, apply
$\Theta_\tau$ to $\psi'(w';\alpha',\beta')$.   Since $\tau$ leaves $M$ fixed, it maps $w'\in N\subset M$ to itself.   Therefore, it maps the fiber of $T^*W$
at $w'$ to itself (not pointwise).   The upshot of this is that $\Theta_\tau$ just maps the brane $\cF'$ to itself, while reversing the orientation of a string
and complex conjugating $\alpha'$ and $\beta'$.  Taking this into account, the Hilbert space inner product $\langle \psi'(w';\alpha',\beta'),\psi'(w;\alpha,\beta)\rangle$
can be computed by a path integral on the rectangle that is sketched in fig. \ref{rectangle}.   The rectangle is constructed by taking the strip of fig.
\ref{strip} and a similar strip with $\cF$ replaced by $\cF'$, and gluing them together after reflecting the second one upside down to account for the action of
$\Theta_\tau$.   (When the second strip is reflected upside down, the wavefunctions $\alpha',\beta'$ at the corners are complex conjugated, to
account for the action of $\Theta_\tau$.)

\begin{figure}
 \begin{center}
   \includegraphics[width=4in]{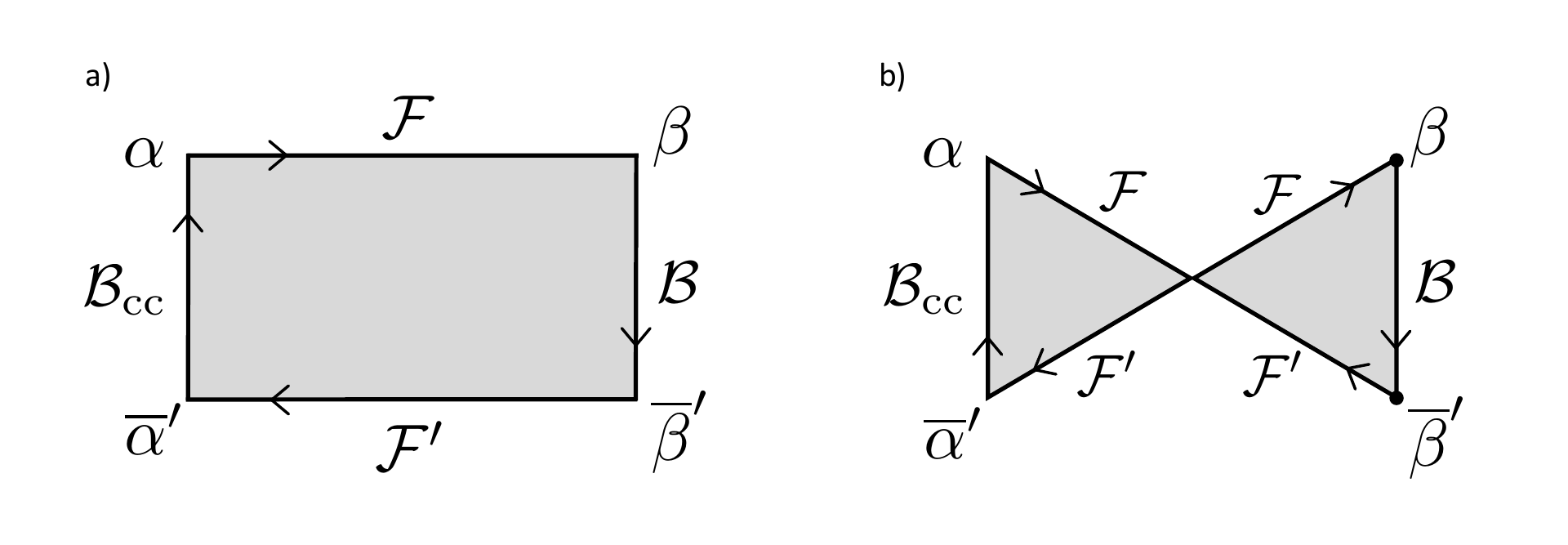}
 \end{center}
\caption{\small (a) One can view the rectangle of fig. (\ref{rectangle}) as describing the propagation from the bottom to the top of the picture
of an element of $\Hom(\B,\B_\cc)$, or the propagation from the left to the right of the picture of an element of $\Hom(\F',\F)$.   Elongating the figure
from left to right suggests the second interpretation.   Because of two-dimensional topological invariance, the metric structure of the rectangle is
immaterial.   (b)  In a limit, the rectangle degenerates to two triangles, glued at a vertex.   Passing through the vertex is a state in $\Hom(\F',\F)$.   The
rectangle can be evaluated by summing over states making up a basis of $\Hom(\F',\F)$, and for each such basis state, evaluating a 
 product of two triangles.   \label{secondone}}
\end{figure}

We claim that  $\langle \psi'(w';\alpha',\beta'),\psi'(w;\alpha,\beta)\rangle=0$ if $w\not=w'$.  
To show this, we
 use the two-dimensional topological invariance of the $A$-model.   We introduced the rectangle by discussing a state in the Hilbert space $\H=\Hom(\B_\cc,\B)$,
propagating from the bottom to the top of the rectangle.   But we can equivalently think of the rectangle as describing the propagation of a state from
left to right.   
As sketched in fig. \ref{secondone}(a), this second viewpoint is natural if, using topological invariance, we stretch the rectangle from left to right. 
The state that propagates from left to right in the figure is an element of $\Hom(\cF',\cF)$.   As $\cF$ and $\cF'$ are Lagrangian $A$-branes supported respectively
on $F$ and $F'$,
the starting point in computing $\Hom(\cF,\cF')$ is the cohomology of the intersection $F\cap F'$.   But if $w\not=w'$, $F\cap F'=\varnothing$.   So $\Hom(\cF,\cF')=0$ and
$\langle \psi'(w';\alpha',\beta'),\psi'(w;\alpha,\beta)\rangle=0$.

Sometimes we will meet a 
 similar rectangle in a situation in which $\Hom(\cF,\cF')$ is not zero.   So for future reference, let us note that the topological field theory path integral on
such a rectangle can be evaluated by letting the rectangle degenerate to a pair of triangles glued at a vertex, as sketched in fig. \ref{secondone}(b).    
The left triangle involves $\B_\cc$ along with two branes of type $(B,A,A)$; we will call such a triangle a holomorphic triangle.   The right triangle
involves three Lagrangian branes $\cF,\cF'$, and $\B$; we will call  such a triangle a Lagrangian triangle.  

Since $\langle \psi'(w';\alpha',\beta'),\psi'(w;\alpha,\beta)\rangle=0$ for $w\not=w'$, it is natural to think that $\psi'(w;\alpha,\beta)$ should be viewed
as a delta function state supported at the point $w\in N$.     If so, we expect  $\langle \psi'(w';\alpha',\beta'),\psi'(w;\alpha,\beta)\rangle$ to be proportional to
$\delta(w,w')$.   Unfortunately, we have not found a convenient way to explicitly exhibit this delta function by studying the path integral on the rectangle.
Instead, we will probe in other ways the interpretation of  $\psi'(w;\alpha,\beta)$ as a delta function.

Let us recall that according to the usual recipe of geometric quantization, the Hilbert space $\H$ obtained in quantizing $M=T^*N$ as a cotangent bundle is
a space of half-densities on $N$ with values in the prequantum line bundle $\frL$. That means that the dual space to $\H$ is a space of half-densities on $N$
with values in the dual of $\frL$.
In the brane construction, $\frL= \L^{-1}$, the inverse of the $\CP$ bundle of $\B$.

If $f$ and $g$ are half-densities on $N$ valued respectively in $\L$ and $\L^{-1}$, then there is a natural bilinear (not hermitian) pairing
$(f,g)=\int_N fg$.   So the dual of $\H$, according to geometric quantization, should consist of half-densities on $N$ valued in $\L$.
Let us see whether we can interpret   $\psi'(w;\alpha,\beta)$ as a half-density on $N$ supported at $w$ and valued in $\L_w$.   To get
started, we note that 
$\psi'(w;\alpha,\beta)$ is proportional to $\beta$, which is valued in $\L_w$.      We still have to understand how to interpret $\psi'(w;\alpha,\beta)$
as a half-density on $N$ supported at $w$.

  The bundle of densities on $N$ is a trivial real line bundle that we will call $\K_N$.
Assume for simplicity that $N$ is orientable and  pick an orientation.  Then $\K_N$ is isomorphic to the bundle of top degree differential forms on $N$.
If $q^1,\cdots, q^n$ are local coordinates on $N$, ordered in a way that is compatible with the orientation of $N$, then $\K_N$ can be trivialized locally
by $\d q^1\d q^2\cdots \d q^n$, which we abbreviate as $\d\vec q$.    Since $\K_N$ is trivial, it has a trivial square root which we will call $\K_N^{1/2}$.  It can
be trivialized locally by a section that we will call $(\d\vec q)^{1/2}$.   

Now let us describe a delta function half-density supported at $w$.  We can choose the $q^i$ to vanish at the point $w$.   Then $\delta^n(q^1,q^2,\cdots, q^n)$,
or $\delta(\vec q)$ for short, is a delta function supported at $w$.   It is naturally understood as a section of $\K_N^{-1}$, since it can be paired with a section of
$\K_N$ to get a number: $1=\int \d \vec q\, \delta(\vec q)$.   
A delta function half-density supported at $w$ is a multiple of $(\d\vec q)^{1/2} \delta(\vec q)$.  This is a section of $\K_N^{1/2}\K_N^{-1}=\K_N^{-1/2}$.   

On the other hand, $\psi'(w;\alpha,\beta)$ was proportional to $\alpha=(\d\vec p)^{1/2}$.   The symplectic structure of $T^*N$ determines a natural measure
$\d\vec p\d \vec q$ on the tangent space to $T^*N$ at $w$, and the square root of this is a natural half-density $(\d\vec p)^{1/2}(\d\vec q)^{1/2}$ on $T^*N$.     Since $(\d\vec q)^{1/2}$ is
valued in $\K_N^{1/2}$,  $(\d \vec p)^{1/2}$
can be interpreted as a section of $\K_N^{-1/2}$, just like $(\d\vec q)^{1/2}\delta(\vec q)$.

Thus $\psi'(w;\alpha,\beta)$ corresponds in a natural way to the following delta function element of  $\H'$:
\be\label{wacko}\psi'(w;\alpha,\beta)\sim (\d \vec q)^{1/2}\delta(\vec q)\cdot \beta.\ee  
This assertion simply means  that given
 the data needed to define $\psi'$, namely the point $w\in N$ and the wavefunctions $(\d\vec p)^{1/2}$ and $\beta$, the object on the right hand side is
canonically determined.   Moreover the identification is consistent with the Hilbert space inner products on the two sides, at least if we only consider inner
products between states supported at distinct points.

To further justify this proposal, we can proceed as follows.   Let $\Psi$ be any state in $\H=\Hom(\B,\B_\cc)$ and define
$f(w,\alpha,\beta)= (\psi'(w;\alpha,\beta),\Psi)$, where $(~,~)$ is the bilinear $A$-model pairing between $\H$ and $\H'$.
Now let
\be\label{oko} \h\Psi(w) =f(w;\alpha,\beta) \sqrt{\d \vec q}\,\beta^{-1}.   \ee
$\h\Psi(w)$ is a half-density on $N$ with values in $\L^{-1}$.   $\h\Psi(w)$ does not depend on any choices that were made, because the explicit factors of
$\sqrt{\d\vec q}$ and $\beta^{-1}$ in the definition of $\h\Psi(w)$ undo the effects of the choices of $\sqrt{\d\vec p}$ and $\beta$
 that were made in the definition of $\psi'(w;\alpha,\beta)$.
So $\h\Psi(w)$ is an element in the Hilbert space $\H_\gq$ that would be defined in geometric quantization, using the real polarization of $M$ that
is given by the fact that it is a cotangent bundle.

Thus we have found a natural map $\H\to \H_\gq$, where $\H$ is the Hilbert space defined in brane quantization of $M$ using the complex manifold $Y$
and $\H_\gq$ is the Hilbert space defined in geometric quantization of $M$ as a cotangent bundle $T^*N$.   We expect that this map is an isomorphism.
To prove this, one would have to show that any state $\Psi$ that is orthogonal to all the $\psi'(w;\alpha,\beta)$ actually vanishes.   We expect this to be the case,
but do not have a direct argument.
As already noted, one would also like a natural way to directly exhibit the delta function in $\langle\psi'(w;\alpha,\beta),\psi'(w';\alpha',\beta')\rangle$.

 In this analysis, we assumed that $N$ is orientable.   If $N$ is unorientable,  then the derivation is modified because $\d\vec p$ is a section not of $\K_N^{-1}$
 but of $\K_N^{-1}\otimes \varepsilon$, where $\varepsilon=\det\,T^*N$.   Related to this, if $N$ is unorientable, the CP bundle of the brane $\B$ is not $\L^{-1}$,
 a unitary line bundle of curvature $-\omega$, but rather is $\frL=\SS_0\otimes \L^{-1}$, where $\SS_0$ is the canonical flat $\spinc$ structure described in 
 section \ref{gq}.    We expect that these subtleties compensate for each other and that even for unorientable $N$, 
 $\H$ is the space of $L^2$ half-densities on $N$ valued in $\L^{-1}$.
 But we will not attempt a careful argument.

For an elementary example of the construction that we have presented, 
consider the case $M=\R^2$.   An affine linear structure on $M$ (a notion of what are linear functions on $M$,
as explained in Section \ref{problem}) gives the information needed to complexify $M$ to  $Y=\C^2$: the real linear functions on $M$ analytically continue to holomorphic linear functions on $Y$.
 For geometric quantization of $M$,
pick conjugate linear functions $p,q$ such that the symplectic form of $M$ is $\d p\,\d q$.  $M$ can be viewed as $T^*\R$, where $\R$ is parametrized by $q$,
and similarly $Y$ can be viewed as $T^*\C$, where $\C$ is parametrized by $q$, now regarded as a complex variable.
So the construction that we have described indicates that geometric quantization of $M$ with this real linear polarization is equivalent to brane
quantization by $\H=\Hom(\B,\B_\cc)$, where $\B$ is a Lagrangian brane supported on $M$.    Since brane quantization did not depend on a choice
of polarization, this shows that the Hilbert spaces obtained by geometric quantization of $M=\R^2$ with any real linear polarization (compatible with 
a given affine linear structure) must be all naturally isomorphic.
The same logic applies for quantization of $\R^{2n}$, once it is given an affine linear structure.   We have recovered a celebrated fact
about quantum mechanics:  different real linear polarizations compatible with an affine linear structure give equivalent quantizations.\footnote{\label{maybe} To
be more precise, these different linear polarizations give quantum Hilbert spaces that are naturally equivalent up to an overall sign. One way to see the appearance of this
sign was explained in footnote \ref{cext}.   It may be that the origin of the sign in the present construction is in the sign of the wavefunction $(\d\vec p)^{1/2}$,
which we did not fix in any natural way.   Another sign entered the above discussion when we used an orientation of $N$ to identify $K_W^{1/2}|_N$ with 
the space of half-densitiies on $N$.}

\subsection{Twisted Cotangent Bundles}\label{shifted}

So far, by considering the case that $Y=T^*W$ with its standard holomorphic symplectic structure, we have been able to compare brane quantization
to geometric quantization, but only in a rather special situation.    Here we will generalize the construction to what is sometimes called a twisted cotangent bundle.\footnote{See
\cite{Ginzburg} for much more detail on this subject in a more general context.}
As we will see, the generalization to twisted cotangent bundles makes possible a much more interesting comparison between brane quantization and geometric
quantization.   

The cotangent bundle 
$Y= T^*W$ has a holomorphic projection $\pi:T^*W\to W$.  The fibers are Lagrangian submanifolds that are copies of $\C^n$, with $n=\frac{1}{2}\dim_\C Y$
Moreover, $W$ can be holomorphically embedded in $T^*W$ as the zero-section.  This embedding $\iota:W\to T^*W$  is a ``section'' of $\pi:T^*W\to W$,
meaning that $\pi \circ \iota =1$ (equivalently, for any $w\in W$, $\iota(w)\in \pi^{-1}(w)$).

We will generalize this by considering
a complex symplectic manifold $Y$ that has all these properties except the last.   In other words, we assume that there is a holomorphic projection $\pi:Y\to W$
for some complex manifold $W$.   The fibers are Lagrangian submanifolds and are isomorphic to $\C^n$.  But the projection $\pi:Y\to W$ may not have a holomorphic
section.

We will call such a structure a holomorphic polarization of $Y$, generically denoted $\varPi$.   The fibers of $\pi:Y\to W$ are the leaves of $\varPi$.

We will construct such a structure for any holomorphic line bundle $\L\to W$.   Pick on $\L$ a unitary connection with curvature $\sff$ of type $(1,1)$; in local coordinates
$\sff=-\i\sum_{ij}f_{i\bar j} \d q^i \d \bar q^{\bar j}$.    Then  $\Omega_0 =\sum_i \d p_i\wedge \d q^i+\sff$ is a two-form on $Y_0=T^*W$
that is closed but, of course, not of type $(2,0)$.   We will explain how to modify the complex structure on $Y_0$ so that $\Omega$ becomes of
type $(2,0)$ and  holomorphic.

An explicit way to describe $\L$ is to cover $W$ with open sets $\U_\alpha$, in each of which one picks  a  trivialization of $\L$.   With this
trivialization, a section of $\L$ over
$\U_\alpha$ is just a holomorphic function $s_\alpha$.   On intersections $\U_\alpha\cap\U_\beta$, the local trivializations 
are glued together with holomorphic transition functions
$e^{\v_{\alpha\beta}}$; a global section $s$ of $\L$ is a family of  local sections $s_\alpha$ that satisfy 
\be\label{comsect} s_\alpha=e^{\v_{\alpha\beta}} s_\beta\ee in
$\U_\alpha\cap \U_\beta$.
The hermitian inner product on $\L$ is then defined in $\U_\alpha$ by  $||s_\alpha||^2=e^{h_\alpha}\bar s_\alpha
s_\alpha$ for some real-valued function $h_\alpha$.   The requirement that $||s_\alpha||^2=||s_\beta||^2$ in $\U_\alpha\cap\U_\beta$ whenever
(\ref{comsect}) is true gives the relation between $h_\alpha$ and $h_\beta$:
\be\label{hermrel} h_\alpha= h_\beta +2\Re\, \v_{\alpha\beta}. \ee
The curvature $\sff$ of $\L$ is defined in the region $\U_\alpha$ by  
\be\label{curvdef} \sff=-\i \partial\bar\partial h_\alpha.\ee  
Eqn. (\ref{hermrel}) implies that this is independent of $\alpha$ in intersection regions, so $\sff$ is globally defined.  
Moreover differentiating (\ref{hermrel}) gives
\be\label{ermrel}\partial_i h_\alpha =\partial_i h_\beta +\partial_i \v_{\alpha\beta}. \ee

The construction we will make
works not just for line bundles but for complex powers of line bundles, so we generalize to this case before proceeding.   Let $\L_s$, $s=1,\cdots, k$
be holomorphic line bundles over $W$, with holomorphic transition functions $e^{\v_{s\,\alpha\beta}}$ and hermitian metrics defined by local functions $h_{s\,\alpha}$,
for $s=1,\cdots, k$.     To generalize the statements in the last paragraph to $\L=\otimes_{s=1}^k \L_s^{\c_s}$, with some complex coefficients $\c_s$,
we define the curvature form $\sff=\sum_{s=1}^k \c_s \sff_s$, and
 set $h_\alpha=\sum_{s=1}^k \c_s  h_{s\,\alpha}$, $\v_{\alpha\beta}=\sum_s\c_s \v_{s\,\alpha\beta}$.    The formulas of the last paragraph
 hold whether the $\c_s$ are integers or not.   In particular, in each $\U_\alpha$, 
eqns. (\ref{curvdef}) and (\ref{ermrel}) hold.   The only thing that goes wrong if the $\c_s$ are not integers is that in triple intersections, $e^{\v_{\alpha\beta}+
\v_{\beta\gamma}+\v_{\gamma\alpha}}$ is a constant but not necessarily 1.   That will not affect any of the following.

The closed two-form $\Omega=\sum_i \d p_i \d q^i+\sff $ is, of course, not holomorphic or even of type $(2,0)$ 
in the natural complex structure of $Y_0=T^*W$.   Shortly
we will modify the complex structure of $Y_0$ to make $\Omega$ holomorphic.  We will call this manifold with its modified complex structure a twisted cotangent bundle,
and denote it as $Y$.   $Y$ will still admit a holomorphic projection $\pi:Y\to W$, with fibers $\C^n$.

 Before explaining this construction, let us note that as $\sum_i \d p_i \d q^i=\d(\sum_i p_i\d q^i)$ is
exact, the cohomology class of $\Omega=\sum_i \d p_i \d q^i+\sff $ is that of $\sff$.   We can express this as a statement about first Chern classes,
since if $[\sff]$ is the cohomology class of a form $\sff$, then $[\sff]/2\pi$ represents the first Chern class $c_1(\L)$.
So
\be\label{jnko} [\Omega]=2\pi{c_1(\L)=2\pi\sum_s v_s c_1(\L_s)}. \ee
Even if the holomorphic projection $\pi:Y\to W$ does not have a holomorphic section, it always has a smooth section.  Moreover,
$Y$ is contractible onto this section, which is a copy of $W$.   So we can view eqn. (\ref{jnko}) as a statement in the cohomology of $W$.

For example, if $W$ is compact, then according to Hodge theory, $c_1(\L)$ is of type $(1,1)$, so eqn. (\ref{jnko}) implies that $[\Omega]$ is of type $(1,1)$.   
Moreover, $c_1(\L_s)$ is real for all $s$, so  if the coefficients $v_s$ are real then
$[\Omega]$ is real.   These are convenient conditions for providing interesting examples of quantization.   For quantization,
we will want to consider a brane supported on a nonholomorphic section $M$ of the fibration $\pi:Y\to W$.   We will want $M$ to be Lagrangian for $\Im\,\Omega$
and symplectic for $\Re\,\Omega$.   The condition $[\Im\,\Omega]=0$ is necessary in order for such an $M$ to exist.    The fact that  $[\Re\,\Omega]$ can be nonzero
is not necessary to give examples, but many of the most interesting  examples have $[\Re\,\Omega]\not= 0$.   

To define the new complex structure, we just say that in  $T^*\U_\alpha$, the functions
\be\label{newc} p_{\alpha\,i}=p_i +\i\partial_i h_\alpha(q,\bar q)  \ee are holomorphic coordinates on the fibers,
and we define a new holomorphic structure in which the functions $p_{\alpha\,i}$ and $q^j$ are holomorphic, rather than $p_i$ and $q^j$.   This definition has been
arranged so that in $T^*\U_\alpha$,
\be\label{ewc} \Omega =\sum_i \d p_{\alpha\,i}\d q^i, \ee
so $\Omega$ is of type $(2,0)$ and holomorphic in $T^*\U_\alpha$. Of course, $\Omega$ is globally defined and this formula holds for all $\alpha$.
  Moreover,  in intersection regions $\U_\alpha\cap\U_\beta$,
we have, by virtue of eqn. (\ref{ermrel}), 
\be\label{newco} p_{\alpha\,i} =p_{\beta\,i}+\i\partial_i \v_{\alpha\beta},\ee
where $\v_{\alpha\beta}$ is holomorphic.  
So although $p_{\alpha\,i}\not=p_{\beta\,i}$, the complex structure in which $p_{\alpha\,i} $ and $q^j$ are holomorphic on $T^*(\U_\alpha\cap \U_\beta)$ is the
same as the complex structure in which $p_{\beta\,i}$ and $q^j$ are holomorphic on that open set.  Hence the complex structures that we have defined
on the various open sets 
$T^*\U_\alpha$ agree on all intersections, and fit together to define a new complex structure on $T^*W$.  
 In this way, we define a new complex manifold $Y$,
equivalent as a smooth manifold to the original cotangent bundle.  On $Y$,  $\Omega$ is holomorphic and of type $(2,0)$, so $Y$ is a complex symplectic manifold.

Eqn. (\ref{newco}) has the same general form as eqn. (\ref{transpo}), with $w=v/\hbar$, so our previous discussion applies.
 Hence if we could assume
 that the transformation law  of the quantum operators $\h p,\h q$ is the same as that of the classical functions $p,q$, we would conclude
 from the discussion of Section \ref{cotangent}
 that  $\Hom(\B_\cc,\B_\cc)$ consists of  holomorphic differential operators acting on 
  $\L=\otimes_{s=1}^k \L_s^{  \lambda_s}$,   with $\lambda_s=c_s/\hbar$.  The geometry of $Y$ is controlled by the parameters $c_s$,
  so  if we keep the geometry of $Y$ fixed while varying $\hbar$, than the exponents $\lambda_s$        
  are of order $1/\hbar$.
    
  This is not the right answer, because it ignores the quantum effect that was analyzed in sections
   \ref{cotangent} and \ref{lagbr}.   Even if  $\v_{\alpha\beta}=0$ in the classical
  formula (\ref{newco}),  as is the case if $Y=T^*W$ with its usual symplectic structure, 
  there is a quantum contribution, such that $\Hom(\B_\cc,\B_\cc)$ consists of holomorphic differential operators acting on sections of
  $K_W^{1/2}$, not on functions.     
  
  It is natural to guess that the classical and quantum effects in the transformation law of $\h p$ should be added.  In that case, $\Hom(\B_\cc,\B_\cc)$ will
  consist of holomorphic differential operators acting on the tensor product  $K_W^{1/2}\otimes \L$. 
    
  To justify this answer, it suffices to observe that the problem has a scaling law under which $q$ is invariant, while $p$, $\hbar$, and $v$ all scale with degree 1.   So 
  in a quantization that preserves the scaling symmetry, the transformation of $\h p$ can only be a linear function of $\hbar$ and $\phi$,
  and hence it must be correct to add the classical and quantum effects in the transformation law of $\h p$
  and hence to take the tensor product of $K_W^{1/2}$ and $\L$.              
 
We conclude by discussing how to classify the $Y$'s that can be constructed this way, for a given $W$.
Suppose that $W$ is compact.  Then, according to Hodge theory, any holomorphic line bundle $\RR\to W$ such that $c_1(\RR)$ vanishes with complex coefficients
admits a flat connection.   For such $\RR$, the holomorphic differential operators that act on $\L$ are the same as those that act on $\RR\otimes \L$.   Therefore,
if $W$ is compact, the complex symplectic manifolds $Y$ that we can make by this construction are completely classified by $c_1(\L)$ with complex coefficients, or in other words (according to
eqn. (\ref{jnko})) by the cohomology class $[\Omega]$.   For noncompact $W$, in general this is not the case; there can be topologically trivial
holomorphic 
line bundles $\RR\to W$ that do not admit a flat connection.  Thus the noncompact case is more complicated in general.

\subsection{An Example}\label{example}

In this section, we will work out in detail a simple example of a twisted cotangent bundle.  We consider deformation quantization only,
deferring quantization to Section \ref{scbq}.

The example we consider is the complex symplectic manifold $Y$ that was already introduced in eqn. (\ref{cosu}).
It is described by the equation\footnote{Later we will set $\j=\hbar j$ and eventually we set $\hbar=1$, replacing $\j$ with $j$.}
\be\label{inco} x^2+y^2+z^2=\j^2 \ee
with the holomorphic symplectic form
\be\label{nco}\Omega =\frac{\d x\,\d y}{z}. \ee

The equation and the symplectic form are invariant under an obvious action of $G= \SO(3,\C)=\PGL(2,\C)$ on the triple $x,y,z$.
   Indeed, $Y$ is a coadjoint orbit of $G$.
It can be regarded as the subspace of the Lie algebra $\g$ of $G$ consisting of elements $T$ with determinant $-\j^2$.   Any such element can be written
\be\label{pok} T=\begin{pmatrix} z & x+\i y \cr x-\i y & -z\end{pmatrix} \ee
for some $x,y,z$ satisfying eqn. (\ref{inco}).   The eigenvalues of any such $T$ are $\j$ and $-\j$, so the eigenvalue problems
\be\label{ok}T\begin{pmatrix} q \cr 1\end{pmatrix} =-\j \begin{pmatrix} q \cr 1\end{pmatrix},~~~T\begin{pmatrix} q' \cr 1\end{pmatrix} =\j \begin{pmatrix} q' \cr 1\end{pmatrix}
\ee
have unique solutions with $q,q'\in \C\cup\infty$:
\begin{align}\label{pola} q& = -\frac{x+\i y}{\j+z}=-\frac{\j-z} {x-\i y}\cr
                     q'& = \frac{x+\i y}{\j-z}=\frac{\j+z}{x-\i y}. \end{align}
Clearly $q$ and $q'$ are exchanged by $\j\leftrightarrow -\j$.  
Either $q$ or $q'$ (plus a point at infinity) parametrizes a copy of $W=\bCP^1$, which is simply the projectivization of the two-dimensional vector
space on which $T$ acts.     Clearly, $G$ will act on $W$ by the familiar fractional linear transformations
 $q\to (aq+b)/(cq+d)$, $q'\to (aq'+b)/(cq'+d)$, with $ad-bc=1$.
 
 There is no point in $Y$ with $q=q'$, since the two eigenvectors of $T$ are always distinct.   That is the only constraint; any distinct pair $q,q'\in \bCP^1$
 determines a unique $T$, which one can find by solving eqns. (\ref{pola}) for $x,y,z$.
    The complement of a point in $\bCP^1$ is  a copy of $\C$.    So the possible values of $q'$ for given $q$ parametrize
 a copy of $\C$; likewise the possible values of $q$ for given $q'$ parametrize a copy of $\C$.    So $Y$ has two holomorphic polarizations, one
 with leaves labeled by the choice of $q$ (and parametrized by $q'$ with the constraint $q'\not=q$) and one with leaves labeled by $q'$ (and parametrized by $q$
 with the constraint  $q\not=q'$).   
  Let us denote the holomorphic polarizations parametrized by $q$ and $q'$
 as $\varPi$ and $\varPi'$, respectively.         
 
 A short calculation gives a formula for the symplectic form in terms of\footnote{The factor $1/(q-q')^2$ in the following formula
 does not really represent a pole in $\Omega$, since $q$ and $q'$ are always distinct, as noted earlier.}    $q$ and $q'$:
 \be\label{shortform}\Omega=2\i\j \frac{\d q\d q'}{(q-q')^2}.\ee
Perhaps surprisingly, the variables conjugate to $q$ and $q'$ can be chosen to be equal, since if 
 \be\label{momform} p=p'=-\frac{2\i \j}{q-q'}, \ee
 then 
 \be\label{canform}\Omega = \d p \d q = \d p'\d q'. \ee
 
 To describe the situation globally, we cover $\bCP^1$ with two open sets $\U$ and $\t\U$, where $\U$ is the 
 complement of the point at infinity in $\bCP^1$ and $\t\U$ is the complement of the origin.   Good variables in $\U$ are $q,q'$, and 
good variables  in $\t \U$ can be obtained by a tractional linear transformation:  $ \t q=-1/q$, $\t q'=-1/q'$.   The conjugate variables are
 \be\label{ranform} \t p = \t p'=-\frac{2\i \j}{\t q-\t q'}. \ee
 From these formulas, it follows that
 \be\label{anform}p=\t q^2 \t p +2\i \j \t q=\t q^2\t p -\frac{2\i \j}{q}. \ee
 To compare this to the general formula (\ref{transp}), we note that in this example, $W=\bCP^1$ has complex dimension 1, so  indices $i,k$ in eqn. (\ref{transp})
 can be omitted.  We have covered $W$ with only two open sets, so $q,p$ correspond to $q_\alpha,p_\alpha$ in (\ref{transp}) and $\t q, \t p$
 correspond to $q_\beta, p_\beta$.   The function $\t q^2$ plays the role of the matrix $M$, and finally to put (\ref{anform}) in the form of (\ref{transp}), we are supposed to write $-2\i \j/q$ as $\i\hbar\partial_q u$
 for some holomorphic function $u$.   So\footnote{The function $u$ is not single-valued; that is the price we pay for covering $W$ with open
 sets whose intersection is not simply-connected.   The procedure of Section \ref{shifted} would tell us to cover $W=\bCP^1$ with a larger number
 of open sets with contractible intersections.  Then we would define transition functions that are always single-valued but do not satisfy
 $S_{\alpha\beta}S_{\beta\gamma}S_{\gamma\alpha}=1$ in triple overlaps unless $2j\in \Z$.  A derivation along those lines would be longer.}       
 $u=-2 j\log q$, where we have set $\j=\hbar j$.   Then we try to interpret $e^u$ as
 the transition function of a line bundle $\L$.   In the present case, $e^u=q^{2j}$.    This is single-valued if and only if $2j\in\Z$,
 so that is the only case in which $\L$ is an honest line bundle; for other values of $j$, $\L$ is really the complex power of a line bundle.
  The line bundle  $\L$ that we get when $2j\in \Z$ is a familiar one.  The section that is $1$ in region $\t\U$ (in the trivialization that is implicit in the above formula for
 $\t p$) is $e^u=q^{2j}$ in region $\U$ (in the trivialization implicit in the formula for $p$).   This section has a zero of order $2j$ at the origin, and no poles,
 so $\L$  is what is usually called $\O(2j)$.      We will use the notation $\O(2j)$ whether $2j\in \Z$ or not; the idea is always that differential operators acting on
 $\O(2j)$ are well-defined -- we will write explicit formulas -- whether or not $\O(2j)$ exists as a line bundle.
 
Now let us discuss the canonical coisotropic $A$-brane $\B_\cc$ on the space $Y$.  According to the general discussion,  $\Hom(\B_\cc,\B_\cc)$ is the
sheaf of holomorphic differential operators acting on $K_W^{1/2}\otimes \L$.  For $W=\bCP^1$, $K_W$ is isomorphic to $\O(-2)$ and $K_W^{1/2}$ is isomorphic
to $\O(-1)$.    So $K_W^{1/2}\otimes \L=\O(2j-1)$.    Hence, we expect $\Hom(\B_\cc,\B_\cc)$ to be the algebra of holomorphic differential operators on $\bCP^1$
acting on $\O(2j-1)$.
At some stages in the derivation that led to this statement, we sheafified $\Hom(\B_\cc,\B_\cc)$ (though this step was arguably
 not entirely natural from a physical point of view),
but in the final statement, no sheafification is necessary.  $\Hom(\B_\cc,\B_\cc)$ is simply the algebra of globally defined holomorphic differential operators
acting on $\O(2j-1)$.    

We note, however, that we started with two equally good holomorphic polarizations of $Y$, differing by $j\to -j$.   If we had used the second one in the last part of the
analysis, we would have identified $\Hom(\B_\cc,\B_\cc)$ with the algebra of globally defined holomorphic differential operators acting on $\O(-2j-1)$.    Therefore,
we expect that these two algebras will be isomorphic.   This is indeed true, as we will see shortly. 
     
The algebra of global differential operators on $\bCP^1$ acting on $\O(2j-1)$ is generated by\footnote{We set $\hbar=1$ in the rest of this discussion.  The formulas
given in eqn. (\ref{throps}) are uniquely determined up to the possibility of taking linear combinations of these operators and adding constants to them.   We have made
a convenient choice.}
\begin{align}\label{throps}    t_- & = -\frac{\d}{\d q} \cr
                     t_3 & = q\frac{\d}{\d q} -j+\frac{1}{2} \cr
                      t_+ & = q^2 \frac{\d}{\d q}-(2j-1)q . \end{align}  
                      These operators satisfy the familiar commutation relations of the $\mathfrak{sl}(2)$ Lie algebra $\g$:
  \be\label{usrel} [t_3,t_\pm]=\pm t_\pm,~~~~ [t_+,t_-]=2t_3. \ee
To verify that these operators are globally holomorphic, we check the behavior at $q=\infty$  by transforming from $q$ to $\t q=-1/q$,
also conjugating by the transition function of $\O(2j-1)$.   For example, $t_+=q^2 \frac{\d}{\d q}-(2j-1)q$ maps to $\t q^{2j-1} \left(\frac{\d}{\d \t q} 
+\frac{2j-1}{\t q}\right) \t q^{-(2j-1)}=\frac{\d}{\d \t q}$, which is holomorphic at $\t q=0$.  A polar term has canceled because of the precise coefficient in the
term  $\frac{2j-1}{\t q}$.
 Altogether
\begin{align}\label{agrops} t_-& = -\t q^2 \frac{\d}{\d \t q}+(2j-1)\t q\cr 
                                           t_3& = -\t q\frac{\d}{\d \t q}+j-\frac{1}{2} \cr
                                            t_+ & = \frac{\d}{\d \t q}.  \end{align}
The operators $t_+,t_-$ and $t_3$ are algebraically independent, except for a single relation which equates the quadratic Casimir operator of $\g$ 
to a $c$-number:
\be\label{eqrel} \frac{1}{2}(t_+t_-+t_-t_+) +t_3^2= j^2-\frac{1}{4}. \ee
This has the expected symmetry under $j\to -j$.  The shift by $K_W^{1/2}$ was crucial in getting this symmetry.
  
The operators $t_+, t_-, $ and $t_3$ are quantum versions of the generators $x,y,z$ of the algebra of holomorphic functions on $Y$.  For
example, the formula $p=-2\i j/(q-q')=\i(x-\i y)$, together with $p=-\i d/\d q$, shows that $t_-=-\d/\d q$ corresponds to $x-\i y$.   More generally,
working through the above relations between $x,y,z$ and $p,q$, we find
\begin{align} \label{correspo} x-\i y &\to t_- \cr 
                                             x+\i y &\to t_+ \cr
                                             z& \to t_3. \end{align}
The Casimir relation (\ref{eqrel}) is thus a quantum-deformed version of the original classical relation $x^2+y^2+z^2=j^2$.                                            
      
In this discussion, we have considered only deformation quantization.   In Section \ref{scbq}, we will add a Lagrangian $A$-brane and analyze quantization.
In doing so, an antiholomorphic involution of $Y$ is important.   If $\j^2$ is real, then $Y$ has the obvious antiholomorphic involution $\tau:(x,y,z)\to (\b x,\b y, \b z)$.
A look back to the starting point in eqn. (\ref{pola}) shows that $\tau$ exchanges the two holomorphic polarizations $\varPi$ and $\varPi'$ (to be precise, $\tau$ maps
the leaf of $\varPi$ with a given value of $q$ to the leaf of $\varPi'$ with $q'=-1/\b q$).    The same complex manifold $Y$,
also for $\j^2$ real, has another antiholomorphic involution $\t\tau:(x,y,z)\to (\b x,-\b y,-\b z)$, also mapping $\Omega$ to $\b\Omega$.   To write more transparent formulas,
it  is convenient to map $\t\tau$ back
to $\tau$ by redefining $(x,y,z)\to (x, \i y, \i z)$.   The equation defining $Y$ becomes 
\be\label{nuuxx} x^2-y^2-z^2=j^2. \ee   The symplectic form is still $\Omega=\d x\,\d y/z$.   We consider this manifold $Y'$ with the original antiholomorphic involution
$\tau$ that complex conjugates $x,y,z$.
Of course, purely as a complex manifold, $Y'$ is equivalent to $Y$, so  it still has two holomorphic polarizations.   But there is an interesting difference.   For $j^2>0$,
the two holomorphic polarizations are again exchanged by $\tau$.   But if $j^2=-s^2$ is negative, the two  holomorphic polarizations are each $\tau$-invariant.
The equation $x^2-y^2-z^2=-s^2$ is equivalent to the statement that the matrix
\be\label{uvux}T'=\begin{pmatrix} z & x-y \cr -x-y & -z\end{pmatrix}\ee
has eigenvalues $\pm s$.    Solving the eigenvalue problems \be\label{eipro}T'\begin{pmatrix} q \cr 1\end{pmatrix}  =-s    \begin{pmatrix} q \cr 1\end{pmatrix}  ,
~~~~T'\begin{pmatrix} q' \cr 1\end{pmatrix}=s \begin{pmatrix} q' \cr 1\end{pmatrix} ,\ee    we    get the formulas 
\begin{align}\label{horiz} q & = -\frac{x-y}{z+s} =-\frac{z-s}{x+y}          
   \cr q'& =-\frac{x-y}{z-s}=  -\frac{z+s}{x+y} \end{align}
which define the two holomorphic polarizations $\varPi$ and $\varPi'$.   As these formulas are completely real, $\varPi$ and $\varPi'$ are each
$\tau$-invariant.   To be more precise, $\tau$ acts on $\varPi$ by $q\to \bar q$, leaving fixed the leaves with real $q$ and exchanging others in pairs,
and acts similarly on $\varPi'$.  

\subsection{Application To Quantization}\label{scbq}

Now we will discuss how to apply holomorphic polarizations to quantization.  We consider the situation introduced in Section \ref{amo}: $Y$  is a complex
symplectic manifold with complex structure $I$ and holomorphic symplectic form $\Omega$, along with
an antiholomorphic involution $\tau$ such that $\tau^*(\Omega)=\b\Omega$.  $M$,
a component of the fixed point set of $\tau$, is symplectic with respect to $\omega_J=\Re\,\Omega$ 
(as well as being inevitably Lagrangian for $\omega_K=\Im\,\Omega$).
We assume that $Y$ has a holomorphic polarization $\varPi$, whose leaves are fibers of a holomorphic map $\pi:Y\to W$, for some $W$.
We want to understand the implications of the existence of $\varPi$ for brane quantization of $M$.

  Since $Y$ is foliated by the leaves of $\varPi$, every point $p\in M$ is contained
in a unique leaf $F_p$.  $F_p$  is Lagrangian for $\Omega$ and hence for $\omega_J$.    
So $\omega_J$ vanishes when restricted to $F_p\cap M$.   Therefore
$F_p\cap M$ is at most middle-dimensional in $M$.  If it is middle-dimensional,  it is a Lagrangian submanifold of $M$.     Suppose that this
is the case for every leaf of $\varPi$ that contains a point  of $M$  and moreover suppose that all these Lagrangian submanifolds are copies of $\R^n$.
  Then the Lagrangian submanifolds of $M$ that are of the form  $F_p\cap M$ define a real polarization $\P$ of $M$.  In this situation,  we will show that brane
quantization of $M$ is equivalent to geometric quantization using the real polarization $\P$.     

Let $N$ be the space of leaves of $\P$.   It is the subspace of $W$ that parametrizes leaves of $\varPi$
that intersect $M$. 
A point in $N$ corresponds to a leaf $F$ of $\varPi$ 
 such
that $F\cap M$ is of real dimension $n$, so  $F\cap M$ is middle-dimensional in $F$.  $F\cap M$ is totally
real (since $M$ is), so $F$ is a complexification of $F\cap M$.   Since $F\cap M$ is $\tau$-invariant, and $F$ is a complexification of $F\cap M$,
$F$ is mapped to itself by $\tau$ (though not pointwise).  

In turn,  by dimension counting, one can see that $N$ is middle-dimensional in $W$; the real dimension of $N$ is
$n=\frac{1}{2}\dim_\R\,M=\frac{1}{2}\dim_\C\,Y$, and this is the same as the complex dimension of $W$.   
$N$ parametrizes a middle-dimensional totally real subspace of $W$, so $W$ can be viewed as a complexification of $N$.  $\tau$ acts as an antiholomorphic symmetry
of $W$, leaving $N$ fixed.  Points in $W$ but not in $N$  correspond to leaves of $\varPi$ that do not intersect $M$; at least in a neighborhood of $N$,
such leaves are exchanged pairwise by $\tau$.  In particular the holomorphic polarization $\varPi$ is $\tau$-invariant, though not every leaf is mapped to itself
by $\tau$.

For an elementary example of this situation, let $Y=\C^2$, parametrized by complex variables $p,q$ with $\Omega=\d p \, \d q$.  Let $\tau$ act
by $(p,q)\to (\b p, \b q)$, with fixed point set $M$.   Expand $p,q$ in real and imaginary parts: $p=p_1+\i p_2$, $q=q_1+\i q_2$.   $M$ is a copy of $\R^2$
parametrized by $p_1,q_1$.   $Y$ has a holomorphic polarization $\varPi$  whose leaves are of the form $q=b$, where $b$ is a complex parameter that
specifies the choice of leaf.   Thus the space $W$ of leaves is a copy of $\C$.   $\tau$ maps the leaf with  $q=b$ to the leaf defined by $\bar q=b$ or $q=\bar b$; 
the $\tau$-invariant leaves are the ones with $b$ real.   So $\tau$ acts on $W\cong\C$ by complex conjugation, leaving fixed the real axis $N$.
A point in $N$ corresponds to a leaf of $\varPi$ that intersects $M$ in a copy of $\R$; these copies of $\R$ are the leaves of a real polarization $\P$ of $M$.

For a somewhat more sophisticated example, consider the complex symplectic manifold $Y'$ of eqn. (\ref{nuuxx}), with $j^2<0$ so that $s$ is real.  
$M$ is described by the same equation with $x,y,z$
real.   Eqn. (\ref{horiz}) describes two different holomorphic polarizations $\varPi$ and $\varPi'$ of $Y'$, one parametrized by $q$ and one by $q'$.    $\tau$ maps
$\varPi$ to itself and likewise maps $\varPi'$ to itself.   The leaves of $\varPi$ or $\varPi'$ that have a nonempty intersection with $M$ are those with $q$ or $q'$ real.
Those leaves provide two real polarizations $\P$ and $\P'$ of $M$.     The two projections  $\pi,\pi':Y\to W\cong \bCP^1$ associated to $\varPi$ and $\varPi'$
restrict along $M$ to projections $\pi,\pi':M\to N=\RP^1$, where $N$ is parametrized by $q$ or $q'$, plus a point at infinity,

Now we consider a quite different possibility.   If $F$ is a leaf of a holomorphic polarization $\varPi$, then 
 $F$ and $M$ are both middle-dimensional in $Y$, so it is natural for $F\cap M$ to consist of just a point (it is also natural for the intersection to consist
 of several points, but we will not consider that case).    Suppose that $F\cap M$ is at most a single point for every leaf of $\varPi$.  If so, the projection $\pi:Y\to W$
restricts to an isomorphism of $M$ onto its image $\pi(M)$ (which may be only part of $W$, as we will see in an example).  
Since $W$ is a complex manifold, this gives $M$ a complex structure, which we will call\footnote{The notation 
is motivated by the fact that if the complex  structure of $Y$, which we call $I$,
 extends to a hyper-Kahler structure with additional complex structures $J,K$, then $\J$ is sometimes the restriction 
of $J$ to $M$.}
 $\J$.    In complex structure $\J$, a holomorphic function on $M$ is the  pullback $\pi^*(f)$ of a holomorphic function on $W$.   Note that if $V$ is a vector
 field on $Y$ (not necessarily holomorphic) that is tangent to the leaves of $\varPi$, then $V$ projects to zero on $W$ and therefore the derivative in the $V$ direction
 of a function pulled back from $W$ vanishes:
 \be\label{looj} \mathcal L_V\cdot \pi^*(f)=0. \ee
 In this situation,  $\J$  is a complex polarization of
$M$ in the sense of geometric quantization, since $\omega_J|_M=\Re\,\Omega|_M$ is of type $(1,1)$ with respect to $\J$.   To see this, let $T_pM$
be the tangent space to $M$ at a point $p\in M$, and let $T_{p,\C}M=T_{p,M}\otimes_\R\C$ be its complexification. Consider the decomposition
$T_{p,\C}M=T^{1,0}_{p,\C}M\oplus T^{0,1}_{p,\C}M$, where the two subspaces are respectively the spaces of holomorphic and antiholomorphic complexified
tangent directions
to $M$ (in complex structure $\J$).  As a vector space,   $T^{0,1}_{p,\C}M$ can be naturally identified with $T_pF_p$, the tangent space to $p$ in the leaf $F_p$
of $\varPi$ that contains $p$ (this follows from  eqn. (\ref{looj}), which says that holomorphic functions on $M$ are annihilated by derivatives in the $T_pF_p$ directions).
Since $F_p$ is Lagrangian for $\Omega$, this tells us that the $(0,2)$ part of $\Omega|_M$ vanishes in complex structure $\J$.  
Together with the fact that $M$ is Lagrangian for $\Im\,\Omega$, this implies that 
the $(2,0)$ and $(0,2)$ parts
of $\Re\,\Omega|_M$ both vanish, showing that $\Re\,\Omega|_M$ is of type $(1,1)$.  

The antiholomorphic involution $\tau$ of $Y$ will map $\varPi$ to a holomorphic polarization $\varPi'=\tau(\varPi)$.   
$\tau$ maps a holomorphically varying family of leaves of $\varPi$ to an antiholomorphically varying family of leaves of $\varPi'$, so it
reverses the induced complex structure of $M$.    In particular, $\varPi$ and $\varPi'$ are always different.
If $\omega$ is positive-definite in the complex structure $\J$ induced by $\varPi$, meaning that $\J$ determines a complex Kahler polarization in the sense of geometric
quantization, then it is negative-definite in the complex structure $-\J$ induced by $\varPi'$.

In this situation, we will argue in Section \ref{branecp} that brane quantization of $M$, at least as a vector space, disregarding the Hilbert space structure, agrees with 
geometric quantization using the complex polarization of $M$ that is induced from $\varPi$.  The Hilbert space structures are actually different generically,
as will be clear from the example studied in Section \ref{flatg}.

But first we consider some illustrative examples of complex polarizations of $M$ induced by holomorphic polarizations of $Y$.
For an elementary example, again set $Y=\C^2$, with $p,q,\Omega, \tau,$ and $M$ as before.
But now define a leaf of $\varPi$ by the equation $q+\i p=b$, where $b$ again is a constant that parametrizes the choice of leaf.   Such a leaf intersects $M$
in the unique point $q_1+\i p_1=b$.   So in this case, $M$ acquires a complex structure in which the function $z=q_1+\i p_1$ is holomorphic.   A vector
field tangent to the leaf $q+\i p=b$ is $\frac{\partial}{\partial q}+\i \frac{\partial}{\partial p}$, which along $M$ is interpreted as ${2}\frac{\partial}{\partial \bar z}$.
The $\tau$-conjugate of $\varPi$ is a polarization $\varPi'$ whose leaves are defined by an equation $q-\i p=b$, again with $b$ as a parameter.   The induced
complex structure is the one in which $\bar z=q_1-\i p_1$ is holomorphic.

For a somewhat more subtle example, consider the complex manifold $x^2+y^2+z^2=\j^2$, studied in Section \ref{example}, with $\j^2>0$.
$M$ is the submanifold with $x,y,z$ real.   For real $x,y,z$, the matrix $T$ defined in eqn. (\ref{pok}) is hermitian.   A $2\times 2$ hermitian matrix with
eigenvalues $\j,-\j$ is uniquely determined if one of the eigenvectors is specified, which we may do by specifying $q$ or $q'$.   Thus any leaf of $\varPi$ or $\varPi'$
intersects $M$ in a unique point.   The $\SO(3)$ symmetry of the construction ensures that the induced complex structure on $M$ is, up to sign, the natural complex
structure on $M\cong S^2\cong \bCP^1$.    But in fact, the signs are opposite.   One way to see this is to observe
that the variety $Y$ admits a holomorphic involution $\zeta: (x,y,z)\to (-x,-y,-z)$.   This involution (which changes the sign of the holomorphic symplectic form
$\Omega=\d x\,\d y/z$) exchanges $\varPi$ and $\varPi'$, since it reverses the sign of the matrix $T$ of eqn. (\ref{pok}), thus exchanging the two eigenvalues.
On the other hand, on the two-sphere $M$ defined by $x,y,z$ real, the map $(x,y,z)\to -(x,y,z)$ reverses the orientation and
acts antiholomorphically.   Explicitly, $\varPi$ induces on $M$ a complex
structure in which $M$ is parametrized holomorphically by $q$, and $\varPi'$ induces a complex structure in which $M$ is parametrized holomorphically by $q'$.
It follows from eqn. (\ref{pola}) that
if $j,x,y$, and $z$ are all real, then $q'$ is related to $q$ by
\be\label{inverso} q'=-\frac{1}{\bar q}.  \ee
Thus the two complex structures induced by $\varPi$ and $\varPi'$ are opposite; a holomorphic function in one is antiholomorphic in the other.

 For a somewhat similar example with an interesting twist, consider again the example $Y'$ of eqn. (\ref{nuuxx}), but
now with $\j^2>0$ so that $s$ is imaginary.  The fixed point locus of $\tau$ is again parametrized by real $x,y,z$.   The real algebraic manifold $
x^2-y^2-z^2=\j^2$ with $\j^2>0$ has two components, with $x>0$ or $x<0$.    We call these respectively $M$ and $M'$.  Each of   $M$ and $M'$ is a copy of
$\SO(2,1)/\SO(2)=\SL(2,\R)/\UU(1)$, which can also be realized as the complex upper half-plane $\mathcal H$.   Leaves
of $\varPi$ or $\varPi'$ with real $q$ or $q'$ do not intersect either $M$ or $M'$ at all. (This follows from eqn. (\ref{eipro}); since $s=\pm \sqrt{-\j^2}$ is imaginary
for $\j^2>0$, this equation clearly has no solutions for real values of $x,y,z$ and $q$ or $q'$.)   The projections $\pi$ and $\pi'$ from $Y$ to $W\cong \bCP^1$ map
$M$ to an upper or lower half plane in $W$ and $M'$ to the opposite half-plane.  The $\SO(2,1)$ symmetry ensures that the induced complex structures
on $M$ and $M'$ are  (again up to a choice of orientation) equivalent to the standard complex structure of $\mathcal H$.

Holomorphic polarizations of the two types that we have discussed are useful for quantization, as we will see.   Most likely some
other types  are similarly useful.   For example,
 in geometric quantization it is possible to have a polarization intermediate between
 real and complex polarizations -- roughly, a polarization that is real in some directions and complex in others.   It is likely that conditions can be placed
on a holomorphic polarization of $Y$ so that brane quantization of $M$ can be related to geometric quantization with a polarization of such intermediate type.

\subsection{Comparing to Geometric Quantization With A Real Polarization}\label{branerp}

In this section and the next one, we generalize the comparison of brane quantization and geometric quantization that was made in Section \ref{ccb}.
Here we consider to general case that $Y$  admits a holomorphic polarization $\varPi$ that restricts to a real polarization $\P$ of $M$; in Section \ref{branecp},
we consider the case that $\varPi$ restricts to a complex polarization.

The resulting story is very similar to the discussion in Section \ref{cotangent}, where we assumed that $M=T^*N$.  However,
the structure is more general than we assumed previously.   The polarization $\varPi$ determines on $M$ a projection
$\pi:M\to N$, whose fibers are the leaves of $\P$.    Topologically, it is possible to pick an embedding $\iota: N\to M$ satisfying $\pi\circ \iota=1$.
Once this is done, $M$ can be identified as $T^*N$.   However, in contrast to Section \ref{cotangent}, there is no natural embedding and we cannot
assume that the symplectic form $\omega_J$ is the standard $\sum_i \d p_i \d q^i$.   All we can say is the following: since the fibers of $\pi$ are Lagrangian,
$\omega_J$ has no terms $\d p_i \d p_j$; since $\omega_J|_M$ is assumed to be everywhere nondegenerate, fiber coordinates $p_i$ can be picked
so that the $\d p \d q$ terms are the usual $\sum_i \d p_i \d q^i$; but we learn nothing about the $\d q^i \d q^j$ terms except that they do not depend on $p$
(since $\d\omega_J=0$), and are closed.     Thus the general form of $\omega_J$ is $\omega_J=\sum_i \d p_i \d q^i+\pi^*(\alpha)$, where $\alpha$ is 
a closed two-form on $N$. 

This is the general form of a symplectic structure that is assumed in geometric quantization in order to quantize a symplectic manifold $M$ using
a real polarization.  So one can hope to compare brane quantization to geometric quantization in this greater generality.   The analysis in Section \ref{ccb} assumed
$\alpha=0$.

In geometric quantization, the quantum Hilbert space $\H_{\gq}$ is defined as
follows.   Consider a point $w\in N$.  The prequantum line bundle $\frL\to M$ is flat when restricted to $\pi^{-1}(w)\cong\R^n$.   So one can define
a line bundle $\frL_N\to N$ whose fiber at the point $w$ is the space of covariantly constant sections of $\frL$ over $\pi^{-1}(w)$.    This line bundle
does not have a natural connection, but it does have a natural hermitian metric, inherited from that of $\frL$.   
The Hilbert space $\H_{\gq}$ is defined as the space of $\ltwo$ half-densities on $N$ valued in $\frL_N$ with the obvious inner product
\be\label{nho}\langle\chi,\psi\rangle =\int_N \bar\chi \psi. \ee

To try to recover such a description from brane quantization, we proceed just as in Section \ref{cotangent} and define a class of candidate delta function states in
$\H'=\Hom(\B_\cc,\cF)$. 
As before, we let $F$ be the leaf of $\varPi$ that contains a given point $w\in N$,
and we let $\cF$ be a brane of type $(B,A,A)$ supported on $F$ with trivial $\CP$ bundle.  And we pick an element in
$ \Hom(\B_\cc,\cF)$ associated to $\alpha=(\d \vec p)^{1/2}$.   We also need a corner in $\Hom(\cF,\B)$.   As before, this is the space of covariantly constant
sections of $\L=\frL^{-1}$ over $F_M=F\cap M=\pi^{-1}(w)$.    Let $\beta$ be such a section.   It is naturally valued in the dual of the fiber of $\frL_N$ at $w$, 
so $\beta^{-1}$ is valued in the fiber of $\frL_N$ at $w$.   
With this data, the construction of fig. \ref{strip} gives a vector $\psi'(w;\alpha,\beta)\in \H'$.   Inner products between two such states vanish for $w\not=w'$ by
the same argument as before, suggesting that
such states should be regarded as delta function states.        Let $\Psi$ be any state in $\H$.  The inner product $f(w;\alpha,\beta)=(\psi'(w;\alpha,\beta),\Psi)$ is
naturally defined, and as in eqn. (\ref{oko}), we can define a half-density on $N$ valued in $\frL_N$ by 
\be\label{rno} \h \Psi(w) = f(w;\alpha,\beta) \sqrt{\d q}\,\beta^{-1}. \ee
This gives a natural map from $\Psi\in \H=\Hom(\B,\B_\cc)$ to $\h\Psi\in \H_{\gq}$.    We expect this map to be an isomorphism.

For an example of a twisted cotangent bundle quantized in this way, one can take $M$ to be the real algebraic variety $x^2-y^2-z^2=j^2$, which
we quantize by complexifying it to the complex manifold $Y'$ defined by the same equation with complex $x,y,z$ (eqn. (\ref{nuuxx})).   $M$ has two real polarizations $\P$ and $\P'$
defined in eqn. (\ref{horiz}) with $q,q'$ real.   These two polarizations are induced by the two holomorphic polarizations $\varPi$ and $\varPi'$ of $Y'$ that
are defined by the same formulas, with complex variables.   So geometric quantization of $M$ with either of the real polarizations $\P$ and $\P'$ will
give an equivalent Hilbert space.   

To make the resulting two quantizations of $M$ explicit, we have to describe the line bundle $\frL_N$ over $N\cong \RP^1$.
Since $M\cong T^*\RP^1\cong T^*S^1$ is not simply-connected, the prequantum line bundle $\frL\to M$, which is defined to be a unitary line bundle with
a connection with 
curvature $\omega$, is not unique; we could modify it by multiplying the holonomy around $S^1$ by an arbitrary element of $\UU(1)$.   The choice we make
below is the unique choice such that the action of $\PGL(2,\R)$ on $M$ lifts to an action of the same group (rather than a cover of it) on $\frL$.

 We cover $N$ with two
open sets $\U$ and $\t \U$, with $\U$ parametrized by a real variable $q$ and $\t U$ by $\t q=-1/q$.   The symplectic form is $\omega=\d p \d q=\d\t p\d \t q$.
The relation between $p$ and $\t p$ was found in eqn. (\ref{anform}); with $\j=\i s$, this relation is
\be\label{relp} p = \t q^2\t p+\frac{2s}{q}. \ee
We describe a  prequantum line bundle as follows.   In region $\U$, $\frL$ has a trivialization such that
the connection form
becomes the one-form  $A= p \d q$; in region $\t \U$, there is a trivialization that leads to the one-form  $\t A=\t p \d \t q$.  
These obey $\d A=\d \t A=\omega$, as expected for a connection on $\frL$.   
 The corresponding covariant
differentials are $\d_A=\d+\i A$, $\d_{\t A}=\d+\i \t A$.
 The two connection forms differ by a gauge transformation:
\be\label{elp} A= \t A+2s\frac{\d q}{q}=\t A+2s\,\d \log q, \ee
so $\d_A=q^{2\i s} \d_{\t A} q^{-2\i s}$.

Consider a section of $\frL$ that in region $\U$ is represented, relative to the trivialization used in writing $A=p \d q$, by a function $\psi(q)$,
and in region $\t \U$ is represented, relative to the trivialization used in writing $\t A=\t p \d \t q$, by a function $\t\psi(\t q)$.   Eqn. (\ref{elp}) suggests 
that in the region $\U\cap \t \U$ these should be related by $\psi(q)= q^{-2\i s}\t\psi(\t q)$. But we have to be careful because $\U\cap \t\U$ is not
connected; it consists of two components, one  with $q>0$ and one with $q<0$.  
We could generalize the gluing law to $\psi(q)=e^{\i\alpha}q^{-2\i s}\t\psi(\t q)$ with a real constant $\alpha$.   If we use the same constant for positive
and negative $q$, this will not matter, but 
if we use different constants, this does matter; the possibility of doing this corresponds to the existence of a one-parameter family of choices of $\frL$.
We fix the choice of $\frL$ by specifying that $\psi(q)=|q|^{-2\i s}\psi(\t q)$ for both positive and negative $q$.

  For a convenient way to describe this, first  note that the measure $\d q$ satisfies 
  \be\label{measures}\frac{ \d q}{q^2} = \d \t q.\ee   Since measures (as opposed to one-forms)
  are positive, it makes sense to raise this formula to any complex power.    We find that the gluing law for $\psi$ and $\t\psi$ is equivalent to 
\be\label{convenone} \psi(q)(\d q)^{\i s} =\t\psi(\t q)(\d\t q)^{\i s}. \ee
So we conclude that a section of $\frL \to N\cong \RP^1$ can be understood as a density on $\RP^1$ of weight $\i s$.

The Hilbert space $\H=\Hom(\B,\B_\cc)$ consists of half-densities on $\RP^1$ with values in $\frL$, so it is the space of $\frac{1}{2}+\i s$ densities on $\RP^1$.
The Hilbert space inner product in the quantization of a twisted cotangent bundle was given in eqn. (\ref{nho}).  As always in brane quantization, the algebra 
$\A=\Hom(\B_\cc,\B_\cc)$ acts on $\H$ in a natural way.   In the present example, this algebra is
generated by the operators $t_-, t_+,$ and $t_3$ of eqn. (\ref{agrops}).   They are all antihermitian as operators on $\H$.  
On $\H$, these operators generate
 a representation of $\PSL(2,\R)$ for any complex $s$.   For real $s$, this representation is manifestly unitary with the Hilbert space structure (\ref{nho}).

We could have started with the other holomorphic polarization $\varPi'$ of $Y$.   This would have led to the same formulas with a substitution $s\to -s$.
So brane quantization predicts an isomorphism between  the unitary representations of $\PSL(2,\R)$ associated to densities of weight $\frac{1}{2} +\i s$
and $\frac{1}{2}-\i s$.  This isomorphism is well-known and is usually proved via explicit elementary formulas, or simply by noting that the two representations
have no highest or lowest weight vector and have the same values of the quadratic Casimir operator.

For a more general choice of $\frL$, we would get $\frac{1}{2}+\i s$ densities on $\RP^1$ with values in a flat line bundle that has a monodromy $e^{\i \veps}$ in
going around $\RP^1\cong S^1$, with $\veps\in \R$.    
This monodromy does not matter topologically; the various line bundles $\frL_N\to N$ are isomorphic as unitary line bundles.   But the quantum theories
do depend nontrivially on $\veps$.  The same algebra $\A$  acts as before, regardless of $\veps$, by the same formulas (\ref{agrops}).  
 But the representation generated by these operators depends on $\veps$.    For $e^{\i\veps}=-1$,  one gets, for real $s$,
 a family of unitary representations of $\SL(2,\R)$; for other values $\veps$, one gets a family of representations of a different cover of $\PSL(2,\R)$,
generically the universal cover.  In all cases the equivalence $s\leftrightarrow -s$ remains.

\subsection{Comparing to Geometric Quantization With A Complex Polarization}\label{branecp}

Now we consider brane quantization of $M$ in the case that $Y$ admits a holomorphic polarization $\varPi$ that induces a complex structure $\J$ on $M$.
As discussed in Section \ref{scbq}, holomorphic polarizations come in pairs $\varPi$, $\varPi'$ that induce opposite complex structures $\J$, $-\J$ on $M$.
These cannot both be Kahler, so we briefly discuss geometric quantization with complex polarizations that are not necessarily Kahler.  

If $M$ has prequantum line bundle $\frL$ and a polarization defined by a complex structure $\J$, then in general what geometric quantization
associates to this data is
 the cohomology $H^r_\J(M, K_M^{1/2}\otimes \frL)$ for $0\leq r\leq n=\dim_\C M$.   
The notation $H^r_\J$ refers to sheaf cohomology with complex structure $\J$ (and when necessary, we write $K_{M,\J}$ for $K_M$).
A complex polarization in general is characterized by an integer $r_0$, defined as the number of complex directions, in the tangent space at any point $m\in M$,
in which the symplectic form $\omega$ is negative.   (This number is the same for any point in $M$, since $\omega$ is everywhere nondegenerate.)
A polarization is Kahler if $r_0=0$.    This is the most important and most-studied case.   
In the case of a Kahler polarization, if one is sufficiently close to a classical limit (that is, if the periods of the symplectic form $\omega$ are large enough),
the cohomology $H^r_\J(M,K_M^{1/2}\otimes \frL)$ vanishes for $r>0$.   In this case, the recipe of geometric quantization reduces to 
 saying that the Hilbert space of quantum states is
\be\label{defq} \H_\gq=H^0_\J(M,K_M^{1/2}\otimes \frL)\ee   with the natural hermitian inner product
\be\label{mellow}\la \chi,\psi \ra=\int_M\d\mu \bar\chi\psi ,\ee
where $\mu$ is the symplectic measure on $M$.     In the case
of  a complex polarization that is not Kahler, one has to consider $H^r_\J(M,K_M^{1/2}\otimes \frL)$ with $r\not=0$.
Provided that one interprets $H^r_\J(M,K_M^{1/2})$ as the space of $\ltwo$ harmonic $(0,r)$-forms on $M$ with values in $K_M^{1/2}\otimes \frL$,
the definition (\ref{mellow}) of a Hilbert space structure on $H^r_\J(M,K_M^{1/2}\otimes \frL)$ makes sense for any $r$.   As long as there is precisely one value
of $r$ for which $H^r_\J(M,K_M^{1/2}\otimes \frL)$ is nonzero, this space is a reasonable candidate as a quantum Hilbert space and thus
geometric quantization gives a reasonable answer for quantization of $M$.     If one is sufficiently close to a classical limit, one can hope
that the relevant $\ltwo$ cohomology is nonzero only for one value of $r$, namely  $r=r_0$.
 If the  cohomology is nonzero for more than one value of $r$, 
then it is doubtful that geometric quantization with the complex polarization $\J$
is really giving a reasonable answer for quantization.

A special case  is that $\J$ defines a Kahler polarization, and we want to quantize $M$ with the complex structure $-\J$.   With respect to $-\J$,
$\omega$ is negative definite so $r_0$ is equal to $n=\dim_\C M$.    Hodge theory identifies $H^n_{-\J}(M, K_{M,-\J}^{1/2}\otimes \frL)$ with
$H^0_\J(M,K_{M,\J}^{1/2}\otimes\frL)$, so geometric quantization with complex structure $-\J$ is equivalent to geometric quantization with complex structure $\J$.
More generally, for any value of $r$, Hodge theory gives an isomorphism
$H^r_\J(M,K_{M,\J}^{1/2}\otimes \frL)\cong  H^{n-r}_{-\J}(M,K_{M,-\J}^{1/2}\otimes \frL)$, 
 so geometric quantization based on $-\J$ is always equivalent to geometric quantization based on $\J$.
 
 Our goal in this section is to compare brane quantization to geometric quantization under certain hypotheses.   Under certain conditions, we will
 show that (regardless of the value of $r_0$) the cohomology groups considered in geometric quantization admit the action of the algebra of observables $\A=\Hom(\B_\cc,\B_\cc)$ that
 acts in brane quantization.   We will also show more directly that, in the case of a Kahler polarization, there is a natural identification of $\H=\Hom(\B,\B_\cc)$ with
 $\H_\gq$ {\it as vector spaces}.   This identification is {\it not} guaranteed to extend to an equivalence between the Hilbert space structures.
When an appropriate antilinear symmetry is present,  brane
 quantization leads to a definition of a Hilbert space structure on $\H$, but there is no evident reason that this Hilbert space structure will be equivalent
 to an elementary formula such as the one in eqn. (\ref{mellow}).  An example that will be considered in Section \ref{flatg} suggests that nothing as simple
 as that is true.

To carry out this program, we assume that $Y$ is presented with  a holomorphic polarization $\varPi$, 
and a projection $\pi:Y\to W$, for some $W$.   More specifically, we  assume that $Y$ is a twisted cotangent
bundle,  constructed as in Section \ref{shifted}  starting with a holomorphic line bundle $\L\to W$, or more generally a formal tensor product of complex
powers of line bundles,\footnote{Here and later we set $\hbar=1$.}  $\L=\otimes_{s=1}^k \L_s^{\lambda_s}$.    

Let us first consider the case that $M$ is compact. In this case, $\pi(M)$ must be all of $W$ (rather than a proper open set in $W$, which would not
be compact).    So $M$ is a nonholomorphic section of $\pi:Y\to W$, which ensures that when restricted to $M$,
$\pi$ is an isomorphism between $M$ and $W$.    Since the fibers of $\pi$ are contractible, this implies that $Y$ is contractible onto $M$.
   To define a Lagrangian brane $\B$ supported on $M$,  first of all $M$ should be Lagrangian, meaning that $\Im\,\Omega$ vanishes
when restricted to $M$.   Since $Y$ is contractible onto $M$, this implies vanishing of the cohomology class $[\Im\,\Omega]$.
We also need a prequantum line bundle $\frL\to M$ whose
curvature is $\Re\,\Omega$.  This implies that  $c_1(\frL)= [\Re\,\Omega]/2\pi$.  Combining the two statements, we need $[\Omega]/2\pi = c_1(\frL)$
for an actual line bundle $\frL$.  A look back to eqn. (\ref{jnko}) shows that we can satisfy these
conditions if and only if $\L$ exists as in actual line bundle, in which case we should set $\frL=\L$.   To be more precise,
$\L$ was only defined up to the possibility of tensoring with a flat line bundle.   In defining $\frL$, we have to make a specific choice.

It is also possible for $W_0=\pi(M)$ to be a proper open subset of $W$.   In that case, we get less restrictive conditions: we require only that 
when restricted to $\pi^{-1}(W_0)$, $[\Im\,\Omega]$ must vanish and $\L$ must exist as an actual line bundle.

We can illustrate the difference between these two cases by considering again the examples that were discussed in Section \ref{example}.  
First consider the complex manifold $Y$ defined by $x^2+y^2+z^2=j^2$, with complex symplectic form $\Omega=\d x\d y/z$.   We assume that $j^2>0$
and define $M$ by the same equation $x^2+y^2+z^2=j^2$ with real $x,y,z$.   As shown in Section \ref{scbq}, $\pi:M\to W$ is an isomorphism from
$M\cong S^2$ to $W\cong \bCP^1$.     Thus $Y$ is contractible onto $M$.   A short calculation gives
\be\label{shortc} \int_M \Omega =4\pi j. \ee
Thus the condition to have a line bundle $\L\to W\cong \bCP^1$ with $c_1(\L) =[\Omega]/2\pi$ is that $j\in\frac{1}{2}\Z$.   In this case, the line bundle $\L$ exists and
is $\L=\O(2j)$.  For the prequantum line bundle of $M$, we pick $\frL=\L$; we define a brane $\B$ supported on $M$ whose
$\CP$ bundle is  $\frL^{-1}$.   

Let us compare this to the superficially similar example of the complex manifold $Y'$ defined by $x^2-y^2-z^2=j^2$, again with $\Omega=\d x\d y/z$
and $j^2>0$.   We define $M$ by the same equation $x^2-y^2-z^2-j^2$, now with real $x,y,z$, with the additional condition $x>0$.    
The difference from the previous case is that  $M$ is contractible; as explained in Section \ref{scbq},
the map $\pi:M\to W$ identifies $M$ with an open upper half-plane (or upper hemisphere) in 
$W\cong \bCP^1$. So the condition for $\L$ to exist as a line bundle when restricted to $M$ is trivial, and
we can take $\L=\O(2j)$ for any real $j$.   The prequantum line bundle is then also $\frL=\O(2j)$.  This would not make sense as a line bundle over $\bCP^1$,
but it does make sense over $M$.   

A test of the claim that in examples of this kind, brane quantization and geometric quantization predict the same spaces of physical states  (but possibly
not the same Hilbert space structure) is as follows.   The algebra  that acts naturally on
$\H=\Hom(\B,\B_\cc)$ is $\A=\Hom(\B_\cc,\B_\cc)$.  If we sheafify along $W$, we can think of this as a sheaf of algebras, not just a single algebra.
  In Section \ref{shifted}, we showed that $\A$ is the sheaf of holomorphic differential operators (in complex
structure $\J$, of course) 
acting on $K_M^{1/2}\otimes \frL$.     Obviously, this algebra acts on holomorphic sections of $K_M^{1/2}\otimes \frL$, and almost equally obviously,
since holomorphic differential operators commute with the $\bar\partial $ operator, the same algebra acts on all of the the $\bar\partial $ cohomology groups
with values in $K_M^{1/2}\otimes \frL$.   Thus, the space of physical states as defined in geometric quantization does admit an action of the algebra
of observables that one would expect in brane quantization.

For the simple examples involving $Y$ or $Y'$, we can make these statements completely explicit.   In the case of $Y$, the algebra of observables
is generated by the three operators $t_+, t_-$, and $t_3$ of eqn. (\ref{throps}).   For $j\in\Z/2$, the physical Hilbert space $\H=H^0(\bCP^1,\O(2j-1))$
is of dimension $2j$; the $2j$ states furnish an irreducible representation of $\SU(2)$ of spin $j-1/2$.   In the representation of eqn. (\ref{throps}),
the states are the polynomials in $q$ of degree $\leq 2j-1$; the operators $t_+, t_-$ and $t_3$ act on these states as the familiar $\frak{su}(2)$ generators.
(For example, $t_3 $ has eigenvalues $j-1/2,j-3/2,\cdots, -j+1/2$.)  Clearly, all this only makes sense because $2j$ is an integer.
The story is similar in the case of $Y'$, but now the operators $t_+,t_-,t_3$ generate
a unitary representation of $\PSL(2,\R)$ (or a cover of this group, depending on the value of $j$), rather than $\SU(2)$, acting on holomorphic sections of
$\O(2j-1)$ over $M\cong \H$. 

If $\J$ is a Kahler polarization, we can argue in a more direct way that $\H_{\gq}$ is the same as $\H=\Hom(\B,\B_\cc)$.   We  imitate the construction
of Section \ref{ccb}.  Let $w$ be a point in $M$, and $F$ the fiber of $\varPi$ that contains $w$.  As usual, we define a brane $\F$ supported on $F$
with trivial $\CP$ bundle, and find an element of $\H'=\Hom(\B_\cc,\B)$ by composing suitable elements of $ \Hom(\B_\cc,\F)$ and $ \Hom(\F,\B)$.   
As before, $\Hom(\B_\cc,\F)\cong H^0(F,K_F^{1/2})$.   
After picking local coordinates $\vec q$  and $\vec p$ parametrizing $M$ and $F$ near $w$, such that $\Omega=\sum_i \d p_i \d q^i$,
we can pick an element of  $ H^0(F,K_F^{1/2})$ that restricts to $\alpha=(\d\vec p)^{1/2}$ at $w$.  Since $\F$ and $\B$ meet at the unique point $w$, 
$\Hom(\F,\B)$ is one-dimensional
and is a copy of $\frL^{-1}|_w$, the fiber of the $\CP$ bundle of $\B$ at the point $w$.  So there is an element of $\Hom(\F,\B)$ for any $\beta\in \frL^{-1}|_w$.  
Composing these elements of $\Hom(\B_\cc,\F)$ and $\Hom(\F,\B)$, we get an element $\psi'(w;\alpha,\beta)\in \H'=\Hom(\B_\cc,\B)$.   $\psi'$ varies
holomorphically with $w$; this will be explained in Section \ref{flatcon}.   As in Section \ref{ccb}, if $\Psi$ is any element of $\H$, we can define the $A$-model 
bilinear pairing
$f(w;\alpha,\beta)=(\psi'(w;\alpha,\beta),\Psi)$.   Then we multiply by $\sqrt{\d\vec q}\,\beta^{-1}$ to undo the dependence on the choices of $\alpha$ and $\beta$.
So we define
\be\label{zofo}\h\Psi = \sqrt{\d \vec q}\,\beta^{-1} f(w;\alpha,\beta). \ee
Because of the factors of $\sqrt{\d \vec q}$ and $\beta^{-1}$, $\h\Psi$  is valued in the fiber at $w$ of $K_M^{1/2}\otimes \frL$.   Moreover, $\h\Psi$
is holomorphic in $w$ because $\psi'(w;\alpha,\beta)$ varies holomorphically with $w$.   So in short,  this construction
defines a map from the physical state space $\H$ of brane quantization to its analog $\H_\gq$ in geometric quantization.    We expect that this map
is nonzero precisely when the complex polarization of $M$ that is induced by $\varPi$  is Kahler, and in that case is an isomorphism.

\begin{figure}
 \begin{center}
   \includegraphics[width=4.5in]{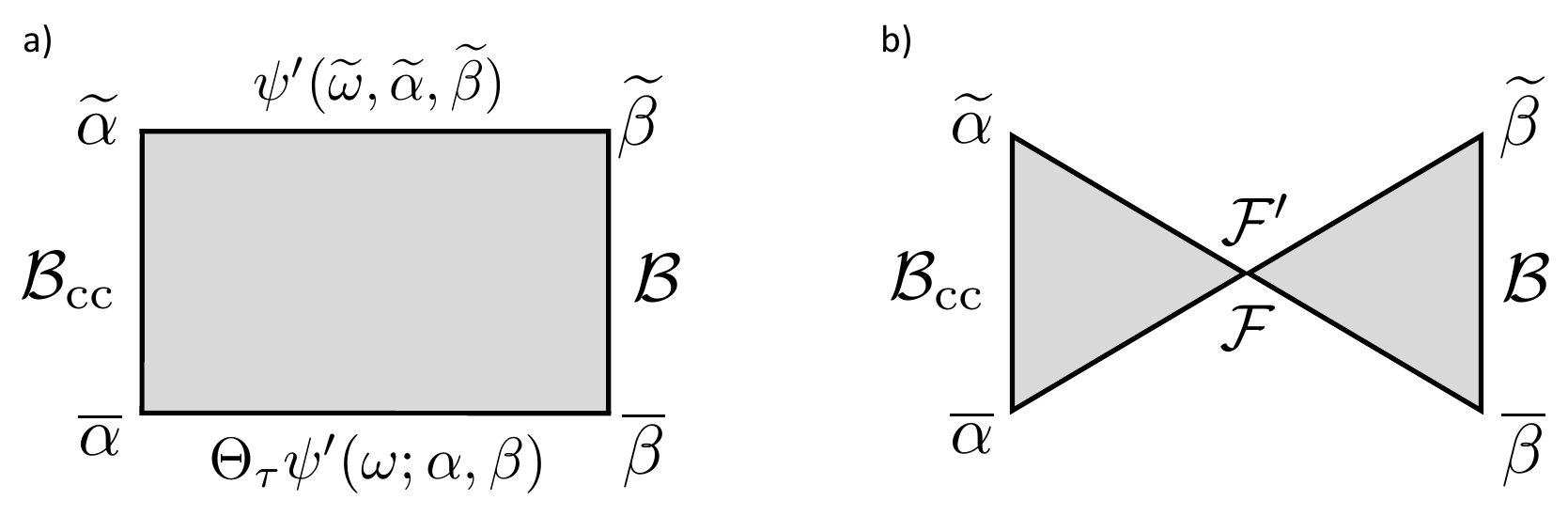}
 \end{center}
\caption{\small (a) A path integral on this rectangle, with the indicated boundary conditions, can be used to compute an inner product.  (b)  The rectangle can
be reduced to this union of triangles, joined at a vertex.   In evaluating the corresponding path integral, one has to sum over states
propagating through that vertex.    \label{newrectangle}}
\end{figure}

Brane quantization gives a recipe to define the hermitian structure of $\H$, but it is hard to reduce this recipe to an explicit formula.   The general recipe 
was explained in Section \ref{hermitian}.   To make it somewhat more explicit in the present context, we can proceed as follows.  First, it is equivalent
to define a hermitian structure on $\H$ or on its dual space $\H'$, and it will be convenient to do the latter.   We will define the inner product between
 $\psi'(w;\alpha,\beta)$ and a similar state $\psi'(\t w;\t\alpha,\t \beta)\in\H'$ defined with another set of choices $\t w, \t\alpha, \t\beta$.
The general recipe says the hermitian inner product is obtained from the $A$-model bilinear pairing after applying the antilinear operator $\Theta_\tau$ to
one of the two states: $\la\psi'(w;\alpha,\beta),\psi'(\t w;\t \alpha,\t\beta)\ra=\bigl(\Theta_\tau \psi'(w;\alpha,\beta),\psi'(\t w;\t \alpha,\t\beta)\bigr)$. The antilinear
mapping $\Theta_\tau$ exchanges the holomorphic polarization $\varPi$ with its conjugate $\varPi'$.   in particular,
$\Theta_\tau$ maps $\F$ to a brane $\F'$ supported on the fiber $F'$ of $\varPi'$ that contains $w$.   $\Theta_\tau$ likewise
maps the elements of $\Hom(\B_\cc,\F)$ and $\Hom(\F,\B)$ that we composed to get $\psi'(w;\alpha,\beta)$ to
elements of $\Hom(B,\F')$ and $\Hom(\F',\B_\cc)$ whose composition is $\Theta_\tau\psi'(w;\alpha,\beta)\in \H=\Hom(\B,\B_\cc)$.   The  $A$-model
pairing between this state and $\psi'(w;\alpha,\beta)$ 
can be represented, as in examples discussed earlier, by a path integral on a rectangle with $\psi'(\t w;\t\alpha,\t\beta)$ at the top
and $\Theta_\tau \psi'(w;\alpha,\beta)$ at the bottom (fig. \ref{newrectangle}).  Unfortunately,
 the path integral on the rectangle is difficult to evaluate.   It reduces to a sum of products of triangles (fig. \ref{newrectangle}(b)), but in that
form it is likely still not possible to get any general formula.  

A noteworthy point is that, in contrast to the discussion of cotangent bundles, the states $\psi'(w;\alpha,\beta)$ are normalizable.   No singularity
develops in fig. \ref{newrectangle}  if we set $w=\t w$.   One should not expect to run into delta function states in the present context, because for example
if $M$ is compact, then the physical state spaces are finite-dimensional and there would be no such thing as a delta function state.

\subsection{What Happens When $Y$ is Hyper-Kahler and $M$ is Compact}\label{hyperm}

We should not leave this subject without explaining that in an important special case, there is a more simple way to partially compare brane quantization
to geometric quantization.

Throughout this analysis, we have merely treated $Y$ as a complex symplectic manifold.   We have not assumed that the complex symplectic
structure of $Y$ extends to a complete hyper-Kahler structure in the sense described in Section \ref{amo}.  This assumption is sufficient to ensure
that a good $A$-model of $Y$ exists, and it holds in many interesting examples. But it is very likely not necessary.

If we do assume that the complex symplectic structure of $Y$ extends to a hyper-Kahler structure, then
a more simple way to partially compare brane quantization to geometric quantization presents itself.  This was actually already discussed in \cite{GW} (Sections 2.3 and
4).    First of all, the symmetry $\tau$ of $Y$, which is antiholomorphic in complex structure $I$, actually acts holomorphically in complex structure $J$ (see 
eqn. (\ref{tabush})).
So $M$, as a component of the fixed point set, is holomorphic in complex structure $J$.   Since $\omega_J$ is Kahler with respect to $J$, this means that
complex structure $J$ provides a complex Kahler polarization of $M$, in the sense of geometric quantization.

Geometric quantization of  $M$ with this polarization leads to the Hilbert space
\be\label{hgq} \H_\gq=H^0(M,K_M^{1/2}\otimes \frL),\ee where as
usual $\frL$ is the prequantum line bundle.  (For simplicity we assume that the $\CP$ curvature $F=\omega_J$ of $\frL$ is sufficiently positive so that
the higher cohomology vanishes.).   The Hilbert space structure is given by the inner product
\be\label{ngg}(\psi,\chi)=\int_M \d\mu \bar\psi\chi,\ee
where $\d\mu$ is the measure derived from the symplectic structure $\omega_J$.

Let us see if we can justify a relationship between $\H_\gq$ and what one would get from brane quantization.     First, we note that in this situation, both
$\B_\cc$ and the Lagrangian brane $\B$ supported on $M$ are branes of type $(A,B,A)$, that is, they are $A$-branes with respect to symplectic
structures $\omega_I$ and $\omega_K$,
and $B$-branes with respect to complex structure $J$.   This can be seen as follows.\footnote{The supercharges
that serve as
the differentials of the $A$-models of $\omega_I$ and $\omega_K$ and the $B$-model of $J$ are linearly dependent.   Hence it suffices
to check invariance under two of these differentials; invariance under the third follows.   It seems illuminating to check all three invariances directly, so we do so.}

  In the case of $\B$, we simply observe that as $\omega_I$ is $\tau$-odd (eqn. (\ref{tabush})), the
$\tau$-invariant submanifold $M$ is Lagrangian for $\omega_I$, just as it is for $\omega_K$.  So $\B$ is an $A$-brane for $\omega_I$ just as it is
for $\omega_K$.    On the other hand, we have noted that $M$ is a complex
submanifold in complex structure $J$.    The $\CP$ bundle of $\B$ satisfies $\sF+\sB=0$.   In particular, $\sF+\sB$ is of type $(1,1)$ in complex
structure $J$.    A brane
supported on a complex submanifold with $\sF+\sB$ of type $(1,1)$  is a $B$-brane, so $\B$ is a $B$-brane in complex structure $J$.   So in short it
is a brane of type $(A,B,A)$.

For $\B_\cc$, its support is all of $Y$, which is a complex manifold in complex structure $J$, and its $\CP$ curvature satisfies $\sF+\sB=\omega_J$, which is of type
$(1,1)$ in complex structure $J$.
Again these conditions ensure that $\B_\cc$ is a $B$-brane with respect to $J$.     
$\B_\cc$ is an $A$-brane with respect to $\omega_K$ because  $\omega_K^{-1}(\sF+\sB)=I$ is an integrable complex structure,
and similarly it is an $A$-brane with respect to $\omega_I$ because $\omega_I^{-1}(\sF+\sB)=-K$ is an integrable complex structure.  So again $\B_\cc$ is a brane
of type $(A,B,A)$.   

The answer that brane quantization gives for the quantization of $M$ with our usual choices of $\CP$ bundles (trivial for $\B_\cc$, $\frL^{-1}$ for $\B$)
is the space of $(\B,\B_\cc)$ strings in the $A$-model of $\omega_K$.   We have been calling
this simply $\Hom(\B,\B_\cc)$,  but
 in the present context it is better to write
$\Hom_{A_K}(\B,\B_\cc)$, where the subscript $A_K$ is meant to remind us that we consider the physical string states in the $A$-model of type $\omega_K$.
The answer (\ref{hgq})  that geometric quantization gives for quantizing $M$ with the Kahler polarization $J$ is the same as $\Hom_{B_J}(\B,\B_\cc)$, in other
words, the space of $(\B,\B_\cc)$ strings in the $B$-model of complex structure $J$.

Thus brane quantization using the complex structure $Y$ agrees with geometric quantization using the complex Kahler polarization $K$ if and only if
$\Hom_{A_K}(\B,\B_\cc)$ coincides with $\Hom_{B_J}(\B,\B_\cc)$.     In general, this is probably too much to hope for, though a clear counter-example is
not immediately apparent.
But as noted in \cite{GW}, if $M$ is compact,  a very simple argument  gives an isomorphism between 
$\Hom_{A_K}(\B,\B_\cc)$ and $\Hom_{B_J}(\B,\B_\cc)$ purely as vector spaces, ignoring their hermitian inner products.   The argument is simply 
 that if $M$ is compact, then instead of considering the space of $(\B,\B_\cc)$ strings in one or another twisted topological field theory,
we can simply consider the supersymmetric ground states of the $(\B,\B_\cc)$ system.   This will coincide, purely as a vector space, with the space
of physical states in any of the possible twisted supersymmetric field theories, and this observation implies the equivalence of vector spaces 
$\Hom_{A_K}(\B,\B_\cc) \cong \Hom_{B_J}(\B,\B_\cc)$.

This relationship, however, does {\it not} extend to a relation between hermitian inner products.   There is no reason to expect the hermitian inner product
of $\Hom_{A_K}(\B,\B_\cc)$ to agree with that  of $\H_{\gq}$, described in eqn. (\ref{ngg}).
 As for $\Hom_{B_J}(\B,\B_\cc)$, for it to carry a natural hermitian structure,
$Y$ should have a $\Z_2$ symmetry that acts antiholomorphically in complex structure $J$.   The existence of such a symmetry does not follow from our
assumptions, so in general $\Hom_{B_J}(\B,\B_\cc)$ does not have a hermitian structure.

In Section \ref{branecp}, we explained that when $M$ has a complex structure $\J$ that is induced from a holomorphic polarization $\varPi$ of $Y$,
$\Hom_{A_K}(\B,\B_\cc)$ is the same, as a vector space, as $H^0_\J(M,K_M^{1/2}\otimes \frL)$.    This description did come with a recipe for
describing the hermitian inner product of $\Hom_{A_K}(\B,\B_\cc)$, but this recipe was fairly inexplicit, involving a holomorphic triangle that is
in general hard to evaluate.  One does not expect that this recipe will in general reduce to something as simple as the usual definition (\ref{ngg}) that
geometric quantization would suggest.   

We thus have two strategies to compare brane quantization of $M$ via a complexification $Y$ to quantization using a complex Kahler polarization.  One
strategy uses a holomorphic polarization of $Y$ and one uses a hyper-Kahler structure of $Y$.

 Instead of further discussion of the sort of elementary
examples that we have analyzed in this article, we will discuss the two approaches in the context of a more sophisticated example, which was discussed
from the hyper-Kahler point of view in \cite{GW}, Section 4 (where further detail was given beyond what we will say here).  

\subsection{Quantizing The Moduli Space of Flat $G$-Bundles}\label{flatg}

Let $G$ be a compact Lie group with
complexification $G_\C$.  We write $\g$ and $\g_\C$ for the corresponding Lie algebras.    $G_\C$ has an antiholomorphic involution $\varrho$ with fixed point set $G$.
Let $C$ be a compact closed oriented two-manifold, and let $M$ be the moduli space of flat $G$-bundles\footnote{In this
discussion, we will ignore technical issues involving singularities of $M$.  One can entirely avoid such issues by, for example, taking $G$ to be $\SO(3)$
and specifying that the bundle $E$ should have $w_2(E)\not=0$.}   $E\to C$ of a specified topological type.   $M$ has a natural
symplectic structure $\omega_0$ that we normalize so that its smallest nonzero period is $2\pi$.   For $G=\SU(N)$, it can be defined by the formula \cite{ABott}
\be\label{defform}\omega_0=\frac{1}{4\pi}\int_C\Tr \,\delta A\wedge \delta A, \ee
where $\Tr$ is the trace in the fundamental $N$-dimensional representation of $G$, and $\delta A$ represents the variation of a flat connection on $E$.
The condition on its periods means that  $\omega_0$ is the curvature of a line bundle
$\L\to M$ (which generates the Picard group of $M$ modulo torsion). One wishes to quantize $M$ with symplectic structure\footnote{The level of the
Chern-Simons and WZW models that will be mentioned shortly, as conventionally defined, is not $\kappa$.  Rather, assuming for simplicity that $G$ is simply-connected,
it is  $k=\kappa-h$, where $h$ is the dual Coxeter number of $G$.  
For simply-connected $G$, $K_M^{1/2}\sim \L^{-h}$.   In some approaches, it is natural to refer to what we call $\frL \otimes K_M^{1/2}$, rather than $\frL$,  as the prequantum
line bundle, and formulas are written directly in terms of $k$ rather than $\kappa$.}
$\omega =\kappa\omega_0$ and prequantum  line bundle $\frL=\L^\kappa$, for some integer $\kappa$.
There is no loss of generality to assume $\kappa>0$, since the sign of $\kappa$ is reversed if one reverses the orientation of $C$.   

$M$ is the phase space of three-dimensional Chern-Simons gauge theory with gauge group $G$ on $\R\times C$, where $\R$ parametrizes the ``time''  \cite{WittenJ}.
The goal of quantization is to construct a Hilbert space associated to the spatial manifold  $C$.   The symplectic structure of $M$ is defined using no structure
of $C$ other than an orientation.  Thus formally one can hope that quantization can be carried out in such a way that the orientation-preserving diffeomorphisms
of $M$ will act naturally on the quantum Hilbert space. 
   In fact, this is necessary if one is going to get a topological field theory from the three-dimensional Chern-Simons theory.

For brane quantization, the first step
is to pick a nice complexification of $M$.   Indeed, $M$ has a nice complexification $Y$,
 the moduli space of flat $G_\C$-bundles $\h E\to C$.   This is a complex symplectic manifold; the complex structure comes from the complex structure
 of $G_\C$, and the holomorphic
 symplectic structure, which is defined by the formula (\ref{defform}) with the $\g$-valued  flat connection $A$ replaced by a $\g_\C$-valued flat
 connection $\cA$,  is an analytic continuation of the symplectic structure of $M$.
   A typical holomorphic function on $Y$ is the trace, in some holomorphic representation, of the holonomy of $\cA$ around
a given oriented loop in $C$.   To be coherent with the notation in the present article, we will call this complex structure $I$, though unfortunately in 
Hitchin's work on Higgs bundles \cite{Hitchin}, whose relevance will be described momentarily, and in applications of that work to geometric Langlands, it is usually
called $J$.   $Y$ has an antiholomorphic involution $\tau$ that leaves $M$ fixed.   $\tau$ can be defined by applying $\varrho$ to $\cA$, or to its monodromies.

The choice of $Y$ did not spoil the diffeomorphism symmetry of $C$; orientation-preserving diffeomorphisms of $C$ act on $Y$, preserving its complex
symplectic structure.   So formally the procedure of brane quantization gives a diffeomorphism-invariant answer for the quantization of $M$:
it is $\H=\Hom_{A_K}(\B,\B_\cc)$.

We can try to make this concrete by either the choice of a holomorphic polarization or the choice of a hyper-Kahler metric. (The latter approach was
described in \cite{GW}, with more detail on some points than we will give here.) 
 Either approach will involve an auxiliary choice
that appears to break the diffeomorphism symmetry.    But the equivalence to  $\Hom_{A_K}(\B,\B_\cc)$ indicates that the output that comes from
the holomorphic polarization or the hyper-Kahler metric does have the diffeomorphism symmetry.   

The starting point in either of the two approaches to making brane quantization of $M$ more concrete is to pick a complex structure on $C$.  Of course,
there is not a natural choice; the possible choices, up to diffeomorphisms that are connected to the identity, make up a Teichm\"{u}ller space $\T$.   For $t\in \T$,
let $\I_t$ be the corresponding complex structure on $C$.

Once $\I_t$ is given, 
 it is possible to make brane quantization concrete by using either a holomorphic
polarization of $Y$ or a hyper-Kahler structure.   First we describe the approach with a holomorphic polarization. 

Once $\I_t$ is given, a flat bundle has a holomorphic structure, defined by the $(0,1)$ part of the flat connection, and an antiholomorphic structure,
defined by the $(1,0)$ part.   To define a holomorphic polarization of $Y$, place on $Y$ an equivalence relation in which two flat $G_\C$ bundles are considered
equivalent if they have the same holomorphic structure.   Every point in $Y$ is in a unique equivalence class.    Consider an equivalence class that
contains the flat connection $\cA=\d \bar z \cA_{\bar z}+\d z \cA_z$.   Any other connection in the same equivalence class has (up to gauge transformation) the
same $(0,1)$ part, so it takes the form
$\cA'=\cA +\W$, where $\W=\d z \W_z$ is a $(1,0)$-form  valued in $\ad(\h E)$,  the adjoint bundle of $\h E$.   If $\cA$ is flat, the
condition that $\cA'$ is also flat is that $\W$ is holomorphic, in other words it is valued in $H^0(C,K_C\otimes \ad(\h E))$.   An index theorem shows
that the dimension of this vector space is $n=\frac{1}{2}\dim_\C Y$.    Thus the equivalence classes are copies of $\C^n$, so this construction defines
a holomorphic polarization $\varPi_t$ of $Y$.   

One can see as follows that 
$\varPi_t$ induces on $M$ a complex polarization.  The definition of $\varPi_t$ makes clear that each point of $M$ is contained in a unique leaf; what
remains is to show that each leaf contains at most one point in $M$.
This statement  is part of the Narasimhan-Seshadri theorem, which establishes a correspondence
between stable holomorphic $G_\C$ bundles over $C$ and flat  $G$ bundles.  The intersection $F\cap M$ is the space of flat unitary $\frak g$-valued connections on
a given holomorphic $G_\C$ bundle.   The Narasimhan-Seshadri theorem says that $F\cap M$ is a single point if $E$ is stable and is empty otherwise.

In terms of the expansion  $\cA=\cA_{\bar z}\d \bar z+\cA_z\d z$ of  a complex flat connection, 
the space of leaves is parametrized holomorphically by $\cA_{\bar z}$.  After restricting to $M$, the expansion becomes $A=\d \bar z A_{\bar z}+ \d z A_z$,
where $A_z=-A_{\bar z}^\dagger$ (assuming one's convention is that a unitary flat connection is anti-hermitian).    
So the complex structure on $M$ that is induced by $\varPi$ is the one
in which $A_{\bar z}$ varies holomorphically. 
  In this complex structure, which we will denote as $\J_t$, $M$ can be interpreted as the moduli space of holomorphic $G_\C$ bundles on $C$.   
If we write $\delta A=\d z \delta A_z+\d\bar z \delta A_{\bar z}$, then the formula (\ref{defform}) is bilinear in $\delta A_z$ and $\delta A_{\bar z}$, showing
that the symplectic form is of type $(1,1)$ with respect to $\J_t$, so that $\J_t$ is a complex polarization of $M$.   This is expected from general arguments in 
Section \ref{scbq}.   With a little more care, one can show that $\J_t$ is a Kahler polarization of $M$.

The arguments of Section \ref{branecp} are applicable to this example, and show that  the Hilbert space $\H$ of brane quantization is, as a vector space, $\H=
H^0_{\J_t}(M,K_M^{1/2}\otimes \frL)$.   This equivalence also comes with a not very explicit description of the 
Hilbert space structure of $\H$, involving a holomorphic triangle.

Applying to $\varPi_t$ the antiholomorphic involution $\tau$, we get another holomorphic polarization $\varPi'_t$.   The leaves of this polarization consist of flat $G_\C$
bundles that all have the same antiholomorphic structure.   $\varPi'_t$ induces the opposite complex structure $-\J_t$ on $M$.      
In general, as explained in Section \ref{branecp}, 
geometric quantization using a complex polarization is invariant under changing the sign of the complex structure.   Accordingly, geometric quantization with
$\varPi'_t$ gives equivalent results to those that come from $\varPi_t$. We will not discuss $\varPi'_t$ further.

The alternative road to making brane quantization of $Y$ concrete is as follows.   Once  the two-manifold $C$ is given the complex 
structure  $\I_t$, one can formulate and solve Hitchin's equations \cite{Hitchin}, giving the flat $G_\C$ bundle $\h E$ the structure of a Higgs bundle.
The space $Y$ gets a new complex structure $J_t$  in which it parametrizes Higgs bundles, and  the complex symplectic structure of $Y$ is extended to
a hyper-Kahler structure.  ($J_t$ is called $I$ in Hitchin's work, and much subsequent literature.)    Concretely, Hitchin's equations involve a reduction of
the structure group of $\h E$ from $G_\C$ to $G$.   Relative to this decomposition, the complex flat connection $\cA$ decomposes as 
 $\cA=A+\i \phi$, where $\cA$ is a $G_\C$-valued flat connection, $A=\d \bar z A_{\bar z}+\d z A_z$ is
a $G$-valued connection, and $\phi=\d\bar z\phi_{\bar z}+\d z\phi_z$ is an adjoint-valued one-form on $C$.   $A$ and $\phi$ obey Hitchin's equations.  
The complex structure $J_t$ is the one in which the holomorphic variables are $A_{\bar z}$ and $\phi_z$.

The discussion in Section \ref{hyperm} shows that $\H$ can be identified as a vector space 
with $H^0_{J_t}(M,K_M^{1/2}\otimes \frL)$.   This identification does not lead to a description of the Hilbert space structure of $\H$.

The complex structure $J_t$ that we get by solving Hitchin's equations is defined on all of $Y$, while the complex structure $\J_t$ that we get by forgetting
the antiholomorphic structure of a flat bundle is defined only on $M$.   But  $J_t$, when restricted to $M$, is the same as $\J_t$. 
To show this, we just observe that $M$ is embedded in $Y$ as the locus $\phi=0$.   So the complex structure $J_t$, in which $A_{\bar z}$ and $\phi_z$
are holomorphic, restricts on $M$ to the complex structure $\J_t$, in which $A_{\bar z}$ is holomorphic.  

Therefore, the
description of $\H$ that we get from the holomorphic polarization is the same as the one that we get by solving Hitchin's equations.

Neither the description in terms of holomorphic polarizations nor the description in terms of solving Hitchin's equations makes it clear why or in what
sense $H^0_{\J_t}(M,K_M^{1/2}\otimes \frL)$ is independent of the choice of $t\in \T$.   Yet this is predicted by the relation to brane quantization.   It is
also needed for the topological invariance of three-dimensional Chern-Simons gauge theory \cite{WittenJ}, and to get a satisfactory picture of the holomorphic factorization
of the WZW model of two-dimensional conformal field theory \cite{KZ}.  The usual approach is to view $H^0_{\J_t}(M,K_M^{1/2}\otimes
\frL)$ as the fiber of a vector bundle over $\T$, and to construct a (projectively) flat connection on this vector bundle.   Integrating this flat connection
then gives an equivalence between $H^0_{\J_t}(M,K_M^{1/2}\otimes
\frL)$ for different $t$, restoring topological invariance and leading to satisfactory results in two-dimensional conformal field theory and three-dimensional
topological field theory.    The  flat connection has been constructed from several points of view \cite{ADW,Hitchtwo,Faltings}; the different approaches
 have been shown to give equivalent
results \cite{And}.  This  flat connection coincides with the natural
flat connection on the space of conformal blocks of the WZW conformal field theory \cite{Las}.
To be compatible with unitarity of Chern-Simons
gauge theory in three dimensions and the expected properties of the WZW model in two dimensions, this flat connection must be unitary; that is, it must
preserve some positive hermitian inner product on the fibers.   The global monodromies of the flat connection then lead to a unitary representation of the mapping
class group (more exactly, a central extension of it, as the connection is only projectively flat).  The known constructions of the flat connection via gauge theory do not lead to anything nearly as simple as the formula (\ref{ngg}) suggested by geometric quantization.  It is believed \cite{El,Gaw,Gaw2} that a correct formula requires
an extra factor and takes the form 
\be\label{nggo}(\psi,\chi)=\int_M \d\mu\, \bar\psi\chi\,  e^{G_\kappa},\ee
where $e^{G_\kappa}$ is the partition function  of a WZW model with target $G_\C/G$ at level $\kappa$.    This factor comes from
integration over orbits of $G_\C$-valued gauge transformations  in reduction from the space of all connections to the space $M$ of flat unitary connections.   
Abstractly, it has been proved that the flat connection is in 
fact unitary \cite{And2} and more concretely,  it is known that the unitary structure can be defined by a formula such as (\ref{nggo}), where for
large $\kappa$ there is an asymptotic expansion  $G_\kappa = F+\sum_{n=1}^\infty \kappa^{-n} F_n$, with the leading contribution
$F$ being the Ricci potential of $M$  \cite{And3}.   At leading order, this result agrees with the prediction based on the WZW
model of $G_\C/G$ \cite{And4}.

Brane quantization gives the answer for the Hilbert space inner product that we described in Section \ref{branecp}.   This description is not very explicit,
since it depends on a holomorphic triangle that is difficult to calculate.  It would be extremely interesting to understand how the expected factor
$e^{G_\kappa}$ arises in this approach.

\subsection{$A$-Branes and $B$-Branes}\label{flatcon}

Here we will describe some basic properties of the moduli of branes of type $(B,A,A)$.   Among other things, this will let us understand better the role of
a holomorphic polarization in arguments such as those of Section \ref{branecp}.

First we consider the $A$-model of a real symplectic manifold $S$ 
with symplectic form $\omega$.     A rank one Lagrangian $A$-brane $\B$ is supported
on a Lagrangian submanifold $L\subset S$, and endowed with a $\CP$ bundle $\L$ with connection $\sA$ and curvature $\sF$ that
obey $\sF+\sB=0$.   To deform $\B$, we can in general
deform both $L$ and $\sA$.   A general deformation of $\sA$ has the form $\sA\to \sA+\varepsilon \sa$, where $\sa$ is a 1-form and $\varepsilon$ is a small parameter.   
To preserve the condition $\sF+\sB=0$, one requires $\d \sa=0$.   However, if $\sa$ is exact, $\sa=\d \phi$ for some $\phi$, then the deformation
$\sA\to \sA+\varepsilon \sa$ is just a gauge transformation.   
  So the nontrivial deformations are by the de Rham cohomology $H^1_{\dR}(M,\R)$.
Let us suppose to begin with that $L$ is compact.   Then the nontrivial deformations correspond, by Hodge theory, to harmonic one-forms $\sa$.   

\begin{figure}
 \begin{center}
   \includegraphics[width=2in]{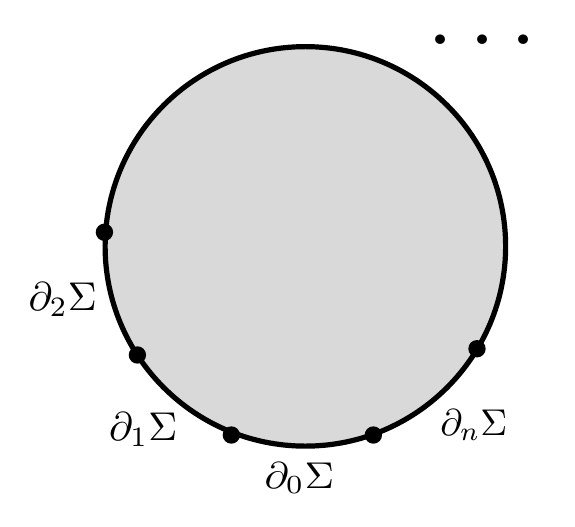}
 \end{center}
\caption{\small A disc with boundary segments labeled consecutively as $\partial_0\Sigma$, $\partial_1\Sigma$, $\cdots, $ $\partial_n\Sigma$.   On each segment
$\partial_\alpha\Sigma$, for $\alpha=0,1,\cdots, n$, the boundary condition is defined by a brane $\B_\alpha$.  Vertex operators are inserted at the junctions between segments.   \label{worldsheet}}
\end{figure}

Now consider instead deformations of $L$.   Locally, we can pick coordinates $p_i,q^j$ such that $\omega=\sum_i \d p_i \d q^i$ and $L$ is described
by $p_1=\cdots = p_n=0$.   Any 1-form $b=\sum_i b_i\d q^i$ gives a possible deformation of $L$ in which the equation $p_i=0$ is deformed
to $p_i=\varepsilon b_i$.    This preserves the fact that $L$ is Lagrangian  if and only if $b$ is closed, $\d b=0$.
However, if $b$ is exact, $b=\d \rho$ for some function $\rho (q)$, then the deformation of $L$ is by an infinitesimal symplectomorphism of $M$ generated by
$\rho$, namely 
$\delta p_i =\{p_i,\rho\}=\partial\rho/\partial q^i$, 
$\delta q_i=\{q_i,\rho\}=0$.   A deformation of $L$ by a symplectomorphism is trivial as a deformation of the $A$-brane $\B$.
Momentarily we will demonstrate this explicitly.   Hence the nontrivial deformations of $L$ in the $A$-model correspond, in first order, to closed one-forms
that are not exact, or in other words to the de Rham cohomology $H^1_{\dR}(M,\R)$.   Again, if $L$ is compact, this is the same as the space of harmonic
one-forms $b$.

Comparing the last two paragraphs, we see an obvious almost complex structure on the moduli space $\M$ of possible deformations of $\B$ as an $A$-brane.
In this almost complex structure, $a+\i b$ is a deformation of type $(1,0)$.   In fact, this almost complex structure is integrable and $\M$ is a complex manifold.

Now let us see explicitly why the exact terms in the one-forms $a$ and $b$ represent trivial deformations of $\B$ as an $A$-brane.   Consider the
$\sigma$-model action for the $A$-model on an oriented two-manifold $\Sigma$ with a map $X:\Sigma\to M$.    
The boundary $\partial\Sigma$  of $\Sigma$ is a union of segments
$\partial_\alpha\Sigma$ that are labeled by branes $\B_\alpha$ with $\CP$ connections $A_\alpha$ (fig. \ref{worldsheet}).   Suppose that one of these segments, say $\partial_0\Sigma$,
is labeled by the brane $\B$ of interest, with its $\CP$ connection $\sA$.  At the junctions between segments, local operators are inserted.
Let $T$ be the set of these junctions.   Most of the $\sigma$-model action is $Q$-exact, where $Q$ is the usual $A$-model
differential.   Spelling out the terms that are not $Q$-exact, we can write the action in the general form
\be\label{genformact} I= \int_\Sigma X^*(\omega) -\i\int_{\partial_0\Sigma} X^*(\sA) -\i\sum_{\alpha\not=0} \int_{\partial_\alpha\Sigma}X^*(\sA_\alpha) +\sum_ {t\in T}c_t+\{Q,\cdots\}. \ee
The  terms  that are not $Q$-exact
 are the contributions of the symplectic form and the $\CP$ connections, along with possible endpoint contributions $c_t$
corresponding to operators inserted at the junctions.  
The contribution of $\partial_0\Sigma$ is important in what follows, so we have separated out this contribution in eqn. (\ref{genformact}).
Likewise, 
the important junctions will be the left and right endpoints of
 $\partial_0\Sigma$; we call them $t_r$ and $t_\ell$.   
If we vary the brane $\B$ by an $a$ or $b$ variation, either $\delta \sA=a$ or else $\delta p_i=b_i, \,\delta q^j=0$, the action $I$ changes by a simple local integral along
$\partial_0\Sigma$:
\be\label{varact}\delta I=-\i\int_{\partial_0\Sigma} (a_i+\i b_i)\d q^i. \ee
We see that this naturally depends on $a$ and $b$ in the combination $a+\i b$, which is a manifestation of the complex structure of the moduli space $\M$.
Now specialize this to what we will call an exact variation, $a+\i b=\d \nu$ for some complex-valued function $\nu$.   In this case, the integral reduces to a
sum of boundary terms,
\be\label{varacto}\delta I=-\i(\nu(t_r)-\nu(t_\ell)) \ee (we abbreviate $\nu(X(t))$ as $\nu(t)$).   
So we can compensate for an exact variation by a suitable transformation of the insertions at the endpoints of $\partial_0\Sigma$:
\be\label{endvar}\delta c_r=\i \nu(t_r),~~~\delta c_\ell=-\i \nu(t_\ell). \ee
Finally we can formulate more precisely the statement that $A$-branes related by an exact deformation are equivalent.   To assert that an exact
deformation of $\B$ is in some sense trivial, 
 we need to know what to do with $\Hom(\B,\B')$ and $\Hom(\B',\B)$, for any other $A$-brane $\B'$, when we make the exact deformation of $\B$.   Eqn.
 (\ref{endvar}) answers this question: it tells us how the insertions at the endpoints of $\partial_0\Sigma$ have to be modified when we make an exact
 deformation of $\B$ to get a  deformation that is truly trivial.   

Now let us specialize to the case that the symplectic manifold $S$ is actually
 a complex symplectic manifold $Y$ and $\B$ is a brane of type $(B,A,A)$, supported on a complex Lagrangian
submanifold $L$.   Thus $L$ is a complex manifold in some complex structure $I$ and is Lagrangian for a holomorphic symplectic form $\Omega=\omega_J+\i
\omega_K$.  To deform $\B$ as a $B$-brane, we can deform $L$ as a complex submanifold and deform the flat $\CP$ bundle $\L$ of $L$ as a holomorphic
line bundle.   Both types of deformation of $\B$ are parametrized by complex moduli.   If $L$ is a compact Kahler manifold, then the moduli space
$\M$ that parametrizes deformations of $\B$ is the same whether one regards $\B$ as a $B$-brane of type $I$ or as an $A$-brane of
type $\omega_J$ or $\omega_K$.     This is so for a reason essentially explained in Section \ref{hyperm}: the tangent space to $\M$ is the
space of supersymmetric ground states of the $(\B,\B)$ system, which can be defined at the level of the physical $\sigma$-model, without
specializing to a particular twisted topological field theory.
However,  $\M$ has different complex structures depending on whether considers it as a moduli space of
$B$-branes in complex structure $I$ or of $A$-branes in symplectic structure $\omega_J$ or $\omega_K$ (or a linear combination thereof).   
On a compact Kahler manifold $M$, the Hodge decomposition expresses a harmonic one-form as the sum of pieces of types $(1,0)$ and $(0,1)$,
giving expansions $a=a^{(1,0)}+a^{(0,1)}$, $b=b^{(1,0)}+b^{(0,1)}$.   If $\M$ is considered to parametrize $B$-branes of type $I$, then it has
a complex structure in which the holomorphic directions are parametrized by $a^{(0,1)}$ and $b^{(1,0)}$.  If it is considered to parametrize $A$-branes 
in a symplectic structure $\J_\zeta$ that is a linear combination of $\omega_J$ and $\omega_K$, then the holomorphic directions are parametrized by
$a^{(0,1)}+\zeta b^{(0,1)}$ and $a^{(1,0)}-\zeta^{-1} b^{(0,1)}$, where $\zeta$ depends on the linear combination considered (conventions can be
chosen so that $\zeta=1$ for $\omega_J$ and $\zeta=\i$ for $\omega_K$).  At $\zeta=0$, $\J_\zeta$ reduces to the complex structure that is appropriate in the
$B$-model of type $I$, and at $\zeta=\infty$, it reduces to the opposite complex structure, appropriate in the $B$-model of type $-I$.

If $L$ is not compact, in general there is no such relation between  the moduli of $\B$ as a $B$-brane and as an $A$-brane.  Instead of being general, 
let us discuss what happens for the example that has actually been important in our analysis.    This is the case of a brane $\F$ of type $(B,A,A)$ that
is  supported on a leaf $F$ of a holomorphic polarization $\varPi$, with trivial $\CP$ bundle.
Since $F\cong \C^n$, $F$ is simply-connected, and $\F$ actually has no moduli as an $A$-brane.
On the other hand, $\F$ does have moduli as an $B$-brane, since different leaves of $\varPi$ support inequivalent $B$-branes.   Locally, we can parametrize
the choice of leaf by holomorphic variables $q^i$, and parametrize the leaves by holomorphic variables $p_i$, such that $\Omega=\sum_i \d p_i \d q^i$.
Then a particular leaf is determined by specifying the values of the $q^i$, say $q^i=w^i$.   Different choices of the $w^i$ label different $B$-branes,
but, roughly speaking -- we make a more accurate statement shortly -- these branes are all the same as $A$-branes.

The claim that the $A$-brane $\F$ labeled by a given choice of $\vec w=(w^1,w^2,\cdots, w^n)$ is independent of $\vec w$ may seem perplexing, because
the dependence on $\vec w$ has actually been important in our analysis.   The point is that we have always considered the brane $\F$ together with a distinguished
element of $\Hom(\B_\cc,\F)$ associated to the wavefunction $\alpha=(\d\vec p)^{1/2}$.  The brane $\F$ together with this particular element of
$\Hom(\B_\cc,\F)$ does depend nontrivially -- and holomorphically, as we will see -- on $\vec w$.    To make a trivial variation of $\F$, we would want to accompany
a change of $\F$ with a change in the endpoint couplings $c_\ell, c_r$ as given in eqn. (\ref{endvar}).   This is different from using the wavefunction
$\alpha=(\d\vec p)^{1/2}$, independent of $\vec w$.

Let us see how this works out for the case that we vary the brane $\F$ to a nearby leaf of the same holomorphic polarization, for example by shifting $w^i$
by some complex constants $b^i$.   We recall that in the situation relevant to quantization, the complexified  symplectic form is $\omega_\C=-\i\Omega$ 
and that  the $\CP$ bundle of $\F$ is trivial.  The action (see eqns. 
(\ref{genformact}) and (\ref{toldo})) becomes 
\be\label{specformact} I=-\i \int_\Sigma X^*(\Omega) +c_\ell + c_r +\cdots +\{Q,\cdots\},\ee
where we have omitted boundary terms and couplings outside the interval $\partial_0\Sigma$, since these are unaffected by the deformation
that we are about to make.   To describe the change in the brane $\F$ by changing $\vec w$, we vary the map $X$ near $\partial_0\Sigma$ by
$\delta p_i=0$, $\delta q^i=b^i$.   With $\Omega=\sum_i \d p_i \d q^i$, we get 
\be\label{acttrans} \delta I =\i\int_{\partial_0\Sigma} b^i\d p_i =\i b^i p_i(t_r)-\i b^i p_i(t_\ell). \ee 
The compensating endpoint variations to get a truly trivial deformation are then 
\be\label{nactra} \delta c_r=\i b^i p_i(t_r), ~~~\delta c_\ell =-\i b^i p_i(t_\ell). \ee

We can interpret these results as follows,   First of all, this variation of $\F$ is holomorphic,
 since the variation of the action is proportional to $\delta w^i =b^i$ with no term proportional to 
$\delta \bar {w^{ i}}=\bar{b^i}$.   This confirms a claim that was made in Section \ref{branecp}.   Second, the specific variation $\delta c_\ell$, $\delta c_r$
that are needed to compensate for a displacement of $\F$ can be interpreted  intuitively as follows.   The displacement of $\F$ was by $b^i\frac{\partial}{\partial q^i}$,
which generates $q^i\to q^i+b^i$.    Quantum mechanically, $b^i\frac{\partial }{\partial q^i}$ corresponds to $\i b^i p_i$.    The correction to a vertex operator
representing an element of $\Hom(\B',\F)$ or $\Hom(\F,\B')$, for some other brane $\B'$, involves multiplication by $\pm \i b^i p_i$.

How $b^i p_i$ acts on $\Hom(\B',\F)$ or $\Hom(\F,\B')$ depends on the kind of brane considered.   Suppose that $\B'=\B_\cc$.  Then  $\Hom(\B_\cc,\F)$
and $\Hom(\F,\B_\cc)$ are identified with $H^0(F,K_F^{1/2})$.   A general element is of the form $f(\vec p)(\d\vec p)^{1/2}$, and $b^i p_i$ acts
as a multiplication operator.   On the other hand suppose that $\B'$ is a Lagrangian $A$-brane whose support $L$ intersects $F$ at a single point $w$.
Then acting on $\Hom(\F,\B')$ or $\Hom(\B',\F)$, $p_i$ can be replaced by its value at $w$.

Finally, consider what is happening when we vary the leaf of a holomorphic polarization $\varPi$.   Nearby leaves of of $\varPi$ are equivalent if when we change
the leaf by $\delta q^i=b^i$, we transform $\Hom(\F,\B')$ and $\Hom(\B',\F)$ by $\pm \i b^i p_i$, as just explained.   This can be regarded as a flat connection
on a family of branes parametrized by $W$.   But it is an unintegrable flat connection.   To integrate this flat connection, we would have to exponentiate
the operator $\pm \i b^i p_i$ acting at the corners between $\F$ and other branes.   For example, to claim that two fibers with $q^i$ differing by
a noninfinitesimal shift $b^i$ are equivalent, we would have to multiply their corners with $\B_\cc$ by $\exp(\pm \i \sum_i b^i p_i)$.   Such exponential
functions are not allowed; we have allowed  polynomial functions of $p$ only.   The inability to integrate the flat connection is a feature, not a bug.
In our applications, it would not be natural to claim that the different fibers of $\varPi$ are equivalent.

\section{Symmetries and Correspondences}\label{symq}

\subsection{Overview}\label{oversym}

As explained in Section \ref{problem}, it is impossible to quantize a real symplectic manifold $M$, with prequantum line bundle $\frL$, 
in such a way that  the natural symmetry group $\G$ of classical mechanics is realized as a group of unitary transformations of the quantum Hilbert space $\H$.
(We recall that an element of $\G$ is a symplectomorphism $\varphi:M\to M$ together with a lift of $\varphi$ to a symmetry of $\frL$.)
  Given this, we may ask instead the following
question:  for a given method of quantizing $M$, what subgroup of $\G$  is realized quantum mechanically as a group of unitary
symmetries?

In geometric quantization, this question has a standard answer.    Geometric quantization of $M$ depends on the choice of a polarization $\P$ of $M$.
For a given choice of $\P$, the subgroup of $\G$ that is realized quantum mechanically as a symmetry
group   is the subgroup $\G_\P$ that preserves $\P$.   In special cases, which are sometimes important, 
the Hilbert space $\H$ defined in geometric quantization admits a unitary
action of a larger subgroup of $\G$, but the generic answer from geometric quantization is that the subgroup of classical
symmetries that acts in the quantum theory is $\G_\P$.

For example, we can describe $\G_\P$ rather explicitly if $M=T^*N$ is a cotangent bundle with its standard symplectic structure and  $\P$ is
the real polarization whose leaves are the fibers
of the cotangent bundle.   The group $\G_\P$ of symplectomorphisms that leaves $\P$ fixed is generated by two types of symplectomorphism.   First, every diffeomorphism
of $N$ extends to a diffeomorphism of $T^*N$ that preserves the symplectic structure.    These diffeomorphisms permute the leaves of $\P$.  In addition,
there are symplectomorphisms of $T^*N$ that map each leaf to itself.   If $q^i$ are local coordinates on $N$ and $p_i$ are corresponding linear functions
on the fibers of the cotangent bundle, a symplectomorphism of $T^*N$ that maps each leaf to itself  leaves $q^i$ fixed and shifts the $p_j$ by
$p_j\to p_j+\partial f/\partial q^j$
for some function $f(q^1,q^2,\cdots,q^n)$ (called the generating function of the canonical transformation). 

What is the symmetry group in brane quantization?    In brane quantization, the first step is to complexify a real symplectic manifold $M$, with symplectic
form $\omega$, to a complex symplectic manifold $Y$, with holomorphic symplectic form $\Omega$.   $Y$ must be such that if we view $Y$ as a real
symplectic manifold with symplectic form $\Im\,\Omega$, then its $A$-model is well-defined.   (For example, $Y$ could be a complete hyper-Kahler manifold,
as discussed in Section \ref{amo}.)
Once a suitable $Y$ is chosen, one defines a Lagrangian brane $\B$ with $\CP$ bundle $\L$, and the quantum Hilbert space of brane quantization
is $\H=\Hom(\B,\B_\cc)$. 

There is an obvious candidate answer for the symmetry group of brane quantization.   The natural symmetries of the $A$-model of $Y$ 
are holomorphic symplectomorphisms, that is, invertible holomorphic maps of $Y$ to itself that preserve the symplectic form $\Omega$.
A holomorphic symplectomorphism of $Y$ is potentially a symmetry of the brane $\B$ if it maps $M$ to itself, and its action on $M$ lifts to an action on $\frL$.
This gives us a natural candidate for the symmetries of brane quantization.  Let $\G_Y$ be the subgroup of $\G$ corresponding to elements whose
action on $M$ can be analytically continued to a holomorphic symplectomorphism of $Y$.   Then $\G_Y$ is the natural candidate for the subgroup of $\G$
that acts unitarily on $\H=\Hom(\B,\B_\cc)$.

Although this proposal may be correct, there is actually a technical difficulty in justifying it.  To make the $A$-model concrete, one usually picks a Riemannian metric
$g$ on $Y$.   This may be a complete hyper-Kahler metric, as discussed in Section \ref{amo}.   At a minimum, $g$ should, together with the real symplectic
form $\omega_K=\Im\,\Omega$, define an almost complex structure $K$ such that $\omega_K$ is of type $(1,1)$ and positive.  Moreover, $g$ should
be such that the $A$-model of $Y$ actually exists.   The $A$-model of $Y$ is independent of $g$ as long as
 $g$ varies within a class of allowed metrics, such that the
$A$-model exists.   Unfortunately, this last condition is difficult and its full import is not easy to understand.

Now let $\varphi$ be a holomorphic symplectomorphism of $Y$.   Generically $g$ will not be $\varphi$-invariant, but will be transformed by $\varphi$ to some
other metric $g^\varphi$.   The $A$-model of $Y$ with the metric $g^\varphi$ may in general not be well-defined.   Even if it is, to argue that
the metrics $g$ and $g^\varphi$ lead to equivalent $A$-models of $Y$, we would need to know that it is possible to interpolate between $g$ and $g^\varphi$
within the class of allowed metrics.  It is always possible to interpolate between $g$ and $g^\varphi$ by a family of metrics that are associated
to suitable almost complex structures.\footnote{The argument for this is similar to an argument given shortly concerning hyper-Kahler metrics.   The symplectic
structure $\omega_Y$ of $Y$ reduces the structure group of the tangent bundle of $Y$ from $\GL(4n,\R)$ to $\Sp(4n,\R)$.    A metric relative
to which $\omega_Y$ is positive and of type $(1,1)$ corresponds to a further reduction of the structure group to a maximal compact subgroup $\UU(2n)$ of $\Sp(4n,\R)$.
It is possible to interpolate between any two such reductions, since the quotient $\Sp(4n,\R)/\UU(2n)$ is   contractible.}   
But it is not obvious that this interpolation can be made within the class of metrics for which the $A$-model exists.  

Physically, one prefers to view the $A$-model as a twisted version of an ultraviolet-complete $\sigma$-model.   Trying to consider an $A$-model based
on a metric such as $g^\varphi$ would in general take us out of that world, and arguments based on interpolation from $g$ to $g^\varphi$ might take us even farther afield.

To further orient the reader, we will describe some concrete questions for which this issue is or is not problematic.   Then we will describe an alternative approach
that enables one to study symplectomorphisms within an ultraviolet-complete world.

We will illustrate both sides
of the story in the context of quantization of $M=\R^{2n}$
with real-valued coordinates $\vec x = (x^1,x^2,\cdots , x^{2n})$ and 
 a standard symplectic structure $\omega=\frac{1}{2}\sum_{i,j=1}^{2n} \omega_{ij} \d x^i \d x^j$.  First, we will try to interpret in terms of brane quantization the
 standard result\footnote{This also follows from arguments in Section \ref{comparing} showing that brane quantization of $\R^{2n}$, which does not depend on
 a choice of polarization, is equivalent to geometric
 quantization of $\R^{2n}$ with a real linear polarization. } that quantization of $M$ can be chosen to be invariant under the group of affine linear symplectomorphisms $\vec x\to A\vec x+\vec b$,
 as discussed in Section \ref{problem}.    We begin in the familiar way, complexifying $M$ to a complex symplectic manifold $Y=\C^{2n}$ with the $x^i$ now as holomorphic coordinates on $Y$.  The  resulting complex structure and complex symplectic structure of $Y$ are of course
 translation-invariant, that is invariant under $\vec x\to \vec x+\vec b$ with a complex constant $\vec b$.  
The translation-invariant complex symplectic structure of $Y$ can be extended, in the sense of Section \ref{amo}, to a translation-invariant hyper-Kahler structure,
by picking a suitable translation-invariant Riemannian metric $g$ on $Y$.   The space of such $g$'s can be described as follows.  Because of the assumed
translation invariance, we can make the analysis at a point  $y\in Y$.   The tangent space to $y$ in $Y$ has dimension $4n$, so the group of linear transformations
of this tangent space is a copy of $\GL(4n,\R)$.   The complex symplectic structure of $Y$ is invariant under a subgroup $\Sp(2n,\C)\subset \GL(4n,\R)$.
A  translation-invariant hyper-Kahler metric on $Y$ that extends its complex symplectic structure
is invariant not under  $\Sp(2n,\C)$ but under a maximal compact subgroup thereof.
Such a subgroup is isomorphic to the compact real form of the symplectic group, which we will call $\Sp_c(2n)$.
Any $\Sp_c(2n)$ subgroup of $\Sp(2n,\C)$ is the subgroup preserved by some hyper-Kahler metric, and all such subgroups are conjugate in $\Sp(2n,\C)$.
   The space of translation-invariant hyper-Kahler metrics on $Y$ that extend its complex
symplectic structure is hence a copy of the homogeneous space $\sW=\Sp(2n,\C)/\Sp_c(2n)$.    Any of the  hyper-Kahler metrics parametrized by $\sW$
can be used to define the $A$-model of $Y$.     $\sW$ is contractible and in particular is connected and simply-connected.
So we can interpolate in $\sW$ between any two points $w_1,w_2\in \sW$, and moreover this interpolation can be made in a unique way, up to homotopy.
(Uniqueness is important because homotopically distinct interpolations from $w_1$ to $w_2$ might have led to  equivalences between quantum theories
that would differ by a unitary transformation.)
Therefore, if we define the $A$-model of $Y$ using a translation-invariant hyper-Kahler
metric, it will not matter which one we use.

Now we can understand in a new way why the brane quantization of $M$ is invariant under affine linear tranformations of $M$.
An affine linear symplectomorphism $\vec x\to A\vec x +\vec b$ of $M$ can be analytically continued to a holomorphic symplectomorphism
of $Y$, defined by the same formula.   A generic affine linear symplectomorphism will not preserve a translation-invariant hyper-Kahler metric
that is used to define the $A$-model of $Y$.   But it will transform this translation-invariant hyper-Kahler metric to another translation-invariant
hyper-Kahler metric, and the $A$-model of $Y$ does not care which translation-invariant hyper-Kahler metric we use.
So quantization of $M$ via the $A$-model is invariant under affine linear symplectomorphisms; the group of affine linear symplectomoprhisms acts
on the quantum Hilbert space $\H$.

To see the limitation of  this kind of argument, we  note the following.   The group of symplectomorphisms of $M=\R^{2n}$ that extend to holomorphic
symplectomorphisms of $Y$ is much larger than the group of affine linear symplectomorphisms.     In this discussion, we will only consider polynomial
symplectomorphisms, since we have only allowed polynomial functions in the $A$-model.    To construct a very large group of polynomial
symplectomorphisms, first split  the variables $x^1,\dots, x^{2n}$ into Poisson-commuting
coordinates $q^1,\cdots , q^n$ and conjugate momenta $p_1,\cdots,p_n$.  For any polynomial function $f(q^1,\cdots, q^n)$, consider the symplectomorphism that
maps $p_i,q^j$ to
\begin{align}\label{worfo} p'_i & = p_i+\frac{\partial f}{\partial q^i} \cr q^i{}'&=q^i. \end{align}
Symplectomorphisms of this kind together with linear symplectomorphisms\footnote{\label{enough}
It is enough to consider here linear symplectomorphisms $\vec x\to A\vec x$
rather than affine linear ones $\vec x\to A\vec x+\vec b$, because constant shifts of the $p$'s are already included in eqn. (\ref{worfo}), and once
we include linear symplectomorphisms (which can exchange the $p$'s and $q$'s), constant shifts of the $q$'s are included as well.}
generate a group that is known 
as the group of tame symplectomorphisms of $\R^{2n}$.   We will denote it as $\TAut(\R^{2n})$.   
To see how enormous  is $\TAut(\R^{2n})$, imagine trying to associate a Lie algebra to this group.  (This is technically problematic in this
infinite-dimensional situation.)
Such a Lie algebra would have to include all polynomial functions of the $p$'s and $q$'s as generators. Indeed, the transformations in eqn. (\ref{worfo})
have generators $f(q^1,\cdots, q^n)$.   Since $\TAut(\R^{2n})$ includes linear symplectomorphisms, which in particular can exchange the $p$'s and $q$'s,
a hypothetical Lie algebra that includes  arbitrary polynomial functions of the $q$'s as generators also has to include arbitrary polynomial functions of the $p$'s.
But then to get a class of functions that is closed under Poisson brackets, we have to allow arbitrary polynomials in the $p$'s and $q$'s.
   So $\TAut(\R^{2n})$ is big enough that if it could be treated as a Lie group, its Lie algebra would include all the polynomials
in the canonical variables.

   For $n=1$, it is known that $\TAut(\R^{2n})$ coincides
with the full group $\Aut(\R^{2n})$ of polynomial symplectomorphisms of $\R^{2n}$.    It is not known whether this is true for $n>1$.   (Note that at the Lie algebra
level, one would not see a distinction between $\TAut(\R^{2n})$ and $\Aut(\R^{2n})$, since a Lie algebra of $\TAut(\R^{2n})$ would have to have arbitrary polynomial
generators, as was just explained.)

In deformation quantization, instead of $\Aut(\R^{2n})$, we can consider the group $\Aut(\C^{2n})$ of all holomorphic polynomial symplectomorphisms of
$\C^{2n}$.  Likewise, instead of $\TAut(\R^{2n})$, we can consider the analogous tame subgroup $\TAut(\C^{2n})$ 
generated by $\vec x\to A\vec x+\vec b$ along with the symplectomorphisms in eqn. (\ref{worfo}), where now $A$, $\vec b$, and $f$ can all be complex.
These groups are symmetries of the classical ring $\A_0$ of polynomial functions on $\C^{2n}$, along with its Poisson bracket.
Deformation quantization replaces $\A_0$ with a noncommutative ring $\A$, sometimes called the Weyl algebra (defined by generators $x_i$ with
relations $[x_i,x_j]=\i\hbar \Omega_{ij}$).  
A quantum analog of the classical symmetry groups $\TAut(\C^{2n})$ and $\Aut(\C^{2n})$
 is the group of automorphisms of $\A$.  How are $\TAut(\C^{2n})$ and $\Aut(\C^{2n})$ related to their quantum analogs?   
   
Kanel-Belov and Kontsevich \cite{BKK} established the striking result that $\TAut(\C^{2n})$ is a subgroup of the automorphism group of the Weyl algebra,
and conjectured that the possibly larger group $\Aut(\C^{2n})$ is the automorphism group of the Weyl algebra.   This has been subsequently proved by
Kanel-Belov, Elishev, and Yu \cite{BEY}.      (These proofs make heavy use of reduction modulo a prime $p$; the Weyl algebra has quite unusual properties
after such reduction.)    As will be explained at the end of Section \ref{symmeq}, the results of \cite{BKK,BEY} have immediate analogs for quantization of $\R^{2n}$
(as opposed to deformation quantization of $\C^{2n}$).   In particular, a central extension of $\Aut(\R^{2n})$ acts as a group of unitary symmetries
of the Hilbert space obtained by quantization $\R^{2n}$.  There is no quantum deformation.
 On the one hand, we might expect this, because this central extension is the group $\G_Y$ introduced
at the outset of this discussion.   On the other hand, the result is remarkable because nothing similar is true at the Lie algebra level.
As explained earlier, a Lie algebra of $\Aut(\R^{2n})$ or even $\TAut(\R^{2n})$ would have to contain all the polynomial functions on $\R^{2n}$.
Because of the usual anomaly  in the passage from classical mechanics to quantum mechanics,  discussed in Section \ref{problem} and Appendix \ref{anomaly},  
there is no way for such a Lie algebra to act on the quantum Hilbert space without quantum deformation.

It is reasonable to aim to use brane quantization to recover some of the results of \cite{BKK,BEY}.  
 However, a general polynomial symplectomorphism $\varphi$ of $Y$ will
transform a translation-invariant metric $g$ on $Y$ into a metric $g^\varphi$ that is far from being translation-invariant.   In general,
$g^\varphi$ may not lead to an ultraviolet-complete $\sigma$-model, and even if it does
 we will not
know how  to interpolate between $g$ and $g^\varphi$ in a class of metrics on $Y$ that lead to such models.     So to recover
from brane quantization some of the results of \cite{BKK,BEY} will require a different approach.   In Section \ref{corresp}, we will introduce an approach
that will enable us to study general polynomial symplectomorphisms while remaining in the ultraviolet-complete world.    Another advantage of this approach
is that it provides a good framework to study operators that are more general than symplectomorphisms.

\subsection{Interfaces And Operators}\label{corresp}

A symplectomorphism $\varphi$ of  a symplectic manifold $Y$ is an invertible map $\varphi:Y\to Y$ that preserves the symplectic structure $\omega$ of $Y$,
in the sense that $\varphi^*(\omega)=\omega$.   Concretely, if $x^i$ are local coordinates on $Y$, then $\varphi$ can be defined by expressing
$x'{}^{i}=\varphi^*(x^i)$ as functions of the $x^k$:
\be\label{expfn} x'{}^{i}=f^i(x^k) . \ee
Saying that $\varphi$ is invertible means that these relations can be inverted,
\be\label{nexpfn} x^i=k^i(x'{}^{k})\ee
The condition that $\varphi$ is a symplectomorphism (and not just a diffeomorphism) of $Y$ means that $\varphi^*(\omega)=\omega$ or in more
detail
\be\label{nexo}\sum_{i,j}\omega_{ij}(x')\d x'{}^{i}\d x'{}^{j}=\sum_{i,j}\omega_{ij}(x)\d x^i \d x^j. \ee
We can of course impose further conditions.  If $Y$ is a complex manifold and $\varphi$ is a holomorphic mapping, then we can choose the $x^i$ to be
local holomorphic coordinates and then the functions $h^i$, $k^i$ are holomorphic.   If there is a distinguished class of polynomial functions on $Y$, then
$\varphi$ is a polynomial mapping if the functions $h^i$ and $k^i$ are polynomials.

The condition (\ref{nexo}) for a symplectomorphism has an interesting interpretation.   Consider the product of two copies of $Y$, say $Y\times Y'$,
with the $x^i$ viewed as coordinates on $Y$ and the $x^{i'}$ viewed as coordinates on $Y'$.      The conditions $x'{}^{i}=f^i(x^k)$ or equivalently
$x^i=k^i(x'{}^{k})$ define a middle-dimensional submanifold $L\subset Y\times Y'$.   Consider $Y\times Y'$ as a symplectic manifold with the symplectic
structure $\omega\boxplus (-\omega)$, that is, the symplectic form of $Y\times Y'$ is the sum of $\omega$ on the first factor and $-\omega$ on the second.\footnote{\label{boxes} In more detail,
if $\pi:\h Y\to Y_1$ and $\pi':\h Y\to Y'$ are the projections to the two factors of $\h Y=Y\times Y'$, and $\alpha,\alpha'$
are two-forms on  $Y$ and $Y'$, one defines
$\alpha\boxplus \alpha'=\pi^*(\alpha)+{\pi'}^*(\alpha')$.   Similarly, if $\RR,\RR'$ are line bundles over the two factors,
one defines $\RR\boxtimes \RR'=\pi^*(\RR)\otimes \pi'^*(\RR')$.   This is the line bundle over $\h Y$ that is $\RR$ on the first factor
and $\RR'$ on the second.} 
Eqn. (\ref{nexo}) means that $L$ is a Lagrangian submanifold of $Y\times Y'$ for this real symplectic structure.

In general, consider a product $Y\times Y'$ of two symplectic manifolds $Y,Y'$.    A Lagrangian submanifold $L\subset Y\times Y'$ is called a Lagrangian
correspondence between $Y$ and $Y'$.   A Lagrangian correspondence $L$ that is associated, as above, to a symplectomorphism between $Y$ and $Y'$
is a very special case in which, if $y_0,y_0'$ are points in $Y$ and $Y'$, respectively, the intersection of $L$ with $y_0\times Y'$ or $Y\times y_0'$ is a single point.

\begin{figure}
 \begin{center}
   \includegraphics[width=3.5in]{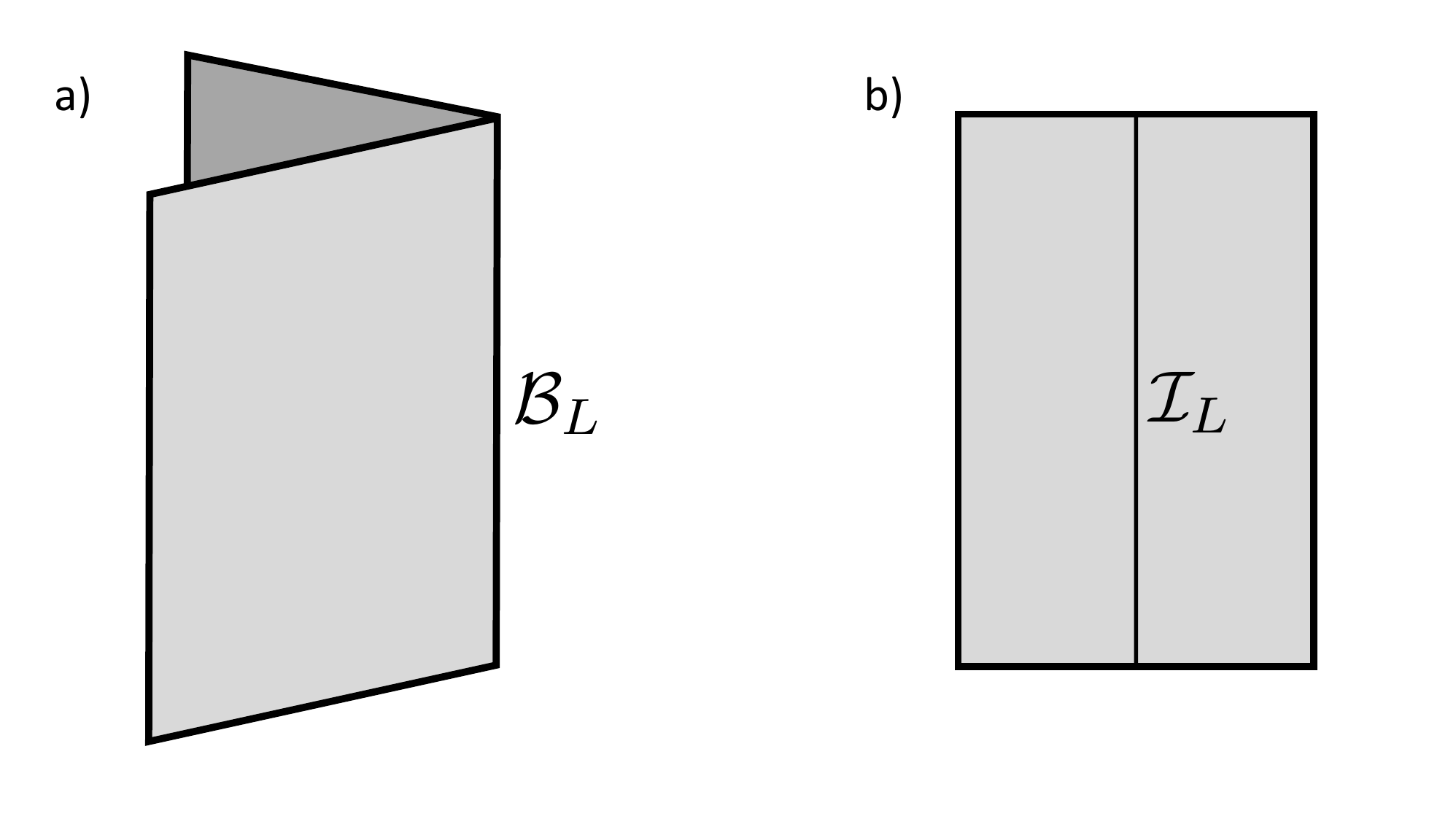}
 \end{center}
\caption{\small  The $\sigma$-model of a product $Y\times Y'$ can be understood as a quantum field on a two-manifold $\Sigma$ that has two identical sheets.
One sheet is mapped to $Y$ and one to $Y'$.
(a) We consider a case that the two sheets are coupled only via a boundary condition associated to a brane $\B_L$.  We depict this coupling
by gluing the sheets together along the boundary.    (b) After ``unfolding,'' the brane
becomes an interface or domain wall $\I_L$ between $\sigma$-models with targets $Y$ (the left part of the figure) and $Y'$ (the right part).   $\I_L$ is supported
on the central vertical line $\ell$. 
  \label{unfolding}}
\end{figure}

If $L$ is a Lagrangian correspondence in $Y\times Y'$, then upon picking a flat $\spinc$ structure on $L$, we can define a Lagrangian $A$-brane $\B_L$
in the $A$-model of $Y\times Y'$.   In the $\sigma$-model of maps  $\Phi:\Sigma\to Y\times Y'$, where $\Sigma$ is a two-manifold,
 the role of the brane $\B_L$ is to define a possible boundary condition along a portion $\partial_0\Sigma$ of
the boundary of $\Sigma$.
Since $Y\times Y'$ is a simple product, we can think of $\Sigma$ as having two sheets, of which one is mapped to $Y$ and one to $Y'$; the two sheets
are coupled only along the boundary of $\Sigma$.   The boundary condition means that $\Phi$ maps $\partial_0\Sigma$ to $L\subset Y\times Y'$ (with a boundary
coupling that depends on the $\spinc$ structure of $L$).
This construction has a useful ``unfolded'' version  (fig. \ref{unfolding}).   For this, we ``flip over'' or unfold one sheet of $\Sigma$  along $\partial_0\Sigma$.
After unfolding, $\Sigma$ is replaced by a two-manifold $\h\Sigma$ in which $\partial_0\Sigma$ is no longer a boundary but an internal line $\ell$.   To the
``left'' of $\ell$, $\Sigma$ is mapped to $Y$ and to the right it is mapped to $Y'$.   Thus, in the unfolded picture, the brane $\B_L$ is replaced by a domain wall
or ``interface''  $\I_L$ between the $\sigma$-model of $Y$ and the $\sigma$-model of $Y'$.

So far these are general remarks about two-dimensional field theories.  Now let us specialize to the case that $Y'$ is a second copy of $Y$,
and we are studying an $A$-model in which the symplectic form on $Y'$ has opposite sign to that on $Y$.   ``Unfolding'' reverses the orientation of one sheet of
$\Sigma$, and in the $A$-model, reversing the worldsheet orientation is equivalent to reversing the sign of the symplectic form.   Since we started
with opposite symplectic structures on the two copies of $Y$, it follows that after unfolding, the symplectic structure is the same everywhere and
we simply have the ordinary $A$-model of a single copy of $Y$.   What in the previous picture was an $A$-brane $\B_L$ in the $A$-model of $Y\times Y'$
becomes after unfolding a topological domain wall or interface $\I_L$ in the $A$-model of $Y$.   Calling this interface ``topological'' means that it can be freely
moved around in $\Sigma$ without affecting $A$-model observables.\footnote{In particular, we could move a topological interface up to a boundary,
whereupon it would combine with an existing boundary condition to make a new one.   This defines what in mathematical language could be called a functor
on the category of boundary conditions.  This is important in some applications (including geometric Langlands), 
but here we will use topological interfaces to define ordinary quantum operators.}

\begin{figure}
 \begin{center}
   \includegraphics[width=4.5in]{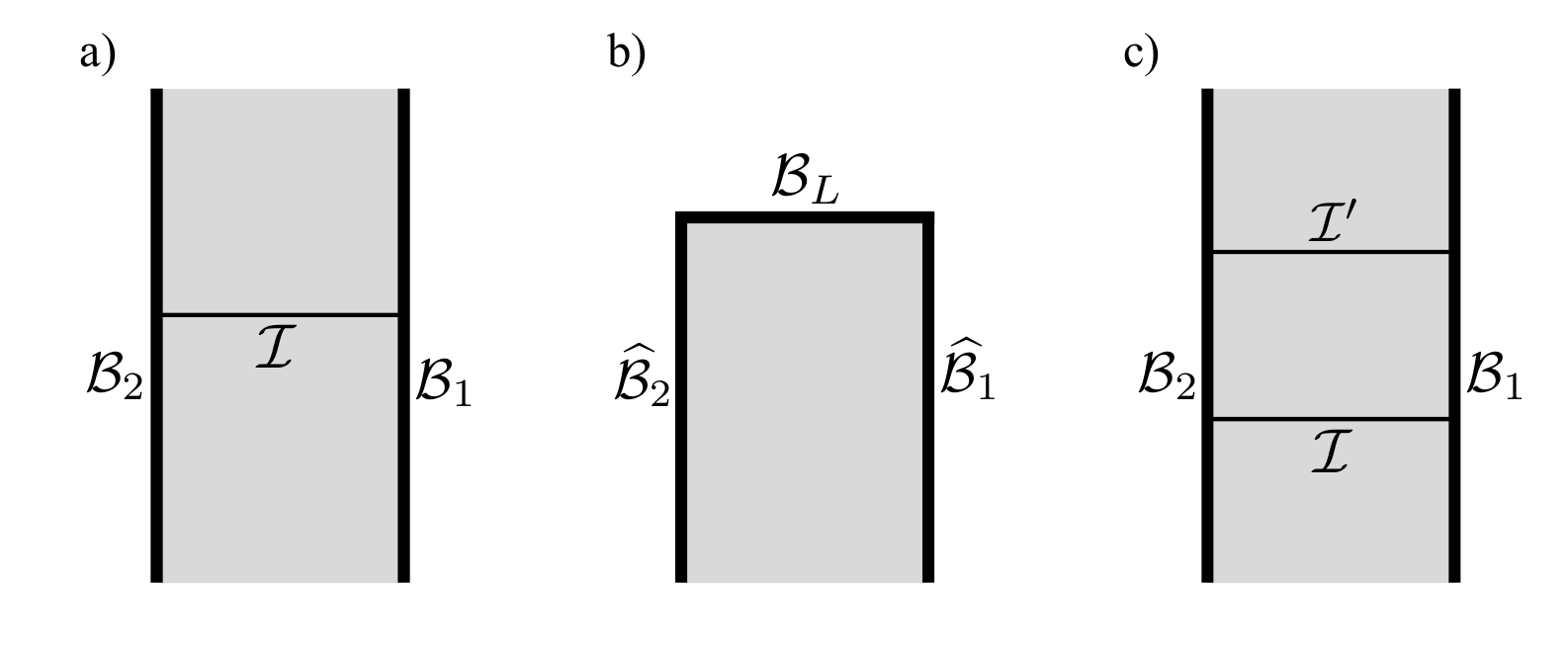}
 \end{center}
\caption{\small  (a) A topological interface $\I$ used to define an operator acting on an $A$-model physical state space $\Hom(\B_1,\B_2)$.   On the  right and left
are branes $\B_1$ and $\B_2$; the horizontal line represents the interface $\I$.   Additional information must be provided at the ``endpoints''
where the horizontal line in the figure meets the boundaries. A physical state enters at the bottom of the strip, interacts with the
interface, and emerges at the top.  If one reads the picture from top to bottom,  it describes the dual or transpose operator on $\Hom(\B_2,\B_1)$.   (b) This picture is obtained from (a) by ``folding'' along the curve $\ell$ that supports the interface $\I_L$.   Folding replaces $\ell$ by
a boundary that is now labeled by a brane $\B_L$.   On the  right and left, $\h \B_1$ and $\h\B_2$ are folded or doubled versions of the branes $\B_1$ and $\B_2$.
The description in the text began with this folded picture, which was then unfolded to get the previous one.
 (c) Composition of operators comes from composition of the associated topological interfaces $\I$ and $\I'$ along with the corresponding endpoints.
 In topological field theory, one can fuse the two interfaces  and use the analog of the operator product expansion for local
operators to combine their endpoints.
  \label{correspondence}}
\end{figure}

Topological interfaces or domain walls can be used to define operators acting on physical state spaces of the $A$-model.   The basic idea is illustrated in fig.
\ref{correspondence}(a).    The $A$-model on a strip $\R\times I$ (where $I$ is a unit interval) with boundary conditions associated to branes
$\B_1,\B_2$ describes $\Hom(\B_1,\B_2)$.    Now we add a topological interface $\I$ running horizontally across the strip.   It is necessary to provide some information
to explain what happens where the interface meets the left or right boundaries of the strip.    Assuming that this has been done in a satisfactory way, the
path integral on the strip defines an operator $\O$ acting on $\Hom(\B_1,\B_2)$.   Reading the picture from bottom to top, a state in $\Hom(\B_1,\B_2)$ enters
at the bottom and interacts with the interface; then a possibly different state in $\Hom(\B_1,\B_2)$ emerges at the top.  (Read  from top to bottom,
the same picture describes the dual or transpose operator acting on the dual vector space $\Hom(\B_2,\B_1)$.)

We will refer to the point where $\I$ ends on the left or right boundary of the strip as an ``endpoint'' of $\I$ with $\B_1$ or $\B_2$.   To understand
the endpoints of interfaces, 
 it is actually convenient to go back to the folded picture.
We do this by folding the worldsheet of fig. \ref{correspondence}(a) along the interface, so that the locus of the interface becomes a boundary (as it was originally
in fig. \ref{unfolding}(a)).    In the folded picture, sketched in fig. \ref{correspondence}(b), the interface $\I$ comes from a brane $\B_L$ in the $A$-model of $Y\times Y'$
and likewise the left and right boundaries are labeled by branes $\h \B_1$ and $\h\B_2$, the folded versions of $\B_1$ and $\B_2$.
   The endpoint of $\I$ on $\B_1$ or $\B_2$ is simply a $(\B_L,\h\B_1)$ or $(\h\B_2,\B_L)$ corner, as studied in Section
\ref{ccb} and fig. \ref{strip}.   In other words, these corners are elements of $\Hom(\B_L,\h\B_1)$ and $\Hom(\h\B_2,\B_L)$.     They are the sort of
thing that we have discussed at length in Section \ref{comparing} and Appendix \ref{details}.

Once corners are chosen, a Lagrangian correspondence defines an operator acting on the physical states of the $A$-model.  In particular, Lagrangian correspondences
can be used to define operators in brane quantization of a real symplectic submanifold $M\subset Y$.  As noted in Section \ref{ques}, this is particularly important
if $Y$ is such that the algebra $\A_0$ of holomorphic functions on $Y$ is relatively ``small,'' too small to effectively characterize the quantization of $M$.  Operators
derived from correspondences are important in such a case.

\subsection{Composing Interfaces}\label{composing}

As explained in Section \ref{problem},  there is no general recipe for quantization, because
quantization does not transform the classical algebra of functions to
a quantum algebra of operators in any simple way.   The usual claim that Poisson brackets are mapped to commutators is only valid to first order in $\hbar$
or for particular classes of functions and operators.   Any known recipe for quantization singles out some class of functions that behave well, either by the
choice of a polarization or (in brane quantization) by the choice of a suitable complexification of the classical phase space.

Defining operators via correspondences rather than functions does not in general avoid this basic difficulty.  
It merely shifts the problem from composing functions at the quantum level to composing interfaces and the associated endpoints.

 In the $A$-model, there certainly is a composition
law for topological interfaces, analogous to the operator product expansion for local operators.   If $\I'$ and $\I$ are two topological interfaces, then bringing
them together gives a new topological interface $\I''=\I'\cdot \I$.    But like the operator product expansion for local operators, the composition law of
topological interfaces cannot just be computed from classical formulas; it is subject to quantum corrections.  

If interfaces $\I$ and $\I'$ are used (together with suitable endpoints) to define operators $ \O_\I$, $\O_{\I'}$, then the composition 
$\O_{\I'}\O_\I$ of the operators can be computed (fig. \ref{correspondence}(c)) by composing the interfaces and the endpoints.   In general, we will run into
quantum corrections both in composing the interfaces and in composing the endpoints.   To get useful results, one needs some way to get these
compositions under control.

In our discussion so far, we have only required $L\subset Y\times Y'$ to be a Lagrangian correspondence for the real symplectic form of the
$A$-model of $Y\times Y'$.     However, a much more specific situation is natural for problems of quantizing a classical phase space.   
Suppose that $Y$ is in the usual sense
a complexification of a real symplectic manifold $M$ that we wish to quantize.    Then to construct an operator in the quantization of $M$, it is natural
to start with a real Lagrangian correspondence $L_r\subset M\times M'$, where similarly to the discussion 
in Section \ref{corresp}, $M'$ is a second copy of $M$ with opposite symplectic structure.  Then we define $L\subset Y\times Y'$ to be the complexification of $L_r$. 
A generic $L_r$ will not have a nice complexification, meaning a complexification $L$
 whose global properties are good enough that rank 1 $A$-branes supported on $L$ exist.
Just as a generic smooth or even real-analytic function on $M$ does not analytically continue to an everywhere-defined holomorphic function on $Y$,
a generic Lagrangian correspondence in $M\times M'$ does not have a nice analytic continuation in $Y\times Y'$.      The correspondences $L_r\subset M\times M'$ that
can be  quantized in brane quantization to give quantum operators are the ones that do have nice analytic continuations in $Y\times Y'$.

A submanifold of $Y\times Y'$
 obtained as the complexification of a Lagrangian correspondence in $M\times M'$ is not just Lagrangian for the real symplectic form of $Y\times Y'$; it
is Lagrangian for the complex symplectic form.   Hence a rank 1 $A$-brane $\B_L$ supported on $L$ is actually
 a brane of type $(B,A,A)$, that is, it is a $B$-brane in the complex structure of $Y$
and an $A$-brane for either the real or imaginary part of the holomorphic symplectic form $\Omega$ of $Y$.

This simplifies the problem of learning how to compose topological interfaces in the $A$-model of $\omega_K=\Im\,\Omega$.   Although
in general there are quantum corrections to the composition of $A$-model interfaces, there are no quantum corrections to the composition of $B$-model
interfaces.   Those compositions can be computed classically.

So as long as we consider branes of type $(B,A,A)$ whose moduli as $B$-branes are the same as their moduli as $A$-branes, the corresponding interfaces can be
composed classically.   However, there is a subtle point here.    
A rank 1 $(B,A,A)$-brane $\B_L$ supported on $L$ has a $\CP$ bundle $\L$ that is a flat complex line bundle
(more precisely a flat $\spinc$ bundle, but this distinction will not be important here).   
To deform $\B_L$ as a $B$-brane, we  deform $\L$ as a holomorphic line bundle, while to deform it as an $A$-brane,
we deform $\L$ as a flat line bundle.   If $L$ is compact, the possible deformations of a flat line bundle $\L\to L$ as a holomorphic line bundle are the same
as its possible deformations as a flat line bundle.   Therefore the $A$-model and $B$-model moduli spaces of $\B_L$ are the same (though this
moduli space carries different complex structures in the $A$- and $B$-models).   Interfaces associated to such $L$'s have the property that composing them
in the $B$-model is the same as composing them in the $A$-model.

In practice, we are usually interested in interfaces associated to non-compact $L$'s.   For example, if $L$ is
the complexification of a real Lagrangian correspondence $L_r\subset M\times M$, it will typically not be compact.      If $L$ is not compact, then in comparing
the moduli of $\B_L$ as a $B$-brane to its moduli as an $A$-brane, it is important to ask if the first Betti number $b_1(L)$ is positive.   If so, $L$ may admit
flat line bundles that are trivial as holomorphic line bundles.   In that situation, $\B_L$ has moduli as an $A$-brane that are invisible in the $B$-model, and
therefore knowing how to compose interfaces of type $(B,A,A)$ in the $B$-model does not immediately tell us how to compose them in the $A$-model.   
We can avoid this situation by assuming that $b_1(L)=0$, in which case $\B_L$ has no $A$-model moduli that are not detected in the $B$-model.   We will
generally do that in what follows.

\subsection{Automorphisms}\label{symmeq}

We return to the theme of  Section \ref{corresp}.   $\I$ is an interface in the $A$-model of $Y$ associated to a Lagrangian correspondence $L\subset Y\times Y'$,
and we want to use $\I$ to define an operator
acting on  on $\H=\Hom(\B,\B_\cc)$.    For this, we need ``endpoints'' at which $\I$ can end on the two branes $\B_\cc$ and $\B$.   In this section, we focus
on $\B_\cc$ and the relation to deformation quantization.  Quantization is discussed in Section \ref{qtz}.

 \begin{figure}
 \begin{center}
   \includegraphics[width=1.5in]{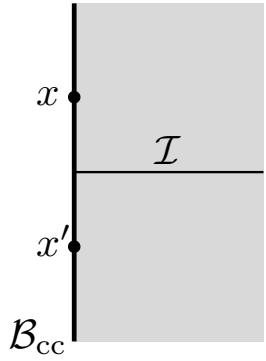}\caption{\small{ Operators inserted above or below the endpoint of an interface $\I$ on the brane $\B_\cc$
   provide, respectively, a left or right action of the algebra $\A$ of boundary operators on the space of endpoints between $\I$ and $\B_\cc$.}  \label{Pictorial}}
 \end{center}
\end{figure}
An endpoint at which $\I$ ends on $\B_\cc$ is interpreted, in the folded picture, as an element of $\Hom(\h\B_\cc,\B_L)$, where $\B_L$ is a Lagrangian
brane supported on $L$, and $\h\B_\cc$ is a product of coisotropic branes on the two factors of $Y\times Y'$.  
 $\Hom(\h\B_\cc,\h\B_\cc)$ is a product of two copies of $\A=\Hom(\B_\cc,\B_\cc)$, 
one for each factor in $Y\times Y'$.
Those two copies act on the space of corners $\Hom(\h\B_\cc,\B_L)$.   So after unfolding, two copies of $\A$ act
on the space of
endpoints between $\I$ and $\B_\cc$.    This statement has a simple pictorial interpretation 
(fig. \ref{Pictorial}); 
an element $ x\in \A$ can be inserted on the boundary of a two-manifold 
 ``below'' or ``above''  a given endpoint.      The insertions below the interface correspond to a right action of $\A$ 
on the space of interfaces, and the insertions above the interface correspond to a left action of $\A$.   We will denote the interface $\I$ equipped
with an endpoint $\alpha$ as $\I(\alpha)$.   For the left or right action of $x\in \A$ on $\I(\alpha)$, we write $\I(\alpha)\to x\I(\alpha)$, or $\I(\alpha)\to \I(\alpha)x'$.   
But as $x$ only acts on $\alpha$, we can also more simply write $ \alpha\to x\alpha$ or $\alpha\to \alpha x'$.

The action of two copies of $\A$ act on the space $\Hom(\h\B_\cc,\B_L)$ of corners can be seen more explicitly as follows.   
From Appendix \ref{details}, the leading approximation to $\Hom(\h\B_\cc,\B_L)$ is
$H^0(L,\L\otimes K_L^{1/2})$, where $\L$ is the $\CP$ bundle of $\B_L$ and for simplicity we  
assume the higher cohomology to vanish.
  Since $L$ is embedded in $Y\times Y'$, one
can multiply an element of $H^0(L,\L\otimes K_L^{1/2})$
 by either a function on $Y$ or a function on $Y'$.  After quantum deformation, this gives the action of two copies of $\A$.   (``Folding'' maps an algebra to the opposite algebra, in which operators are multiplied in the opposite order; this is why one copy of $\A$
acts by a left action and one by a right action.)   

As an important special case, we can consider an interface $\I$ that generates an automorphism of the algebra $\A$.   This means that there is a corner 
$\alpha\in \Hom(\h\B_\cc,\B_L)$ such that every corner can be written  as $x\alpha$ for a unique $x\in \A$, and also as $\alpha x'$ for a unique $x'$ in
$\A$.    This condition  determines an invertible map $\h\varphi$ from $\A$ to itself: for $x\in \A$, $\h\varphi(x)$ is the element of $\A$ such that
$x\alpha= \alpha \h\varphi(x)$.   Moreover, $\h\varphi$ is invertible; for $x\in \A$, $\h\varphi^{-1}(x)$ is the element of $\A$ such that
$\h\varphi^{-1}(x)\alpha =\alpha x$.   Finally, $\h\varphi$ is an automorphism of $\A$, since for $x_1,x_2\in\A$, $x_1x_2\alpha=\alpha\h\varphi(x_1 x_2)$
but on the other hand $x_1 x_2\alpha=x_1\alpha \h\varphi(x_2)=\alpha\h\varphi(x_1)\h\varphi(x_2)$, so
$ \h\varphi(x_1 x_2)=\h\varphi(x_1)\h\varphi(x_2)$.

Complex constants are in the center of $\A$ and also commute with interfaces, so in particular if $\lambda$ is a nonzero complex constant, then
$\lambda\alpha$ generates the same automorphism as $\alpha$.   More generally, suppose that $x_0$ is any invertible element of $\A$.
Then for $x\in \A$, we have $x (x_0\alpha)=x_0 (x_0^{-1} x x_0)\alpha=x_0\alpha \h\varphi (x_0^{-1} x x_0)$, showing that $x_0\alpha$ generates
the automorphism $x\to \h\varphi(x_0^{-1} x x_0)$.    Conversely, if $x_0\alpha$ generates an automorphism, then the corner $\alpha $ is
equal to $x x_0\alpha$ for some $x\in \A$, implying that $x_0$ has the inverse $x$.    Thus if there is any corner $\alpha$ by means of which a given
interface generates an automorphism, then the corners that lead to automorphisms are precisely the
ones of the form $x_0\alpha$, where $x_0\in \A$ is invertible.

In studying automorphisms associated to interfaces, there is consequently some simplification if $Y$ is such that constants are the only invertible (polynomial)
holomorphic functions on $Y$.   This is actually automatic if we assume that $b_1(Y)=0$, an assumption that will be convenient at several points
in the following.\footnote{Suppose that the holomorphic function $f$ on $Y$ is invertible, meaning that it nowhere vanishes.   Then the
closed 1-form $\lambda=\d \log f$ is globally-defined.   If $\lambda$ is exact, $\lambda=\d u$ for a globally-defined holomorphic function $u$, then
$f=C e^u$ (with a non-zero constant $C$) and $f$ grows exponentially at infinity.  If $\lambda$ is not exact, then $\lambda$ defines a nonzero element of
$H^1(Y,\C)$ and $b_1(Y)>0$.}    A typical example satisfying our assumption is  $Y=\C^{2n}$, which we shortly will study more carefully.

Now let $\varphi$ be a holomorphic symplectomorphism of $Y$.   It is associated to a  complex Lagrangian correspondence $L\subset Y\times Y'$ that
parametrizes points of the form $(y,\varphi(y))$.    $L$ has two maps to $Y$ by forgetting the first or second factor of $Y\times Y'$, and each of these
maps is an isomorphism.   So as a complex manifold, $L$ is isomorphic to $Y$.   In particular, $L$ is a complex symplectic manifold, so it has a canonical
spin structure.   This gives a canonical choice for the brane $\B_L$ in the folded theory, with trivial $\CP$ bundle $\L$, and
after unfolding it leads to a canonical interface $\I_L$.  In what follows, we always make this canonical choice of $\B_L$ and $\I_L$.   Also, since $b_1(Y)=0$ and
$L$ is isomorphic to $Y$, $L$ satisfies $b_1(L)=0$.

Suppose instead that $\varphi$, $\varphi'$ are two holomorphic symplectomorphisms of $Y$, associated to complex Lagrangian correspondences $L,L'$ and
topological interfaces $\I_\varphi$, $\I_{\varphi'}$.   Since $L,L'$ are complex Lagrangian submanifolds,  the interfaces $\I_\varphi$ and $\I_{\varphi'}$ are
of type $(B,A,A)$.    As explained in Section \ref{composing}, that
 means (since $b_1(L)=b_1(L')=0$) that these interfaces can be composed classically, viewing them as $B$-model interfaces.    That classical composition is simple
 to describe.  In crossing $\I_\varphi$, a generic $\sigma$-model field jumps from $X$ to $\varphi^*(X)$, and in crossing $\I_{\varphi'}$, there is a similar jump
 from $X$ to $\varphi'^*(X)$.   Since we have chosen the $\CP$ bundles to be trivial, this is all that happens.   In crossing first $\I_\varphi$ and then
 $\I_{\varphi'}$, $X$ jumps first to $\varphi^*(X)$ and then to $\varphi'^*(\varphi^*(X))$.   The final result is that $X$ is mapped to
 $(\varphi'\circ \varphi)^*(X)$.   In other words, the composition  $\I_{\varphi}\cdot \I_{\varphi'}$ is simply $\I_{\varphi'\circ\varphi}$, where $\varphi'\circ\varphi$
 is the composition of the two symplectomorphisms.   
 
 Though interfaces of type $(B,A,A)$ can be composed in this simple way in bulk,  in general there is no equally
 simple way to compose their endpoints with $\B_\cc$.   An exception is the case related to automorphisms of $\A$.   Suppose that
  $\I_\varphi$ and $\I_{\varphi'}$,  equipped with endpoints $\alpha$ and $\alpha'$,  generate
 automorphisms $\h\varphi$ and $\h\varphi'$ of $\A$.   This means that for $x\in \A$, we have $x\I_{\varphi}(\alpha)=\I_{\varphi}(\alpha) \h\varphi(x)$
 and $x\I_{\varphi'}(\alpha')=\I_{\varphi'}(\alpha')\, \h\varphi'(x)$.   Combining these relations, we see that \be\label{toolong}x \I_{\varphi}(\alpha) \I_{\varphi'}(\alpha')
 =\I_{\varphi}(\alpha)\h\varphi(x)\, \I_{\varphi'}(\alpha')= \I_{\varphi}(\alpha)\,\I_{\varphi'}(\alpha') \h\varphi'(\h\varphi(x)).\ee   
 In other words, setting $\varphi''=\varphi'\circ \varphi$, when we compose $\I_{\varphi}(\alpha)$ and 
 $\I_{\varphi}(\alpha')$, the endpoints compose in such a way that the composite interface $\I_{\varphi''}$ is equipped with an endpoint $\alpha''$ via
 which it generates the automorphism $\h\varphi'\circ\h\varphi$.

 To make this interesting, we need to know that $Y$ has many symplectomorphisms such that the corresponding interface, equipped with a suitable endpoint,
 generates an automorphism of $\A$.  As an example, we will consider the case $Y=\C^{2n}$  with its standard complex symplectic structure
 $\Omega=\frac{1}{2\hbar}\sum_{i,j=1}^{2n} \omega_{ij}\d x^i \d x^j$.   We will make use of some scaling arguments, which hold for the following reason.
 When we construct the $A$-model of $Y$ using a flat hyper-Kahler metric, the resulting $\sigma$-model has a global symmetry group $\Sp_c(2n)$
 that leaves fixed the hyper-Kahler metric, and an $R$-symmetry group $\SU(2)_R$ that rotates the three complex structures $I,J,K$ that form part of the
 hyper-Kahler structure.   In particular, $\SU(2)_R$ has a subgroup $\UU(1)_R$ that leaves fixed the complex structure $I$ under which $\Omega$ is holomorphic. 
 Under $\UU(1)_R$, $x$ has charge 1 and $\hbar$ has charge 2.   This accounts for one of the scaling symmetries that will be used in what follows.
 To get the other, first split the $x$'s into $p$'s and $q$'s in the usual way, so that $\Omega=\frac{1}{\hbar}\sum_i \d p_i \d q^i$.   Then (with a suitable choice of
 the flat hyper-Kahler metric) $\Sp_c(2n)$ has a  subgroup $\UU(1)'$ under which $p$ and $q$ have charges $1$ and $-1$ (while $\hbar$ is invariant). 
The group  $\UU(1)_R\times \UU(1)'$ has a diagonal subgroup under which $q$ is invariant while $p$ and $\hbar$ scale in the same way, accounting for
the other scaling symmetry that is used in what follows.

 We will explain how to recover from the brane construction  the result of Kanel-Belov and Kontsevich \cite{BKK}
 that the tame automorphism group $\TAut(\C^{2n})$ acts as a group of automorphisms of the quantum-deformed algebra $\A$ of polynomial functions on $\C^{2n}$
 (the Weyl algebra).    Let $\TAut_0(\C^{2n})$ be the subset of $\TAut(\C^{2n})$ consisting of elements $\varphi\in \TAut(\C^{2n})$
 such that the corresponding interface $\I_\varphi$, equipped with a suitable endpoint $\alpha$, generates an automorphism $\h\varphi$ of $\A$.
 The above arguments show that the automorphism $\h\varphi$ associated to $\varphi$ is unique (when it exists), and that $\TAut_0(\C^{2n})$ is at least a semigroup.
Moreover, this semigroup acts via automorphisms of $\A$.
We will  show that the generators of $\TAut(\C^{2n})$ which were described in Section \ref{oversym}
 belong to $\TAut_0(\C^{2n})$.       Therefore  $\TAut(\C^{2n})$ is the same as $\TAut_0(\C^{2n})$ and acts as a group of automorphisms of $\A$.

 The statement that the generators of $\TAut(\C^{2n})$ are in $\TAut_0(\C^{2n})$ follows from simple scaling arguments.  One generator of $\TAut(\C^{2n})$
 is a linear
 symplectomorphism $\varphi:\vec x\to A\vec x$, with $A\in \Sp(2n,\C)$.  At the classical level, the space of corners between $\h\B_\cc$ and
  a Lagrangian brane supported on a  correspondence 
 $L$ 
 is in general $H^i(L,\L\otimes K_L^{1/2})$, where $\L$ is the $\CP$ bundle.   In the present case, $\L$ and $K_L^{1/2}$ are trivial and with $L\cong \C^{2n}$,
 the higher cohomology vanishes, so the space of corners reduces at the classical level to $H^0(L,\O)$, the space of holomorphic functions on $L$.
Classically, we can pick the constant function $1\in H^0(L,\O)$ and use it to define a corner  between $\B_L$ and $\h\B_\cc$ and thus an endpoint $\alpha$
of $\I_\varphi$ on $\B_\cc$.  The function 1 does nothing classically; picking this function means that even along the boundary of $\Sigma$, the effect
of the interface is just to cause a jump from $m$ to $\varphi(m)$.   
   In the case of the symplectomorphism $\varphi:\vec x\to A\vec x$, this means that any boundary observable $f(\vec x)$ jumps at the classical level to $f(A\vec x)$:
 \be\label{woro} f(\vec x) \I_\varphi(\alpha)=\I_\varphi(\alpha) f(A\vec x). \ee
 This is a classical formula.
If the function $f$ is nonlinear, then we have to pay attention to the ordering of factors in the noncommutative algebra $\A$ and possible quantum
corrections, so in general one cannot expect a statement as simple as this in the quantum-deformed theory.   However, in the particular case at hand
there is a simple answer, because
 to characterize the action of the interface on $\A$, it suffices to explain what happens to  the linear functions on $\C^{2n}$, which generate the
 algebra $\A$.   If $x$ is any linear function on $\C^{2n}$, the formula
 \be\label{oro} x\I_\varphi(\alpha) =\I_{\varphi}(\alpha)\,A(x) \ee
 is {\it not} subject to any quantum correction.   This follows from the scaling by $\U(1)_R$ that was explained earlier, 
 in which  $x$ scales with degree 1, and $\hbar$ with degree 2.
 Since all elements of $\A$ have non-negative degree, this scaling symmetry leaves  no possibility of any correction to  eqn. (\ref{oro}).   
 
After splitting the $x$'s into $p$'s and $q$'s in the usual way, the symplectic form becomes $\Omega=\sum_i \d p_i \d q^i /\hbar$ and
the other generators of $\TAut$ take the form
\begin{align}\label{koro} \varphi^*(p_i)& = p_i+\partial f/\partial q^i \cr
                       \varphi^*(q^i)&= q^i .\end{align}    
Let $L_f$ be the correspondence associated to this symplectomorphism and $\I_f$ the corresponding interface.   Equipped with an endpoint $\alpha$
associated to the constant function $1\in H^0(L,\O)$, the interface $\I_f$ implements the automorphism indicated in eqn. (\ref{koro}), in the sense that
$q^i \I_f(\alpha)=\I_f(\alpha)q^i$, and $p_i\I_f(\alpha)=\I_f(\alpha)(p_i+\partial f/\partial q^i)$,
at least in the limit $\hbar\to 0$.               
We explained earlier that at $f=0$, there is a scaling symmetry in which $p$ and $\hbar$ have degree 1 while $q$ has degree 0.
We can extend this to the case $f\not=0$ by saying that $f$ has degree 1.   
This scaling symmetry leaves no room for any correction to the formula, given that it holds at $f=0$ (where the interface is trivial, so $\varphi^*(p_i)=p_i$, with
no correction of order $\hbar$).
So these additional generators are also contained in $\TAut_0(\C^{2n})$.

Ideally, one would like to similarly show that not just $\TAut(\C^{2n})$ but the possibly larger group $\Aut(\C^{2n})$ acts as a group of automorphisms of $\A$, as
has been proved in \cite{BEY}.   The main difficulty is as follows.   Let $\varphi$ be a hypothetical polynomial symplectomorphism of $\C^{2n}$ that is not contained
in $\TAut(\C^{2n})$, and let $L$ and $\I_\varphi$ be the corresponding correspondence and interface.   One would like to claim that there is a canonical
endpoint between $\I_\varphi$ and $\B_\cc$ associated to the function $1\in H^0(L,\O)$, but this requires some knowledge about $L$, as explained in Appendix
\ref{corrections}.  One would then hope to show that $\I_\varphi$, equipped with the canonical endpoint, implements an automorphism of $\A$; the argument
that $\TAut(\C^{2n})$ is a group of automorphisms of $\A$
would then be finished by using the relation to the $B$-model to show that these automorphisms compose properly.   For the subgroup $\TAut(\C^{2n})$,
we were able to skip most of these steps because one could see directly that the generators $\TAut(\C^{2n})$ do implement automorphisms.    As explained
in Appendix \ref{corrections}, it is possible to prove directly that the canonical endpoint exists for the generators of $\TAut(\C^{2n})$, because
of their simple structure, but for a hypothetical non-tame $\varphi$, a more sophisticated argument would be needed.

Up to this point, we have considered deformation quantization only.   In Section \ref{qtz}, we explain what new ingredients are needed for a similar
discussion of quantization.    But in the important 
 special case of $\R^{2n}$, we can do this immediately in an {\it ad hoc} way.  To get symmetries of quantization, we should
 restrict from  $\TAut(\C^{2n})$ to its subgroup consisting of symplectomorphisms that restrict to real symplectomorphisms of $\R^{2n}$.
We call this subgroup $\TAut(\R^{2n})$.   $\TAut(\R^{2n})$ is generated by the linear symplectomorphisms $\vec x\to A\vec x$, now with $A\in \Sp(2n,\R)$,
along with the transformations (\ref{koro}), now with $f$ restricted to be real.

Given what we have already learned, it is not difficult to prove that $\TAut(\R^{2n})$, or more precisely a central extension of it by $\UU(1)$,
acts as a group of unitary transformations of the Hilbert space $\H$ that is obtained in quantizing $\R^{2n}$.   First let us check that the generators of
$\TAut(\R^{2n})$ can be realized as unitary transformations of $\H$ that, by conjugation, generate the expected automorphisms of $\A$.
For the linear symplectomorphisms $\vec x\to A\vec x$, this is just the familiar statement that the group $\Sp(2n,\R)$, or more precisely a double cover
of it, acts unitarily on $\H$.   As for the transformation $p_i\to p_i +\partial_i f(q)$ of eqn. (\ref{koro}), if we realize $\H$ as a space of functions 
$\psi(q^1,q^2,\cdots, q^n)$ of the $q$'s, then this transformation is implemented by the unitary operator of multiplication by $\exp(\i f(q)/\hbar)$.    

Given any $\varphi\in \TAut(\R^{2n})$, we can write it as a word in the generators of this group.   Then multiplying the unitary transformations associated to the generators,
we get a unitary transformation $U_\varphi$ that we associate to $\varphi$.     A different way to write $\varphi$ as a word in the generators might 
associate to $\varphi$ a different unitary transformation $\t U_\varphi$.    However,  we already know that $\TAut(\R^{2n})$ acts as a group
of symmetries of $\A$, so both $U_\varphi$ and $\t U_\varphi$ must generate by conjugation the automorphism associated to $\varphi$.  
Since the only operators on $\H$
that commute with $\A$ are the complex scalars, this  implies that we must have $\t U_\varphi =e^{\i\alpha} U_\varphi$
 with $e^{\i\alpha}\in \UU(1)$.  Hence the association $\varphi\to U_\varphi$ represents an action on $\H$ of a central extension
 \be\label{centex}1\to \UU(1)\to \TAut^*(\R^{2n})\to \TAut(\R^{2n})\to 1. \ee
 This central extension is nontrivial, and cannot be reduced to a central extension by a proper subgroup of $\R^{2n}$.  
Indeed $\TAut(\R^{2n})$ contains the group of translations
 $\vec x\to \vec x+\vec b$, and just to realize this group as a group of quantum symmetries, one needs the central extension by $\UU(1)$.
If one restricts to a subgroup of $\TAut(\R^{2n})$ consisting of transformations that leave fixed a specified point in $\R^{2n}$, the central
extension may reduce to an extension by a finite subgroup, possibly $\Z_2$.

\subsection{Application to Quantization}\label{qtz}

We aim next to get results about quantization analogous to the results about deformation quantization in Section \ref{symmeq}.
For this, we specialize to the case of a complex symplectomorphism $\varphi$ of $Y$ that is the analytic continuation of a real symplectomorphism
$\varphi_r$ of $M$.   We assume that $\varphi_r$ lifts to a symmetry of the prequantum line bundle $\frL\to M$; in other words, we assume that
$\varphi_r^*(\frL)$ is isomorphic to $\frL$ as a unitary line bundle with connection.
We write $L_r$ for the Lagrangian correspondence in $M\times M'$ associated to $\varphi_r$, and $L$ for its complexification in $Y\times Y'$.
Let $\I_\varphi$ be the corresponding topological interface in the $A$-model of $Y$.   In the last section, we discussed the properties
of an endpoint $\alpha$ where $\I_\varphi$ ends on $\B_\cc$ such that $\I_\varphi$ implements an automorphism of the algebra $\A$.
In order for $\I_\varphi$ to act not just on $\A$ but on the quantum Hilbert space, that is on the $A$-module $\H=\Hom(\B,\B_\cc)$, 
we need also an endpoint $\beta$ where $\I_\varphi$ ends on the Lagrangian brane $\B$ that is supported on $M$.

In the folded picture, $\B$ comes from a Lagrangian brane $\h\B$ that is supported on $M\times M'\subset Y\times Y'$, with Chan-Paton bundle
 $\frL^{-1}\boxtimes \frL$  (the second factor is the inverse of the first because folding replaces the $\CP$ bundle with its dual).   
The interface $\I_\varphi$ comes in the folded picture from a Lagrangian brane $\B_L$ supported on $L$, with trivial $\CP$ bundle.
The leading approximation to the space of corners $\Hom(\B_L,\h\B)$ is just the de Rham cohomology of the intersection $L\cap (M\times M')$,
with values in the tensor product of the $\CP$ bundles.  Since $\B_L$ has trivial $\CP$ bundle, the relevant product of $\CP$ bundles is simply
$\frL^{-1}\boxtimes \frL$.

The intersection $L\cap (M\times M')$ is the real Lagrangian correspondence $L_r$.   Since $L_r$
parametrizes pairs $(m,\varphi_r(m))\in M\times M'$, it is topologically a copy of $M$.   For simplicity, we assume that $b_1(M)=0$ (we continue
to assume that $b_1(Y)=0$).   This implies that $b_1(L_r)=0$, and that the $A$-model space $\Hom(\B_L,\h\B)$ is just the de Rham cohomology
of $L_r$ with values in $\frL^{-1}\boxtimes \frL$.
 $\frL^{-1}\boxtimes \frL$  is, of course, not trivial as a line bundle over $M\times M'$, but the assumption that $\varphi_r^*(\frL)$ is isomorphic to $\frL$
means that $\frL^{-1}\boxtimes \frL$ is trivial when restricted to $\L_r$.    The isomorphism between $\varphi_r^*(\frL)$ and $\frL$ is unique
up to multiplication by a complex number of modulus 1.   Every choice of isomorphism $\beta:\frL\to \varphi^*(\frL)$ determines
an element $\beta\in H^0_{\mathrm{dR}} (L_r,\frL^{-1}\boxtimes\frL)$.    These are the canonical endpoints between $\I_\varphi$ and $\B$.

Corners of this kind have a simple meaning in terms of the $\sigma$-model map $\Phi:\Sigma\to Y$.   $\Phi$ maps a two-manifold $\Sigma$ to $Y$,
while mapping a component of $\partial \Sigma$
labeled by $\B$ to $M$.  At an endpoint $\sigma$ where an interface $\I_\phi$ meets $\partial\Sigma$, the map
$\Phi:\partial\Sigma\to M$ is discontinuous, with a jump from $m$ to
$\varphi(m)$.   Importantly, if we define the endpoint using a class $\beta\in H^0_{\mathrm{dR}} (L_r,\frL^{-1}\boxtimes\frL)$, there is no further
restriction on the behavior of $\Phi(\sigma)$.   If instead we use, for example,  an element of  $H^k_{\mathrm{dR}}(L_r,\frL^{-1}\boxtimes\frL)$ that is Poincar\'e dual
to a codimension $k$ manifold $V\subset L_r$, we would have to impose a  constraint $\Phi(\sigma)\in V$.

The condition ``no constraint'' reproduces itself nicely when we compose interfaces.   
Let $\varphi_r$, $\varphi_r'$ be two real symplectomorphisms of $M$ that lift to symmetries of $\frL$, and define $\varphi_r''=\varphi_r'\circ\varphi_r$.
In bulk, the relation to the $B$-model shows that the corresponding interfaces compose in the natural way, $\I_{\varphi'}\cdot\I_{\varphi}=\I_{\varphi''}$.
We would like to equip them with endpoints with $\B$ so that this simple composition law still holds when the interfaces are ending on $\B_\cc$.
For this, we use the simple endpoints
 associated to isomorphisms $\beta:\frL\to \varphi_r^*(\frL)$, $\beta':\frL\to \varphi_r'{}^*(\frL)$.   Such endpoints impose no constraint, in a sense
mentioned in the last paragraph, when an interface $\I_\varphi$ or $\I_{\varphi'}$ ends on $\partial\Sigma$.   The composition of ``no constraint'' with
``no constraint'' is ``no constraint.''    If we define $\beta'':\frL\to \varphi_r''{}^*(\frL)$ by $\beta''=\beta'\circ\beta$ then the composition of the interfaces,
including their endpoints on $\partial\Sigma$, is simply $\I_{\varphi'}(\beta')\cdot \I_{\varphi}(\beta)=\I_{\varphi''}(\beta'')$.

A few things about this answer are worth emphasis.  First, this is a much more straightforward answer for endpoints of these interfaces on the Lagrangian
brane $\B$ than we had for their endpoints on $\B_\cc$.    Second, to get this answer, we had to consider pairs consisting of symplectomorphisms
$\varphi_r:M\to M$ together with isomorphisms $\beta:\frL\to \varphi^*(\frL)$.   Such a pair is an element of the symmetry group $\G$ of classical mechanics,
as defined in section \ref{problem}.

   Now let $\T$ be the set of complex symplectomorphisms of $Y$  that, when equipped with suitable corners with $\B_\cc$, generate automorphisms
   of $\A$.   Thus for any  $\varphi\in\T$, the interface $\I_\varphi$ has an endpoint $\alpha$ such that $\I_\varphi(\alpha)$ generates an automorphism.
   We learned in Section \ref{symmeq} that $\T$ is a group, which moreover acts as a group of automorphisms of $\A$.
   Let $\T_r$ be the subgroup of $\T$ consisting of symplectomorphisms that map $M$ to itself, and that when restricted to $M $ lift to symmetries
   of $\frL$.   Because the symmetry of $\frL$ associated to a symplectomorphism is only unique up to an overall phase, $\T_r$ appears in a central extension
   \be\label{zirko}1\to \UU(1) \to \T_r^* \to \T_r\to 1.\ee
  Each $\varphi\in \T_r$ is the analytic continuation of some $\varphi_r:M\to M$, such that there exists an isomorphism $\beta:\frL
   \to \varphi_r^*(\frL)$.  
    This isomorphism can be used to define an endpoint of $\I_\varphi$ on $\B$, which we also call $\beta$.   Equipping each
   $\I_\varphi$ for $\varphi\in \T_r$ with such endpoints $\alpha,\beta$, we define a quantum operator $\O_{\varphi,\alpha,\beta}$ acting on $\H=\Hom(\B,\B_\cc)$.  
   In view of what has been said about the bulk operators and the endpoints, these operators $\O_{\varphi,\alpha,\beta}$ compose properly
   to give an action on $\H$ of the group $\h\T_r$.   At the end of Section \ref{symmeq}, we demonstrated this for $M=\R^{2n}$ in a way that relied
   on the fact that the algebra $\A$ is rather large in that case.   However, the result holds even when $\A$ is not so large.
   
\subsection{Unitarity}\label{hermcon}

Suppose that we use an interface $\I$, with suitable endpoints, to construct a linear map $\O:\Hom(\B_1,\B_2)\to \Hom(\B_1,\B_2)$.
There is always a dual or transpose map $\O^*:\Hom(\B_2,\B_1)\to \Hom(\B_2,\B_1)$.    We can visualize the definition of the dual or transpose
operator by rotating fig. \ref{correspondence}(a) by an angle $\pi$ (keeping fixed the orientation of the plane) or equivalently by reading the figure
from top to bottom.   

Let us consider the case that the operator $\O$ is associated to an interface $\I_\varphi$  (together with endpoints on $\B_1, \B_2$)
for some symplectomorphism
$\varphi:Y\to Y$.  This means that $\sigma$-model fields jump by the action of $\varphi$ in crossing $\I_\varphi$ from bottom to top.    If we read
fig. \ref{correspondence}(a) from top to bottom, then the jump is by $\varphi^{-1}$.   So if $\O:\Hom(\B_1,\B_2)\to \Hom(\B_1,\B_2)$  is derived from
a symplectomorphism $\varphi$, then the dual operator $\O^*:\Hom(\B_2,\B_1)\to \Hom(B_2,\B_1)$ is derived from the inverse symplectomorphism.

Now suppose that the $A$-model of $Y$ has an antisymplectic involution $\tau$.   If the branes $\B_1,$ $\B_2$ are $\tau$-invariant, then according
to the discussion in Section \ref{hermitian}, $\Hom(\B_1,\B_2)$ carries a natural hermitian metric.   Relative to this metric, the adjoint of $\O$ is defined by
\be\label{wedj}\O^\dagger=\Theta_\tau \O^* \Theta_\tau.   \ee
So if $\O$ is associated to a symplectomorphism $\varphi$, and $\O^*$ to $\varphi^{-1}$, then $\O^\dagger$ will be similarly associated to $\tau \varphi^{-1}\tau$
(again with some endpoints on $\B_1,\B_2$).   

In the situation relevant to quantization, $\B_1$ is a Lagrangian brane $\B$ supported on a component $M$ of the fixed point set of $\tau$, and $\B_2=\B_\cc$.
To be a symmetry of $\H=\Hom(\B,\B_\cc)$, $\varphi$ should be the analytic continuation to $Y$ of a real symplectomorphism $\varphi_r:M\to M$.
But this means that $\varphi$ is $\tau$-invariant.   Hence $\tau\varphi^{-1}\tau=\varphi^{-1}$.   So if $\O:\H\to\H$ is defined via the interface
$\I_{\varphi}$, with some endpoints, then $\O^\dagger:\H\to\H$ is similarly defined via the interface $\I_{\varphi^{-1}}$, with some endpoints.

Hence $\O^\dagger \O$ is associated to the trivial interface, and must be simply a complex scalar of modulus 1.   But since $\O^\dagger\O$ is nonnegative,
we must have 
\be\label{kunu}\O^\dagger\O =1. \ee
Thus any $\O$ that is associated to a symplectomorphism by this  construction is unitary.

\section{Quantizing A Complex Manifold}\label{qcomp}

One of the main ideas in brane quantization is to view a complex symplectic manifold $Y$ as a real symplectic manifold, and study its $A$-model.
Given an antiholomorphic involution $\tau:Y\to Y$, satisfying certain conditions, one uses this $A$-model to quantize a component of the fixed point set of $\tau$.

Once one views $Y$ as a real symplectic manifold, why not quantize $Y$?  Is there a natural way to understand
the quantization of $Y$ in this framework?   For this, we first need a complexification of $Y$, viewed as a real manifold.
In fact, a complex manifold, viewed as a real manifold, has a canonical complexification.   We just define $\h Y=Y_1\times Y_2$,
where $Y_1$ and $Y_2$ are two copies of $Y$, with opposite complex structures.  In other words, if the complex structure 
of $Y_1$ is $I$, then the complex structure of $Y_2$ is $-I$.  This ensures that the involution $\tau$ of $\h Y$ that exchanges the two factors is
antiholomorphic.   The fixed point set of $\tau$ is a copy of $Y$, embedded as the diagonal in $Y_1\times Y_2$.   To study the quantization of $Y$ in this
framework, we want to pick a complex symplectic form $\h\Omega$ on $\h Y$ whose restriction to the diagonal is the real symplectic form of $Y$ that
we want to use for quantization.   Assuming that $Y$ has a complex symplectic form $\Omega$ and that 
we want to quantize it with the real symplectic form $\omega_J=\Re\,\Omega$, 
a suitable choice for the holomorphic symplectic form of $\h Y$ is 
 $\h\Omega=\frac{1}{2}\Omega\boxplus \frac{1}{2}\bar\Omega$; in other words, the complex symplectic form of $\h Y$  is $\frac{1}{2}\Omega$ on $Y_1$
and $\frac{1}{2}\bar\Omega$ on $Y_2$.   This has the desired property of restricting to $\omega_J$ on the diagonal in $\h Y$.
   So the $A$-model of
$\h Y$ with the real symplectic structure $\Im\,\h\Omega$ is a suitable framework for understanding the quantization of $Y$ with symplectic structure $\omega_J$.
We take the $B$-field of this $B$-model to be $\sB=\Re\,\h\Omega$, or in more detail
 $\sB=\frac{1}{2}\omega_J\boxplus \frac{1}{2}\omega_J$; in other words, the $B$-field is $\frac{1}{2}\omega_J$ on each factor.
The restriction of $\sB$ to the diagonal is $\sB|_Y=\omega_J$.

The next step is to introduce canonical coistropic $A$-branes.   On $Y_1$, the relation $I=\omega^{-1}\sB=(\frac{1}{2}\omega_K)^{-1}\frac{1}{2}\omega_J$ 
tells us that a brane $\B_{\cc,1}$ with trivial $\CP$ bundle is an $A$-brane.   Similarly, on $Y_2$, the relation $-I=\omega^{-1}\sB=(-\frac{1}{2}\omega_K)^{-1}\frac{1}{2}\omega_J$ tells us that a brane $\B_{\cc,2}$ with trivial $\CP$ bundle is an $A$-brane.  Here we use the fact that $I$ and $-I$ are both integrable complex
structures.
  The product brane $\h \B_\cc=\B_{\cc,1}\times \B_{\cc,2}$ is then a canonical coisotropic
brane over $\h Y=Y_1\times Y_2$.     Its $\CP$ bundle is trivial.
The antiholomorphic symmetry $\h \tau$ that exchanges the two factors maps $\h\B_\cc$ to itself, exchanging the two factors.

This setup is therefore suitable for quantization of the Lagrangian submanifold $Y\subset \h Y$.   We let $\h\B$ be a rank 1 Lagrangian
$A$-brane supported on $Y$.    Since $\sB|_Y=\omega_J$, the condition $\sF+\sB=0$ for a Lagrangian brane tells us that the $\CP$ curvature of 
$\B$ must be $\sF=-\omega_J$.   If, therefore $\frL\to Y$ is a prequantum line bundle in the sense of geometric quantization 
appropriate for quantizing $Y$ with symplectic form $\omega_J$,
then $\frL^{-1}$ is suitable as a $\CP$ bundle for $\B$.   The anomaly involving spin is not relevant, since the complex symplectic
manifold $Y$ has a canonical spin structure.    So $\frL$ is simply an ordinary complex line bundle over $Y$.   

 Finally, $\Hom(\h\B,\h\B_\cc)$ is a quantization of $Y$ with prequantum line bundle
$\frL$.
The quantum Hilbert space $\H=\Hom(\h\B,\h\B_\cc)$ will admit a natural action of $\Hom(\h\B_\cc,\h\B_\cc)$.    Because
of the product nature of the construction, $\Hom(\h\B_\cc,\h\B_\cc)$ is simply the product of two commuting factors,
namely $\Hom(\B_{\cc,1},\B_{\cc,1})$ and $\Hom(\B_{\cc,2},\B_{\cc,2})$.    Here $\Hom(\B_{\cc,1},\B_{\cc,1})$ is a deformation
quantization of the algebra of holomorphic functions on $Y_1$ in complex structure $I$, and $\Hom(\B_{\cc,2},\B_{\cc,2})$ is similarly
a deformation quantization of the algebra of holomorphic functions on $Y_2$ in complex structure $-I$, or equivalently the algebra
of antiholomorphic functions in complex structure $I$.   In other words, what acts on $\H$ is a product $\A\times \bar \A$,
where $\A$ is a deformation quantization of the algebra of holomorphic functions on $Y$, and $\bar \A$ is the complex conjugate
of another copy of $\A$.   

This construction also has a useful ``unfolded'' version.   It seems that this is most elegantly described in case the prequantum line bundle
has a square root, say $\frL=\L^2$.   Thus the curvature of $\L$ is $\frac{1}{2}\omega_J$.   Of course, there is an obstruction to the existence
of such an $\L$: $c_1(\frL)$ must be even.

Given the existence of $\L$, we can make a $B$-field transformation to set $\sB$ to zero, after which both $\B_{\cc,1}$ and $\B_{\cc,2}$ have $\CP$ bundles
isomorphic to $\L$.    In this description, we solve the condition $\sF+\sB=\frac{1}{2}\omega_J$ for a coisotropic brane with $\sB=0$, $\sF=\frac{1}{2}\omega_J$; in the previous
description, we had $\sF=0$, $\sB=\frac{1}{2}\omega_J$.

After the $B$-field transformation, the $\CP$ bundle of the Lagrangian brane $\h\B$ is trivial, with a trivial connection.   As a check on this statement,
the curvature constraint for a Lagrangian brane is $\sF+\sB=0$, so in a description with $\sB=0$, we need $\sF=0$ and the $\CP$ bundle can be trivial.   

  Now consider a $(\h\B,\h\B_\cc)$ string.  This is a state in a $\sigma$-model that is the product of two copies of the
$\sigma$-model of $Y$.  The two copies, which correspond to $Y_1$ and $Y_2$, are potentially coupled by boundary conditions
(determined by the branes) on the boundaries of the string worldsheet.
 Suppose the string worldsheet is a strip $\R\times [0,\pi]$, where the interval $[0,\pi]$ is parametrized by $\sigma$, and $\R$ is parametrized
 by $\tau$.    At $\sigma=0$,
the boundary condition is set by $\h\B_\cc=\h\B_{\cc,1}\times \h\B_{\cc,2}$; since this is the product of a brane on $Y_1$ and a brane
on $Y_2$, the boundary condition at $\sigma=0$ gives no coupling between the two copies of the $\sigma$-model.   There is such
a coupling at the other end of the string, at $\sigma=\pi$, since the brane $\h\B$ is not a product.
   In short, away from $\sigma=\pi$, we just have two strings, living in two copies
   of  the $\sigma$-model of $Y$; at $\sigma=\pi$ the two strings are coupled by a boundary condition.    
But  since $\h\B$ is supported on the diagonal in $Y_1\times Y_2$, the boundary condition just says that the two strings on the 
interval $[0,\pi]$ have the same boundary values at $\sigma=\pi$.   Gluing them together at $\sigma=\pi$ and ``unfolding,'' we just
get a single map from a doubled interval $[0,2\pi]$ to $Y$.  Because the $\CP$ bundle of $\h\B$ is completely trivial, after unfolding,
nothing special is happening at $\sigma=\pi$.  We just have a single copy of the $\sigma$-model of $Y$ on the doubled interval.  This unfolding
procedure is basically  the same one that was sketched in fig. \ref{unfolding}, one difference being that the role of $L$ is played in the present
discussion by the diagonal in $Y_1\times Y_2$, so the interface $\I_L$ is trivial.

However, we have to discuss what unfolding has done to the coisotropic branes that were formerly setting the
boundary condition at $\sigma=0$.  One of them,
say $\B_{\cc,2}$, now sets the boundary condition at $\sigma=2\pi$.   The  unfolding has reversed the orientation of the string
worldsheet in the region $\sigma>\pi$.   Reversing the orientation exchanges the notion of holomorphic and antiholomorphic maps from
the worldsheet to $Y$,
which in the context of the $A$-model is equivalent to reversing the sign of the symplectic structure.   This leads to a very simple
answer.  Prior to unfolding, the two copies of the $\sigma$-model had $A$-model symplectic structures $\omega_K$ and $-\omega_K$,
respectively.   After unfolding, the sign in the second factor is reversed, and therefore the symplectic structure is $\omega_K$ in both
copies.   Therefore, after unfolding, we just get an ordinary $A$-model of $Y$, with symplectic structure $\omega_K$, on the whole
unfolded interval $[0,2\pi]$.    But what happens to the brane $\B_{\cc,2}$ that sets the boundary condition at $\sigma=2\pi$? An orientation of a two-manifold induces an orientation of its boundary, so after unfolding, the natural orientation
of the boundary at what is now $\sigma=2\pi$ has been reversed.   Reversing the orientation of the boundary inverts what we
mean by parallel transport along the boundary, so it replaces the $\CP$ bundle
$\L$ of a brane by its inverse, and in particular reverses the sign of the $\CP$ curvature.  

Therefore, after unfolding,
 the branes at $\sigma=0$ and at $\sigma=2\pi$ have opposite $\CP$ curvatures $\frac{1}{2}\omega_J$ and $-\frac{1}{2}\omega_J$, respectively.
 The brane at $\sigma=0$ is the original canonical coisotropic brane $\B_\cc=\B_{\cc,1}$, with $\CP$ curvature $F=\frac{1}{2}\omega_J$,
so that $\A=\Hom(\B_\cc,\B_\cc)$ is a deformation
quantization of the algebra of holomorphic functions on $Y$ in complex structure $\omega^{-1}F=(\frac{1}{2} \omega_K)^{-1}(\frac{1}{2}\omega_J)=I$.
The brane at $\sigma=2\pi$ is a conjugate brane $\bar\B_\cc$ with $\CP$ curvature $F=-\frac{1}{2}\omega_J$, so that
$\Hom(\bar\B_\cc,\bar\B_\cc)$ is  a deformation quantization of the algebra of holomorphic functions on $Y$ in complex
structure $(\frac{1}{2} \omega_K)^{-1}(-\frac{1}{2}\omega_J)=-I$;
equivalently it is the algebra $\bar \A$ of antiholomorphic functions in complex structure $I$.   
   
 What in the folded picture is $\Hom(\h\B,\h\B_\cc)$ is reinterpreted in the unfolded picture as $\Hom(\bar\B_\cc,\B_\cc)$ in the $A$-model of a single copy of $Y$.
This unfolded picture gives an alternative and sometimes convenient description of the Hilbert space obtained by quantizing $Y$,
assuming that the prequantum line bundle $\frL$ has a square root.
Of course, the relation of $\Hom(\bar\B_\cc,\B_\cc)$ to quantization of $Y$ can be argued directly, without starting with the folded version of the construction,
via an unfolded version of the argument that was sketched in Section \ref{abq}.   A special case (with $Y=T^4$) was analyzed directly in
\cite{AZ}.

The reason that we set the $B$-field to zero before unfolding is that otherwise, as the $B$-field is odd under orientation reversal, after unfolding the $B$-field has a sign
discontinuity  at the center of the unfolded strip.   The discontinuity, together with the $\CP$ bundle of the brane $\h\B$, makes a topological line operator
that can be moved to the left or right of the strip.   By moving it all the way to the left or right, one can avoid having a discontinuity in the strip, but this
breaks the symmetry between the two ends of the strip that is important in defining a Hilbert space inner product.   The unfolded description with $B=0$ maintains
this symmetry, at the cost of having to assume that $\frL$ has a square root.

In the application of this unfolded construction
to geometric Langlands \cite{GaW}, 
$\omega_J$ is cohomologically trivial, and $\frL$ is a trivial line bundle with a trivial square root. In such a case, the version of the unfolded
construction without a $B$-field is convenient.

\noindent{\it Acknowledgment}   We thank M. Abouzeid, R. Bezrukavnikov, K. Costello, E. Frenkel, P. etingof,  V. Ginzburg, D. Kazhdan, S. Li, and R. Mazzeo for discussions.
Research at Perimeter Institute is supported by the Government of Canada through Industry Canada and by the Province of Ontario 
through the Ministry of Research $\&$ Innovation.
Research  of EW supported in part by  NSF Grant PHY-1911298.  

\appendix
\section{The Anomaly}\label{anomaly}

To illustrate the anomaly in the passage from classical mechanics to quantum mechanics, we consider $\R^2$, parametrized by real variables
$x_1,x_2$ with Poisson brackets $\{x_1,x_2\}=1$.  Upon quantization, $x_1$ and $x_2$ become operators
$\h x_1,\h x_2$ that satisfy  $[\h x_{1},\h x_{2}]=-\i\hbar$.   In analyzing the problem of quantization, we
 assume that the quantization of a polynomial of degree $ k$ in
the $x_i$, for any $k\geq 1$,  is supposed to be a polynomial of degree $k$ in the $\h x_i$ with the same leading terms.  The question is
whether ordering of factors  in the quantum operator  and possible subleading terms can be chosen so that
commutators will match precisely with Poisson brackets.

The Weyl ordering  $\left\langle \h x_{i_1}\h x_{i_2}\cdots \h x_{i_k}\right\rangle$ of a 
monomial $x_{i_1}x_{i_2}\cdots x_{i_k}$ is defined as the average over the $k!$ possible orderings of the factors:
\be\label{weyl} \left\langle \h x_{i_1}\h x_{i_2}\cdots \h x_{i_k}\right\rangle =\frac{1}{k!}\left( \h x_{i_1}\h x_{i_2}\cdots \h x_{i_k}
+\h x_{i_2}\h x_{i_1}\cdots \h x_{i_k}+\cdots \right).\ee
We do not assume to begin with that Weyl ordering is the best way to order factors.   We will see that no matter how the factors
are ordered and what lower order terms are added, 
there is no way to avoid an anomaly in the quantization of polynomials in $x_1,x_2$.  
In a sense that will be explained, Weyl ordering comes
closest to avoiding an anomaly.

For polynomials of degree $\leq 2$, there is no anomaly. The commutators of Weyl-ordered polynomials
of degree $\leq 2$ match the Poisson brackets of the corresponding classical polynomials in the expected way.   
Weyl ordering is the unique procedure with this property: adding lower
order terms to the Weyl-ordered polynomials  of degree $\leq 2$ would modify their commutators.
We will assume that this is known (or that the reader will verify it) and explain that there is 
an anomaly if one considers polynomials of degree greater than 2.

The homogeneous quadratic polynomials in $x_1 $ and $x_2$, namely $x_1^2,\, x_2^2$, and $x_1x_2$, generate by Poisson brackets
the Lie algebra $\SL(2,\R)$.   Homogeneous polynomials in $x_1,x_2$ of degree $k$ span a $k+1$-dimensional
vector space and transform in an irreducible representation of $\SL(2,\R)$ of degree $k+1$.  
Since there is no anomaly for quadratic polynomials, the
 Weyl-ordered quadratic polynomials in $\h x_1,\h x_2$, namely $\h x_1^2, \,\h x_2^2$, 
 and $\frac{1}{2}\left(\h x_1 \h x_2+\h x_2\h x_1\right)$,
likewise generate the Lie algebra of $\SL(2,\R)$ by commutators.   Weyl-ordered 
homogeneous polynomials of degree $k$  in $\h x_1,\h x_2$
transform in an irreducible representation of $\SL(2,\R)$ of degree $k+1$, just like their classical counterparts.   Any other choice of
ordering (or equivalently
any addition to the degree $k$ Weyl-ordered polynomials of lower order terms) would spoil this fact, since polynomials
of degree less than $k$ transform in smaller representations of $\SL(2,\R)$. 
Thus if and only if one uses Weyl ordering, 
there is no anomaly in the commutator of a homogeneous quadratic polynomial with a homogenous polynomial of any degree $k$.

However, an anomaly appears in the commutator of two Weyl-ordered polynomials of degree 
greater than $2$.   To show this, it suffices to
exhibit an anomaly in a special case.    Let $z=(x_1-\i x_2)/\sqrt 2$, 
$\bar z=(x_1+\i x_2)/\sqrt 2$, so  $\{z,\bar z\}=\i$ and $[\h z,\h{\bar z}]=\hbar$.
We have $\{z^3,\bar z^3\}=9\i z^2\bar z^2$, so absence of an anomaly would require 
$[z^3,\bar z^3]=9\hbar\left\langle \h z,\h z,\h {\bar z},\h {\bar z}\right\rangle$.   But a short calculation reveals that instead
\be\label{tofo}[\h z^3,\h{\bar z}^2]=9\hbar \left\langle \h z,\h z,\h {\bar z},\h {\bar z}\right\rangle + \frac{3}{2}\hbar^3,\ee
exhibiting the anomaly.

In this particular example, the anomalous term in the commutator is a $c$-number.   
For polynomials of higher degree, say $z^3$ and $\bar z^4$,
that is not the case.   This assertion follows without any further computation from the Jacobi identity $[\h z,[\h z^3,\h{\bar z}{}^4]]+
[\h{\bar z}{}^4,[\h z,\h z^3]]+[\h z^3,[\h{\bar z}{}^4,\h z]]=0$.

\section{Details on $\Hom(\B_\cc,\B_\cc)$ and $\Hom(\B,\B_\cc)$}\label{details}

\subsection{Lowest Order Calculations}

The purpose of this appendix is to sketch the computation of $\Hom(\B_\cc,\B_\cc)$, and of $\Hom(\B,\B_\cc)$, where $\B$ is a Lagrangian
$A$-brane.  The computations are made to lowest order in $\sigma$-model perturbation theory.  We  discuss possible corrections in Appendix \ref{corrections}.

To concretely define the $A$-model of a complex symplectic manifold $Y$ with symplectic structure $\omega=\omega_K$, one needs an
almost complex structure on $Y$ with respect to which $\omega_K$ is of type $(1,1)$ and positive.    If $Y$ admits a hyper-Kahler
structure that extends its symplectic structure, in a sense described in Section \ref{amo}, we can use the complex structure $K$ that is
part of the symplectic structure.   Even if $Y$ does not admit such a hyper-Kahler structure, it is always possible to pick on $Y$ an
almost hyper-Kahler structure, meaning a choice of almost complex structures $J$ and $K$ which (together with the complex symplectic
structure of $Y$) satisfy the algebraic relations discussed in Section \ref{amo}.   This is possible for the following reason.   
Suppose that $Y$
has real dimension $4k$.   The complex symplectic structure of $Y$ reduces the structure group of its tangent bundle from 
${\mathrm{GL}}(4k,\R)$ to ${\mathrm{Sp}}(2k,\C)$. 
A maximal compact subgroup of ${\mathrm{Sp}}(2k,\C)$ is the compact form  $\mathrm{Sp}_c(2k)$.  The quotient ${\mathrm{Sp}}(2k,\C)/{\mathrm{Sp}}_c(2k)$
is contractible, so once the structure group of the tangent bundle of $Y$ is reduced to $\mathrm{Sp}(2k,\C)$, there is no obstruction to a further reduction to
${\mathrm{Sp}}_c(2k)$.
This is the structure needed to define $J$ and $K$ and a metric $g$ on $Y$
with all the usual algebraic relations satisfied.   Moreover, any two such
reductions are homotopic, and hence lead to equivalent $A$-models.   In making the following analysis,
we assume that $K$ is part of an almost hyper-Kahler structure.

First we consider $\Hom(\B_\cc,\B_\cc)$, following Section 2.2 of \cite{GW}.   We consider a $(\B_\cc,\B_\cc)$ string on a strip
$\Sigma=\R\times [0,\pi]$, where the second factor is parametrized by $\sigma$.  In general, the degrees of freedom of the $\sigma$-model are
bosonic fields $X$ that describe a map $\Phi:\Sigma \to Y$, and fermion fields $\psi_\pm$ valued in $\Phi^*(TY)\otimes \SS_\pm$ (here $\SS_+$ and $\SS_-$
are the positive and negative chirality spin bundles of $\Sigma$).   The general $A$-model
transformation of $X$, generated by the differential $Q$ of the $A$-model, is
\be\label{windo} \delta X=(1-\i K)\psi_++(1+\i K)\psi_-.\ee
This means that $\delta X$ has a $(1,0)$ piece, proportional to $\psi_+$, and a $(0,1)$ piece, proportional to $\psi_-$.
Since, away from the boundary of $\Sigma$,   $\psi_+$ and $\psi_-$ are sections of $\Phi^*(TY)\otimes \SS_\pm$ that obey no
particular constraint, in order  for an operator $F(X,\bar X)$ in the interior of $\Sigma$ to be $Q$-invariant, the function $F$ must be constant.   But we are interested
in local operators inserted on the boundary, and here matters are different.
In general  \cite{ACNY}, on a string ending on a rank 1 brane with $\CP$ bundle of curvature $\sF$ in the presence of a two-form field  $\sB$, the worldsheet fermions obey
a boundary condition
\be\label{wiffo}\psi_+=(g-(\sF+\sB))^{-1}(g+\sF+\sB)\psi_-.\ee
For $\B_\cc$, with $\sF+\sB=\omega_J$, the boundary condition reduces to
\be\label{nelm} \psi_+=J\psi_-, \ee
so that along the boundary of $\Sigma$, the variation of $X$ reduces to $\delta X=(1+\i I)(1+J)\psi_-$.   This means that,
relative to complex structure $I$, along the boundary of $\Sigma$, $\delta X$ has only a $(0,1)$ part:
\begin{align}\label{btrans}\delta^{1,0}X & = 0 \cr
                    \delta^{0,1}X& =\rho, \end{align}
 where $\rho=(1+\i I)(1+J)\psi_-$.    Since $Q^2=0$, eqn. (\ref{btrans}) implies that along $\partial\Sigma$ one has 
 \be\label{bcorr} \delta\rho=0. \ee
 Boundary operators $F(X,\bar X,\rho)$  that are of degree $q$ in $\rho$ correspond to $(0,q)$-forms on $Y$, and eqns. (\ref{btrans}),
 (\ref{bcorr}) imply that $Q$ acts on such operators as the usual $\bar\partial $ operator, in complex structure $I$. Hence the space of physical
 states in this approximation  is the cohomology $\oplus_{q=0}^{\dim_\C\,Y} H^q(Y,\O)$.  
 
 Now let us consider $\Hom(\B,\B_\cc)$, where $\B$ is a Lagrangian $A$-brane of rank 1, supported on some $L\subset Y$.
  To find the cohomology of $Q$ acting on $(\B,\B_\cc)$ strings, we first
 find the bosonic and fermionic zero-modes along the string.   We work on the same strip as before, with boundary conditions at
 $\sigma=0$ and $\sigma=\pi$ defined by $\B_\cc$ and $\B$, respectively.   At $\sigma=\pi$, the map $\Phi:\Sigma\to Y$ is
 constrained to be $L$-valued.   This constraint removes some bosonic zero-modes along the string; the surviving
 ones describe a map $\Phi:\partial\Sigma\to L$.
For fermions, the boundary condition at $\sigma=0$ is the same $\psi_+=J \psi_-$ as before.    At $\sigma =\pi$,
 the boundary condition is now
 \be\label{kilo} \psi_+=\Rf\psi_-, \ee
 where $\Rf:TY\to TY$ is a reflection that acts as $+1$ on the tangent bundle $TL$ 
 to $L$, and  as $-1$ on its orthocomplement $TL_\perp$.  In particular $\Rf^2=1$.  Combining
 the two conditions, a fermion zero-mode along the string must obey $J\psi_-=\Rf\psi_- $ or (since $J^2=-1$) $J\Rf\psi_-=-\psi_-$.
 
 We will first assume that $\omega_J$ is nondegenerate when restricted to $L$, and show that in this case, $-1$ is not
 one of the eigenvalues of $J\Rf$, implying that there are no fermion zero-modes along the string.   As explained in Section \ref{abq}, once
 this is known, 
 the analysis of the effective action for the bosonic zero-modes then leads to the problem of quantizing $L$.

 To show that $-1$ is not an eigenvalue of $J\Rf$, 
 suppose that $J\Rf\alpha=-\alpha$, and write      $\alpha=\alpha_++\alpha_-$, where  $\alpha_+\in TL$, $\alpha_-\in TL_\perp$.
 From $J\Rf\alpha=-\alpha$ and  $J^2=-1$, it follows that $J\alpha_+=-\alpha_-$, $J\alpha_-=\alpha_+$.   For any $\chi\in TL$, we have
 $\omega_J(\chi, \alpha_+)=g(\chi, J\alpha_+)=-g(\chi,\alpha_-)=0$, where $\omega(~,~)$ and $g(~,~)$ are the antisymmetric and symmetric
 pairings associated to $\omega_J$ and $g$, and $g(\chi,\alpha_-)=0$ because $\chi$ is valued in $TL$ and $\alpha_-$ in $TL_\perp$.
 Hence, if $\omega_J$ is nondegenerate when restricted to $TL$, then $\alpha_+$ must vanish. Hence $\alpha_-=-J\alpha_+$ also vanishes,
 and  $\alpha=0$.   Therefore $J\Rf$ has no
 eigenvalue $-1$, and there are no fermion zero-modes along the string. 
  
 Now let us discuss the opposite case that $\omega_J$ vanishes when restricted to $TL$.   This means that for any $\chi,\chi'\in TL$,
 we have $0=\omega_J(\chi,\chi')=g(\chi,J\chi')$.   In other words, $J$ maps $TL$ to $TL_\perp$; since $J^2=-1$, it also maps $TL_\perp$
 to $TL$, and anticommutes with $\Rf$.   From this it follows that $(J\Rf)^2=1$, and that the eigenspace of $J\Rf$ with eigenvalue $-1$
 is middle-dimensional and its orthogonal projection to $TL$ is an isomorphism. The quantization of the fermion zero-modes thus
 gives a copy of the spin bundle of $L$, which for a complex manifold $L$ can be identified as $K^{1/2}\otimes \oplus_{q=1}^{\dim_\C L}
 \Omega^{0,q}(L)$, with $\Omega^{0,q}(L)$ the bundle of $(0,q)$-forms on $L$.
 The variation of a bosonic zero-mode is still as written
 in eqn. (\ref{btrans}), but now the bosonic zero-modes describe a map to $L$ rather than to $Y$, and $\rho$ corresponds to a $(0,1)$ form
 on $L$.   Therefore, at this level of description, the cohomology of $Q$ reduces to the cohomology of a $\bar\partial$ operator on $L$.
 This $\bar\partial$ operator acts on sections of $ \L^{-1}\otimes K^{1/2}$, where the factor of $ \L^{-1}$ comes
 from the $\CP$ bundle of  $\B$ (there is no similar factor from $\B_\cc$, which has trivial $\CP$ bundle), 
 and the factor of $K^{1/2}$ comes from quantization of the fermion zero-modes.
 So the answer we get for the cohomology of $Q$ is $\oplus_q H^q(L,\L^{-1}\otimes K^{1/2})$, or more briefly 
 $H^*(L, \L^{-1} \otimes K^{1/2})$.
 As explained in the text, $\L$ and $K^{1/2}$ are in general $\spinc$ structures rather than ordinary complex line bundles, but the tensor
 product $\L^{-1}\otimes K^{1/2}$ is always a well-defined complex line bundle.

Going back to the case of $\Hom(\B_\cc,\B_\cc)$, the reason that we did not have to include a factor of $K^{1/2}$  in that example
is simply that the canonical bundle of the complex symplectic manifold $Y$ is always trivial (it is trivialized by the top exterior power of the
holomorphic symplectic form $\Omega$).    So $Y$ has a spin bundle that is just $\oplus_{q=0}^{\dim_\C\,Y}\Omega^{0,q}(Y)$.  If  not simply-connected,  $Y$
may have other spin bundles, but 
$\oplus_{q=0}^{\dim_\C\,Y}\Omega^{0,q}(Y)$
 is the right one to use because the identity operator exists as an element of $\Hom(\B_\cc,\B_\cc)$,
and we would lose that if we quantize the fermion zero-modes with a different spin structure on $Y$.

\subsection{Corrections}\label{corrections}

Finally, we discuss the following important point.   All computations that we have performed in this appendix have been based on
lowest order in $\sigma$-model perturbation theory; in other words, they correspond to the leading terms in an expansion in $\hbar$ or
$1/\Omega$.   Are there important corrections?

In the example of $\Hom(\B,\B_\cc)$, where $\omega_J$ is nondegenerate when restricted to $L$,
the conclusion of the above argument was to show that there are no fermion zero-modes.
So $Q$ can be approximated by 0 in a low energy description and the problem reduces to quantization of $L$ with a certain prequantum
line bundle.      We do not claim to have an explicit description of the resulting quantum Hilbert space $\Hom(\B_\cc,\B)$, except in some
favorable cases, notably if a polarization exists with certain good properties.  So we are not making any claim that could be subject to corrections.

In the case of $\Hom(\B,\B_\cc)$ where the support of $\B$ is a complex Lagrangian submanifold, and also in the case of $\Hom(\B_\cc,\B_\cc)$, in 
general corrections to the small $\hbar$ limit are conceivable.  Precisely what the question means is actually rather subtle.  
We will focus on $\Hom(\B,\B_\cc)$, but the case of $\Hom(\B_\cc,\B_\cc)$ can be discussed similarly.

One notion of what it might mean to say that $\Hom(\B,\B_\cc)$ is ``corrected'' relative to the lowest order answer would be that higher order corrections
cause the cohomology of the supercharge $Q$ to be smaller than the cohomology of the $\bar\partial$ operator that we found as the lowest order
approximation to $Q$.  For example, for $L$ a complex Lagrangian submanifold, one might have $u\in H^0(L, K_L^{1/2})$, $\eta\in H^1(L,K_L^{1/2})$,
with both of them annihilated by $\bar\partial$; but higher order corrections in $\sigma$-model perturbation theory might lead to $Qu=\epsilon \eta$
(with $\epsilon$ a small parameter), removing both $u$ and $\eta$ from the cohomology of $Q$.   This sort of correction is possible only if the cohomology
$H^i(L,K_L^{1/2})$ is nonzero for two adjacent integers $i,i+1$.   

A simple condition that ensures that there is no correction in that sense is that $H^i(L,\L^{-1}\otimes K_L^{1/2})=0$ for $i>0$.    This is actually so in many of the
examples considered in this article.   In any such case, $\Hom(\B,\B_\cc)$ is in some sense uncorrected.

However, it is much more interesting when one can say in a more precise sense that $\Hom(\B,\B_\cc)$ is uncorrected.    Before explaining this, we will
first point out two important classes of examples in which one can assume that $\L^{-1}\otimes K_L^{1/2}$ is trivial, so that the lowest order answer reduces
to $H^i(L,\O)$.   Further we will specialize to the case that $H^i(L,\O)=0$ for $i>0$, so that the lowest order answer actually reduces to $H^0(L,\O)$, that
is, the ring of holomorphic functions on $L$.   

For a first class of examples that satisfies these criteria, suppose that $L\cong \C^n$ is a fiber of a holomorphic polarization of some complex symplectic manifold
$Y$.   Then $K_L$ is trivial,
and as $L$ is simply-connected, the flat line bundle $\L$ and the square root $K_L^{1/2}$ are trivial.  Finally, it is also true that $H^i(\C^n,\O)=0$ for $i>0$.

For a second class of examples, suppose that $L\subset Y\times Y'$ is a correspondence associated to a holomorphic symplectomorphism of $Y$.
Then $L$ as a complex manifold is isomorphic to $Y$, so $K_L$ is trivial, and we can choose $\L$ and $K_L^{1/2}$ to be  also
trivial.   Finally, if $H^i(Y,\O)=0$ (which is the case for many interesting complex symplectic manifolds, though certainly not all), $L$ will have the same property.

In either of these cases, $\Hom(\B,\B_\cc)$ reduces in the leading approximation to the space of holomorphic functions on $L$.    This space contains
a distinguished one-dimensional subspace consisting of functions that do not grow at infinity -- constant functions.    In our discussion in the text of
these examples, corners associated to a constant function played an important role.   So let us discuss carefully what is involved, beyond the assumptions
that we have made so far, in ensuring that there is a canonical element of $\Hom(\B,\B_\cc)$ associated to the constant function 1.

Let us consider what happens in $\sigma$-model perturbation theory, which we will assume is governed by a small parameter $\veps$ (if one rescales the metric
of $Y$ by a constant $t$, then one can take $\veps=1/t$).     In lowest order in $\veps$, $Q$ reduces to $\bar\partial$, which annihilates a constant function.   Suppose
that in order $\veps^k$ (for some $k>0$), one runs into a first nonzero contribution.   This would take the form
\be\label{firstcor} Q\cdot 1 = \alpha ,\ee
where $\alpha$ is a $(0,1)$-form on $L$ that is of order $\veps^k$.
    Since $Q^2=0$, we will have $Q\alpha=0$, a condition that in order $\veps^k$ just reduces to   $\bar\partial\alpha=0$.   To compensate for this effect,
   we choose $w$ such that 
    \be\label{newcor}\bar\partial w =-\alpha .\ee
    Such a $w$ exists, because $\alpha$ is $\bar\partial$-closed and $H^1(L,\O)=0$.     After picking such
    a $w$, we replace 1 with $\psi_1=1+w$ and we see that $Q\psi_1$ vanishes up to order $\veps^{k+1}$.    In higher orders, we may find that $Q\psi_1$ is nonzero
    in order $\veps^r$ for some $r>k$.   If that happens, we follow the same procedure again and make a further correction to $\psi_1$.   
    
    At each stage of this procedure, we have to solve an equation of the general form $\bar\partial w =-\alpha$.  Such a $w$ will always exist (since $\alpha$
    will always be $\bar\partial$-closed) and $w$ is uniquely determined up to the possibility of adding to it a holomorphic function.   
    If it is always possible to pick $w$ in a canonical
    way, then this procedure will give a distinguished $\psi_1$ associated to the constant function 1.   So it will not be true that the constant function 1 ``is'' an
    element of $\Hom(\B,\B_\cc)$, but it will be true that there is an element of $\Hom(\B,\B_\cc)$ that is canonically associated to the constant function 1.
    That statement is what one wants to know in practice for applications.  
    
    How would we specify a canonical solution of the equation $\bar\partial w=-\alpha$?  The obvious way to do this is to ask for $w$ to vanish at infinity along $L$.
    If the equation has a solution such that $w$ vanishes at infinity, then this solution is unique, since any other solution would be obtained by adding to $w$
    a holomorphic function.   If $\alpha$ vanishes fast enough at infinity, then the equation does have a solution that vanishes at infinity.   For example,
    for $L=\C^n$ parametrized by $z_1,\cdots, z_n$ with metric $\sum_{k=1}^n |\d z_k|^2$, if $\alpha$ vanishes at infinity faster than $1/|z|$, then the
    equation for $w$ has a solution that vanishes at infinity.
    
   As an important example of a nice situation, suppose that $Y$ and $L$ are asymptotically conical.   This means in particular that their curvatures
   vanish at infinity.   Since $\sigma$-model corrections to the leading order answer involve the intrinsic and extrinsic curvature of $L$ and the curvature of $Y$,
   in such a case $\alpha$ will vanish at infinity sufficiently rapidly and $w$ can be assumed to vanish at infinity.
It is reasonable to expect the conclusion of the above analysis to hold as an exact statement, though the argument was phrased in perturbation theory.

   We will conclude with a couple of simple examples to illustrate potential pitfalls.
     For our first example, let $Y=\C^2$, parametrized by $p, q$, with symplectic form $\d p\d q$ and flat
   hyper-Kahler metric $|\d p|^2+|\d q|^2$.   Let $Y'$ be a second copy of $Y$ parametrized by $p',q'$, and consider the Lagrangian correspondence
   $L$ defined by
   \begin{align}\label{cordef} p'& =p+q^n \cr q'&=q. \end{align}
   The metric of $Y\times Y'$ is $|\d p|^2+|\d q|^2+|\d p'|^2+|\d q'|^2$.   After setting $q'=q$, the metric of $L$ is $|\d p|^2+|\d p'|^2+2|\d q|^2$, with $p,p',q$
   related by $p'=p+q^n$.    Let $p_\pm =(p'\pm p)/\sqrt 2$.   We then have $L=\C\times L'$, where $\C$ is parametrized by $p_+$ with flat metric 
   $|\d p_+|^2$,  and $L'$ is parametrized by $p_-$, $q$, related by
   \be\label{fleq} q^n=\sqrt 2 p_-,\ee
   and with the metric $|\d p_-|^2+2|\d q|^2$.   Since $Y\times Y'$ and $L$ have an exact symmetry that shifts $p_+$ by a constant, the first factor of
   $L=\C\times L'$ decouples from $\sigma$-model perturbation theory, and in the preceeding discussion, the equation $\bar\partial w = -\alpha$ can be understood
   as an equation on $L'$.    $L'$ only has one infinite ``end,'' with $p_-,q\to\infty$.  This region is conveniently parametrized by $p_-$, with $q\sim p_-^{1/n}$, and the form of eqn. (\ref{fleq}) and of the metric
   of $L'$ show that $L'$ is asymptotically flat, with curvature vanishing as $1/|p_-|^{4-2/n}$.    So there will always be a solution with $w$ vanishing
   at infinity, and this is an example in which there should exist an
   element of $\Hom(\B,\B_\cc)$ canonically associated to the constant function 1.
   
   The generators of the group of tame symplectomorphisms of $\C^{2n}$  (in language explained in Section \ref{oversym}) can be treated similarly for any $n$.   However, let us consider a hypothetical
   non-tame symplectomorphism of $\C^{2n}$.   Parametrizing $\C^{2n}$ by coordinates $x^1,x^2,\cdots, x^{2n}$,   a correspondence $L$ associated to
   a symplectomorphism of $\C^{2n}$ is in general described by equations
   \begin{align}\label{expfno} x'{}^{i}&=f^i(x^k) \cr
 x^j&=k^j(x'{}^{k})\end{align}
 For the case of a general non-tame symplectomorphism, with little known about the functions $f^i$ and the inverse functions $k^j$, it is not
 clear how to describe the region at infinity in $L$ and how to argue that the equation $\bar\partial w=-\alpha$ will have a solution that vanishes at infinity.   Some
 insight about this might have made possible a stronger conclusion in our discussion of symplectomorphisms of $\C^{2n}$ in Section \ref{symmeq}.

 \section{The Opposite Algebra}\label{opposite}
 
 In Section \ref{lagbr},   we explained that if $W$ is a complex manifold with canonical bundle $K$, and $Y=T^*W$ with the standard complex symplectic form 
 $\Omega=\sum_i\d p_i \d q^i$, then  $\A=\Hom(\B_\cc,\B_\cc)$ is the algebra of differential operators acting on sections of $K^{1/2}$.
 This statement also holds after restricting to an open set in $W$.

Here we will explain this along more standard mathematical lines.   
The following  is somewhat similar to the construction of a hermitian inner product in Section \ref{hermitian}, but we use a linear operation rather
than an antilinear one.

If $\A$ is an algebra, the opposite algebra $\A^\op$ is defined as follows: elements $a^\op$ of $\A^\op$ are in one-to-one correspondence with elements $a$ of $\A$,
but they are multiplied in the opposite order (so $a^\op b^\op=(ba)^\op$).  
In the case of deformation quantization of $Y=T^*W$, the algebra $\A=\Hom(\B_\cc,\B_\cc)$ is isomorphic to its own opposite.   This may be seen as follows.
$Y$ has a holomorphic involution $\tau$ that maps $(p,q)$ to $(-p,q)$, and therefore satisfies  $\tau^*(\Omega)=-\Omega$.   In the $A$-model, one can compensate
for reversing the sign of $\Omega$ by reversing the orientation of the two-manifold $\Sigma$ on which the model is defined.   Hence the $A$-model on
$\Sigma$ is invariant under a diffeomorphism of $\Sigma$ that reverses its orientation, together with the action of $\tau$ on the target space.
   Such a diffeomorphism reverses the orientation of $\partial\Sigma$,
and hence reverses the order in which operators are inserted on $\partial\Sigma$.   But reversing the order of operator insertions on the boundary is
equivalent to exchanging $\A$ with $\A^\op$.   The conclusion is that the algebra $\A$ is isomorphic to its opposite algebra $\A^\op$ by an automorphism
that at the classical level acts by $(p,q)\to (-p^\op,q^\op)$.

On the other hand, let $\DD$ be the algebra of differential operators acting on sections of a holomorphic line bundle $\L$ (or possibly a complex power of line
bundles).   As we explain momentarily, the opposite algebra $\DD^\op$ is the algebra of differential operators acting on $K\otimes \L^{-1}$.      Hence for
$\DD$ to be isomorphic to $\DD^\op$ means that $\L$ is isomorphic to $K\otimes \L^{-1}$, in other words $\L=K^{1/2}$.   Hence, if $\A $ is  isomorphic to $\A^\op$,
it must be the algebra of differential operators acting on sections of $K^{1/2}$.

The assertion about $\DD^\op$ is proved by a formal integration by parts.   We recall that on a complex manifold, the exterior derivative has an expansion
$\d=\partial+\bar\partial$, where $\partial$ shifts the degree of a differential form by $(1,0)$.  The canonical bundle $K$ is the same as the bundle of
$(n,0)$-forms.   For $\lambda$ an $(n-1,0)$-form, $\partial\lambda$ is an $(n,0)$-form, that is, a section of $K$.  If $D$ is a differential operator acting on
sections of $\L$, its ``transpose'' $D^\tr$ is a differential operator acting on sections of $K\otimes \L^{-1}$ that is defined as follows.   Let $g$ be a local section
of $\L$ and $f$ a local section of $K\otimes \L^{-1}$.   Then $D^\tr$ is defined by the condition that for any such $f,g$,
\be\label{kilox}f Dg = (D^\tr f) g +\partial \lambda,\ee
where $\lambda$ is is an $(n-1,0)$-form that is 
constructed locally from $f,g$, and their derivatives.   From this it follows immediately that if $D_1, D_2$ are two differential operators
acting on sections of $\L$, then $(D_1 D_2)^\tr=D_2^\tr D_1^\tr$.    Since the order of multiplication is reversed, this shows that the association $D\to D^\tr$
maps $\DD$ to its opposite algebra $\DD^\op$.   Hence if $\DD$ is the algebra of differential operators acting on sections of $\L$, then $\DD^\op$ is the algebra of
differential operators acting on sections of $K\otimes \L^{-1}$.

As an example, suppose that $W$ is a Riemann surface, $\L$ is the trivial line bundle $\O$, and $D$ is a differential operator that can be written
locally as $D=a(z)\partial_z$.   Let $g$ be a local section of $\O$ and $f$ a local section of $K$.   The identity
\be\label{zor}f a\partial_z g =\left(-\partial_z(af)\right) g +\partial_z\left(fag\right) \ee
shows that $D^\tr$ can be defined locally by $D^\tr f=-\partial_z(a(z)f(z))$.   Here the minus sign reflects the fact that at the classical level,
the map from $\DD$ to $\DD^\op$ comes from an involution that acts by $(p,q)\to (-p,q)$.

\section{Anomalies and $\B_\cc$}\label{anombcc}

In lowest order of $\sigma$-model perturbation theory, to
 every Lagrangian submanifold $L$ of a symplectic manifold $Y$, endowed with a flat $\spinc$ bundle, one can associate
an $A$-brane $\B_L$.   At the quantum level, however, disc instanton effects  obstruct the existence of some of these branes.

The analogous question for coisotropic branes has apparently not been considered in the physics literature, though a related question
concerning deformation quantization of algebraic varieties has been studied mathematically  \cite{bezk}.    
Rank 1 coisotropic branes were constructed by Kapustin and Orlov in lowest order of $\sigma$-model perturbation
theory \cite{KO}.  Might there be an anomaly that obstructs the existence of some such branes in higher orders?

Here we will consider this question in the basic example of a coisotropic brane $\B_\cc$ whose support is all of the $A$-model target space $Y$.
In particular, $Y$ will be a complex symplectic manifold with $\omega_Y=\Im\,\Omega$.    A complex symplectic manifold has vanishing first Chern class,
so the usual fermion number anomaly (which is important in analyzing disc instanton obstructions to Lagrangian branes) is absent and we can use fermion
number conservation to constrain the analysis.   Though we only consider coisotropic branes whose support is all of $Y$, it is worth mentioning that at least for
branes of rank 1 (the only case in which coisotropic branes are well-understood), the general case is a mixture of the case we consider and the more familiar  case of a
Lagrangian brane.    That is because a general coisotropic brane of rank 1 looks like $\B_\cc$ in some directions and like a Lagrangian $A$-brane in other directions.

If $\B_\cc$ is a valid $A$-brane, this means that the BRST operator $Q$ of the theory satisfies $Q^2=0$ even in the presence of a boundary labeled by $\B_\cc$.
The alternative is that there might be a boundary contribution to $Q^2$; though $Q^2=0$ in bulk, we might have 
\be\label{unu} Q^2 =F(0), \ee
where $F(0)$ is a local operator inserted at an endpoint where a string ends on $\B_\cc$.   There might also be a contribution at the second endpoint of the
string, but we can focus on the contribution at one endpoint.  The operator $F(0)$ will have conformal dimension 0 in the classical limit (if the model
is not conformally invariant at the quantum level, this scaling might be modified by logarithms in $\sigma$-model perturbation theory).   The operators
with appropriate scaling have the form $F(X,\bar X,\rho)$, in the notation of appendix \ref{details} (here the fields $X,\bar X,\rho$ are to be evaluated at the string
endpoint -- the point ``0'' in eqn. (\ref{unu})).     As noted in the discussion of eqns. (\ref{btrans}) and (\ref{bcorr}),
these operators correspond to $(0,q)$-forms on $Y$, for $q=0,\cdots, \dim_\C Y$, and 
$Q$ acts on them as the usual $\bar\partial$
operator mapping $(0,q)$-forms to $(0,q+1)$-forms.   The condition that $F(0)$ has fermion number 2 means that it is a $(0,2)$-form, and the condition
$Q^2=F(0)$ implies that $[Q,F(0)]=[Q,Q^2]=0$, so $F$ is a $\bar\partial$-closed $(0,2)$-form.    If $F$ is $\bar\partial$-exact, $F=\bar\partial\lambda$ 
for a $(0,1)$-form $\lambda$, then the anomaly can be removed by shifting $Q\to Q-\lambda(0)$.    Thus the potential anomalies that cannot
be removed by a redefining $Q$ correspond to elements of $H^2(Y,\O)$.

Bezrukavnikov and Kaledin, however, show \cite{bezk} (in a different language) that the anomaly associated to $\beta\in H^2(Y,\O)$ can be removed if
there is a closed two-form $b$ on $Y$ such that the part of $b$ of type $(0,2)$, which we denote as $b^{(0,2)}$,
 is equal to $\beta$.   To understand this in our language, we recall that as well as a symplectic form $\omega_Y
=\Im\,\Omega$, the $A$-model under study has a $B$-field $\sB=\Re\,\Omega$.   The reason for this particular choice is that Kapustin and Orlov \cite{KO},
in verifying that the equation $Q^2=0$ receives no boundary contribution, ran into
the condition $(\omega_Y^{-1}(\sF+\sB))^2=-1$, which can be satisfied by $\sB=\Re\,\Omega$, $\sF=0$.   If one shifts $\sB$ to $\sB+b$, for a closed two-form $b$,
then the Kapustin-Orlov condition is violated and $Q^2$ receives a boundary contribution proportional to $b^{(0,2)}$.   But that also means
that an anomaly proportional to $b^{(0,2)}$ can be canceled by shifting $\sB$.

If $Y$ is compact, then by Hodge theory, every class in $H^2(Y,\O)$ has a harmonic representative.  This representative is a closed two-form, implying
that any potential anomaly can always be canceled by shifting $\sB$.  Usually, for quantization one is not interested in the case
that $Y$ is compact, but the argument just stated probably has an analog for a suitable class of non-compact $Y$'s.
    If $Y$ has, for example, a complete hyper-Kahler metric that is asymptotically
conical, then one can hope to prove that any anomaly is cohomologous to a $(0,2)$-form of compact support, and then use a version of Hodge theory
for complete Riemannian manifolds to argue that the anomaly can be eliminated.   However, we will not make any precise argument along these lines.
Most examples studied in the present article  have $H^{0,2}(Y)=0$. The  cotangent bundles $Y=T^*W$ can be exceptions.  
For these examples, it is possible to use the scaling symmetry of the cotangent bundle to show that there is no anomaly.

Bezrukavnikov and Kaledin parametrize the possible deformation quantizations of the sheaf of holomorphic functions on 
$Y$ by what they call a period map, a sort of quantum version of the cohomology
class $[\Omega]$ of $\Omega$.  In doing so, one runs into the fact that $Y$ might have deformations as a complex symplectic manifold that do not change
 $[\Omega]$.   The presence of such deformations complicates the discussion of a period map.  
 
 From our point of view, deformations of $Y$ as a complex symplectic manifold that do not change $[\Omega]$ have the following significance.  Suppose that the
complex structure and complex symplectic form $I,\Omega$ of $Y$ can be deformed to another complex structure and complex symplectic form $I',\Omega'$, such
that $[\Omega]=[\Omega']$.   By applying a suitable diffeomorphism to the pair $I',\Omega'$, one can reduce to the case\footnote{In general, if a real symplectic
form $[\omega]$ is varied in a continuous family keeping $[\omega]$ fixed, then the symplectic forms in this family are all equivalent up to diffeomorphism.
To see this, one observes that a first order variation of $\omega$ keeping $[\omega]$ fixed is of the form $\delta\omega=\d \gamma$ for some 1-form $\gamma$.
This first-order deformation of $\omega$ can be canceled by an infinitesimal diffeomorphism generated by the vector field $V=\omega^{-1}\gamma$.}
  $\Im\,\Omega=\Im\,\Omega'$.
One cannot simultaneously require $\Re\,\Omega=\Re\,\Omega'$, but since $[\Re\,\Omega]=[\Re\,\Omega']$, one does have $\Re\,\Omega=\Re\,\Omega'-\d\lambda$
for some 1-form $\lambda$.    We can regard $\sA=\lambda$ as a connection form on a trivial line bundle $\L$; 
the curvature of $\L$ is
$\sF=\d\lambda$.
In addition to the familiar solution of the Kapustin-Orlov condition $(\omega_Y^{-1}(\sF+\sB))^2=-1$ with
$\omega_Y=\Im\,\Omega$, $\sB=\Re\,\Omega$, $\sF=0$, we have another solution with the same $\omega_Y$ and $\sB$ but $\sF=\d\lambda$ (so $\sF+\sB=
\mathrm{Re}\,\Omega'$).  
Therefore, in addition to the familiar coisotropic brane $\B_\cc$ with trivial, flat
$\CP$ bundle,  the $A$-model with $\omega_Y=\Im\,\Omega$ and $\sB=\Re\,\Omega$ 
has a new and equally good coisotropic brane $\B_\cc'$ with $\CP$ bundle $\L$.   More generally, if the complex symplectic structure
$I,\Omega$  that leads to the existence of the brane $\B_\cc$ can be varied in a family while keeping $[\Omega]$ fixed, then the $A$-model with
fixed $\omega_Y$ and $\sB$ has a family of coistropic branes labeled by the same data.  (This is a classical description; the anomaly that can be potentially
canceled by shifting $\sB$ may complicate the picture in some cases.)   

The possibility of deforming the pair $I,\Omega$ while keeping $[\Omega]$ fixed can be described as follows.   Consider the exact sequence of sheaves
\be\label{tolimo} 0 \to \C\to \O\to \O/\C\to 0,\ee
where $\C$ is the sheaf whose local sections are complex constants.   Associated to this short exact sequence is an exact sequence in cohomology that reads in part
\be\label{polimo} \cdots\to H^1(Y,\C)\stackrel{j}{\to} H^1(Y,\O)\to H^1(Y,\O/\C)\stackrel{\delta}{\to} H^2(Y,\C)\stackrel{j}{\to} H^2(Y,\O) \to \cdots.\ee
We can interpret $\O$ as the sheaf of (holomorphic) Hamiltonian functions on $Y$, and $\O/\C$ as the sheaf of holomorphic functions mod constants.
Dividing by constants is appropriate for analyzing the deformations of $Y$ as a complex symplectic manifold, because the Hamiltonian vector field
$\Omega^{-1}\d f$ associated to a Hamiltonian function $f$ is invariant under shifting $f$ by a constant.   Accordingly, first order deformations of $Y$ as a 
complex symplectic manifold are classified by $H^1(Y,\O/\C)$.   For a deformation corresponding to $x\in H^1(Y,\O/\C)$, the corresponding deformation of
$[\Omega]$ is $[\Omega]\to [\Omega]+\delta(x)$.   Thus the deformations of $Y$ as a complex symplectic manifold without changing $[\Omega]$ are
classified by the kernel of $\delta$.   Exactness of the sequence (\ref{polimo}) means that the kernel of $\delta$ is the
same as the quotient $H^1(Y,\O)/j (H^1(Y,\C))$, which parametrizes $\bar\partial$ closed $(0,1)$ forms  modulo those of the form $c^{(0,1)}$ for
a closed 1-form $c$.   That quotient parametrizes the family of  first order deformations of $\B_\cc$ as a coisotropic $A$-brane, 
in a fixed $A$-model with no change in $\omega_Y$ or $\sB$.

\bibliographystyle{unsrt}

\begin{thebibliography}{99}



\bibitem{G46} H. G. Groenewald, ``On The Principles of Elementary Quantum Mechanics,''  Physica {\bf 12} (1946) 405-460.

\bibitem{V51} L. Van Hove,  ``Sur les probl\`{e}mes des relations entre les transformations
unitaires de la m\'{e}canique quantique et les transformations canoniques
de la mecanique classique,'' Acad. Roy. Belgique Bull. Cl. Sci {\bf 37} (1951) 610-20.

\bibitem{Todorov}I. Todorov, ``Quantization is a Mystery,'' arXiv:1206.3116.

\bibitem{Ko} B. Kostant, ``Quantization and Unitary Representations,''
Lect. Mod. An. Appl. {\bf III} (1970) 87-208.

\bibitem{So} J.-M. Souriau,  ``Quantification Geometrique,'' Commun. Math. Phys.
{\bf 1} (1966) 374.

\bibitem{Wo} N. Woodhouse, {\it Geometric Quantization} (Oxford University Press, Oxford, 1980).

\bibitem{Snia}J. \'{S}niatycki, {\it Geometric Quantization And Quantum Mechanics} (Springer-Verlag, New York, 1980).

\bibitem{Hall}Brian C. Hall, {\it Quantum Theory For Mathematicians} (Springer, New York, 2013).


\bibitem{GW}S. Gukov and E. Witten, ``Branes and Quantization,'' ATMP {\bf 13} (2009) 1445-1518, arXiv:0809.0305.

\bibitem{KO}A. Kapustin and D. Orlov, ``Remarks on A-branes, Mirror Symmetry, and the Fukaya
Category,'' J. Geom. Phys. {\bf 48} 84 (2003), arXiv:hep-th/0109098.

\bibitem{BrS} P. Bressler and Y. Soibelman, ``Mirror Symmetry And Deformation Quantization,''
arXiv:hep-th/0202128.



\bibitem{Ka}  A. Kapustin, ``A-Branes And Noncommutative Geometry,''  arXiv:hep-th/0502212.

\bibitem{Pe}
V. Pestun, ``Topological Strings In Generalized Complex Space,'' arXiv:hep-th/0512189.

\bibitem{Gu} M. Gualtieri, ``Branes On Poisson Varieties,'' aarXiv:0710.2719.

\bibitem{AZ}M. Aldi and E. Zaslow, ``Coisotropic Branes, Noncommutativity, and the Mirror Correspondence,''
JHEP {\bf 0506} (2005) 019, arXiv:hep-th/0501247.


\bibitem{Many}
F. Bayen, M. Flato, C. Fronsdal, A. Lichnerowicz, and D. Sternheimer, ``Deformation Theory and Quantization,''
Ann. Phys. {\bf 111} (1978) 61-151.

\bibitem{WL}
M. De Wilde and P. B. A. Lecomte, ``Existence of Star-Product and of Formal Deformations In Poisson
Lie Algebra of Arbitrary Symplectic Manifold,'' Lett. Math. Phys. {\bf 7} (1983) 487-96.

\bibitem{Fedosov}
B. V. Fedosov, ``A Simple Geometrical Construction Of Deformation Quantization,'' J. Diff. Geom. {\bf 40} (1994) 213-38.



\bibitem{Kon}M. Kontsevich, ``Deformation Quantization of Poisson Manifolds, I,''    
Lett. Math. Phys.  {\bf 66} (2003) 157-216,
arxiv.org/abs/q-alg/9709040.

\bibitem{CF}
A. S. Cattaneo and G. Felder, 		``A Path Integral Approach to the Kontsevich Quantization Formula,''
math.QA/9902090, Commun. Math. Phys. {\bf 212} (2000) 591-612.

\bibitem{CF2}
A. S. Cattaneo and G. Felder, ``Poisson Sigma Models and Deformation Quantization,''  Mod. Phys. Lett. {\bf A16} (2001) 
179-190, arXiv:hep-th/0102208.

\bibitem{SL}
R. E. Grady, Qin Li, and Si Li, ``Batalin-Vilkovisky Quantization And The
Algebraic Index,'' arXiv:1507.01812.


\bibitem{KW}A. Kapustin and E. Witten, 
``Electric-Magnetic Duality and the Geometric Langlands Program,''   Commun. Num. Theor. Phys.
{\bf 1} (2007) 1-236,
arXiv:hep-th/0604151.

\bibitem{Herbst}
M. Herbst, ``On Higher Rank Coisotropic $A$-Branes,'' J. Geom. Phys. {\bf 62} (2012) 156-69, arXiv:1003.3771.

\bibitem{EFK}
P. etingof, E. Frenkel, and D. Kazhdan, ``An Analytic Version Of The Langlands Correspondence For Complex Curves,'' 
arXiv:1908.09677.

\bibitem{EFK2}
P. etingof, E. Frenkel, and D. Kazhdan, ``Hecke Operators and Analytic Langlands Correspondence For Curves Over Local Fields,''
arXiv:2103.01509.

\bibitem{EFK3}
P. etingof, E. Frenkel, and D. Kazhdan, ``Analytic Langlands Correspondence for $\mathrm{PGL}_2$ on ${\Bbb P}^1$ With Parabolic
Structure Over Local Fields,'' arXiv:2106.05243.


\bibitem{T}
J. Teschner, ``Quantization Conditions Of The Quantum Hitchin System and the Real Geometric Langlands Correspondence,'' in J. E. Dancer 
et. al., eds., {\it Geometry and Physics}, vol. 1 (Oxford University Press, 2018),
arXiv:1707.07873.



\bibitem{GaW}
D. Gaiotto and E. Witten, ``Gauge Theory and the Analytic Form of the Geometric Langlands Program,'' arXiv:2107.01732.

\bibitem{Gotay}
M. J. Gotay, ``A Class Of Non-Polarizable Symplectic Manifolds,'' Mh. Math.{\bf 103} (1987) 27-30.

\bibitem{FW} D. S. Freed and E. Witten, ``Anomalies In String Theory With $D$-Branes,'' 
Asian  J .Math. {\bf 3} (1999) 819-852,
arXiv:hep-th/9907189.

\bibitem{W}
E. Witten, ``A New Look At The Path Integral of Quantum Mechanics,''  Surv. Diff. Geom. {\bf 15} (2010) 345-420, arXiv:1009.6032.





\bibitem{K2} V. G. Knizhnik and A. B. Zamolodchikov, ``Current Algebra and Wess-Zumino Model In Two Dimensions,'' Nucl. Phys. {\bf B247} (1984) 83-103.

\bibitem{TY}
G. Tian and S.-T. Yau, ``Complete Kahler Manifolds With Zero Ricci Curvature, I,'' J. Amer. Math. Soc. {\bf 3} (1990) 578-609.




\bibitem{ABott}
M. F. Atiyah and R. Bott, ``The Yang-Mills Equations Over Riemann Surfaces,'' Phil. Trans. R. Soc. Lond.
{\bf A308} (1983) 523-615.

\bibitem{GR}
D. Gaiotto and M. Rap$\check{{\mathrm c}}$\'{a}k, ``Vertex Algebras at the Corner,'' JHEP {\bf 01} (2019) (160),
arXiv:1703.00982.

\bibitem{Ginzburg}
V. Ginzburg, ``Lectures on ${\mathcal D}$-Modules,'' available at \url{http://people.math.harvard.edu/~gaitsgde/grad_2009/Ginzburg.pdf}.
\bibitem{Hitchin}
N. Hitchin, ``The Self-Duality Equations On A Riemann Surface,'' Proc. London Math. Soc.  {\bf 55}  (1987) 59-126.

\bibitem{WittenJ}
E. Witten, ``Quantum Field Theory and the Jones Polynomial,'' Commun. Math. Phys. {\bf 121}   (1989) 351.

\bibitem{KZ}
V. G. Knizhnik and A. B. Zamolodchikov, ``Current Algebra And Wess-Zumino
Model In Two Dimensions,'' Nucl. Phys. {\bf B247} (1984) 83-103.

\bibitem{ADW}
S. Axelrod, S. DellaPietra, and E. Witten, ``Geometric Quantization Of Chern-Simons Gauge Theory,'' J. Diff. Geom. {\bf 33} (1991) 787.

\bibitem{Hitchtwo}
N. Hitchin, ``Flat Connections And Geometric Quantization,'' Commun. Math. Phys.
{\bf 131}  (1990) 347-380.

\bibitem{Faltings}
G. Faltings, ``Stable $G$-Bundles and Projective Connections,'' J. Alg. Geom. {\bf 2} (1993) 507-68.

\bibitem{And}
J.  E. Andersen, ``Hitchin's Connection, Toeplitz Operators, and Symmetry Invariant Deformation Quantization,''  Quantum Topol. {\bf 3} (2012) 293-325.

\bibitem{Las}Y. Laszlo, ``Hitchin's and WZW Connections Are The Same,'' J. Diff. Geom. {\bf 49} (1998) 547-76.

\bibitem{El}S. Elitzur, G. Moore, A. Schwimmer, and N. Seiberg,
``Remarks On The Canonical Quantization of the Chern-Simons-Witten
Theory,'' Nucl. Phys. {\bf B326} (1989) 108-34.

\bibitem{Gaw}K. Gawedzki, ``Non-Compact WZW Conformal Field Theories,'' arXiv:hep-th/9110076.


\bibitem{Gaw2}K. Gawedzki, ``$\SU(2)$ WZNW Path Integral At Higher Genera From Gauge Field Functional Integral,'' Algebra i Analiz. {\bf 6} (1994) 107-17,
arXiv:hep-th/9312051.

\bibitem{And2}
J. E. Andersen and K. Ueno, ``Construction of the Witten-Reshetikhin-Turaev TQFT From Conformal Field Theory,'' Inv. Math. {\bf 201}
(2015) 519-559, arXiv:1110.5027.

\bibitem{And3}
J. E. Andersen, ``Mapping Class Group Invariant Unitarity of the Hitchin Connection Over Teichm\"{u}ller Space,''
arXiv;1206.2635.

\bibitem{And4}
J. E. Andersen and N. S. Poulsen, ``An Explicit Ricci Potential For The Universal Moduli Space of Vector Bundles,'' arXiv:1609.01242.


\bibitem{BKK}
A. Kanel-Belov and M. Kontsevich, ``Automorphisms of the Weyl Algebra,'' Lett. Math. Phys. {\bf 74} (2005) 181-99, arXiv:math/0512169.

\bibitem{BEY}
A. Kanel-Belov, A. Elishev, and J.-T. Yu, ``Automorphisms of Weyl Algebra and a Conjecture of Kontsevich,'' arXiv:1802.01225.

\bibitem{bezk}
R. Bezrukavnikov and D. Kaledin, ``Fedosov Quantization In An Algebraic Context,''  Mosc. Math. J. {\bf 4} (2004) 559-92, arXiv:math/0309290.

\bibitem{ACNY}
A Abouelsaood, C. G. Callan, Jr., C. R. Nappi, and S. A. Yost, ``Open Strings In Background Fields,'' Nucl. Phys. {\bf B280} (1987) 599-624.

\end{thebibliography}

\end{document}